\documentstyle[12pt]{article}
\def\hybrid{\topmargin -20pt    \oddsidemargin 0pt
        \headheight 0pt \headsep 0pt
        \textwidth 6.25in       % A4 paper
        \textheight 9.5in       % A4 paper
        \marginparwidth .875in
        \parskip 5pt plus 1pt   \jot = 1.5ex}

\hybrid

\def\baselinestretch{1.2}

\catcode`\@=11

\def\marginnote#1{}
\newcount\hour
\newcount\minute
\newtoks\amorpm
\hour=\time\divide\hour by60
\minute=\time{\multiply\hour by60 \global\advance\minute by-\hour}
\edef\standardtime{{\ifnum\hour<12 \global\amorpm={am}%
        \else\global\amorpm={pm}\advance\hour by-12 \fi
        \ifnum\hour=0 \hour=12 \fi
        \number\hour:\ifnum\minute<10 0\fi\number\minute\the\amorpm}}
\edef\militarytime{\number\hour:\ifnum\minute<10 0\fi\number\minute}
\def\draftlabel#1{{\@bsphack\if@filesw {\let\thepage\relax
   \xdef\@gtempa{\write\@auxout{\string
      \newlabel{#1}{{\@currentlabel}{\thepage}}}}}\@gtempa
   \if@nobreak \ifvmode\nobreak\fi\fi\fi\@esphack}
        \gdef\@eqnlabel{#1}}
\def\@eqnlabel{}
\def\@vacuum{}
\def\draftmarginnote#1{\marginpar{\raggedright\scriptsize\tt#1}}

\def\draft{\oddsidemargin -.5truein
        \def\@oddfoot{\sl preliminary draft \hfil
        \rm\thepage\hfil\sl\today\quad\militarytime}
        \let\@evenfoot\@oddfoot \overfullrule 3pt
        \let\label=\draftlabel
        \let\marginnote=\draftmarginnote
   \def\@eqnnum{(\theequation)\rlap{\kern\marginparsep\tt\@eqnlabel}%
\global\let\@eqnlabel\@vacuum}  }

\def\preprint{\twocolumn\sloppy\flushbottom\parindent 2em
        \leftmargini 2em\leftmarginv .5em\leftmarginvi .5em
        \oddsidemargin -.5in    \evensidemargin -.5in
        \columnsep .4in \footheight 0pt
        \textwidth 10.in        \topmargin  -.4in
        \headheight 12pt \topskip .4in
        \textheight 6.9in \footskip 0pt
        \def\@oddhead{\thepage\hfil\addtocounter{page}{1}\thepage}
        \let\@evenhead\@oddhead \def\@oddfoot{} \def\@evenfoot{} }

\def\numberbysection{\@addtoreset{equation}{section}
        \def\theequation{\thesection.\arabic{equation}}}

\def\underline#1{\relax\ifmmode\@@underline#1\else
        $\@@underline{\hbox{#1}}$\relax\fi}

\def\titlepage{\@restonecolfalse\if@twocolumn\@restonecoltrue\onecolumn
     \else \newpage \fi \thispagestyle{empty}\c@page\z@
        \def\thefootnote{\fnsymbol{footnote}} }

\def\endtitlepage{\if@restonecol\twocolumn \else \newpage \fi
        \def\thefootnote{\arabic{footnote}}
        \setcounter{footnote}{0}}  %\c@footnote\z@ }

\catcode`@=12
\relax

\def\figcap{\section*{Figure Captions\markboth
        {FIGURECAPTIONS}{FIGURECAPTIONS}}\list
        {Figure \arabic{enumi}:\hfill}{\settowidth\labelwidth{Figure
999:}
        \leftmargin\labelwidth
        \advance\leftmargin\labelsep\usecounter{enumi}}}
 \relax
\def\tablecap{\section*{Table Captions\markboth
        {TABLECAPTIONS}{TABLECAPTIONS}}\list
        {Table \arabic{enumi}:\hfill}{\settowidth\labelwidth{Table
999:}
        \leftmargin\labelwidth
        \advance\leftmargin\labelsep\usecounter{enumi}}}
 \relax
\def\reflist{\section*{References\markboth
        {REFLIST}{REFLIST}}\list
        {[\arabic{enumi}]\hfill}{\settowidth\labelwidth{[999]}
        \leftmargin\labelwidth
        \advance\leftmargin\labelsep\usecounter{enumi}}}
 \relax
\makeatletter
\newcounter{pubctr}
\def\publist{\@ifnextchar[{\@publist}{\@@publist}}
\def\@publist[#1]{\list
        {[\arabic{pubctr}]\hfill}{\settowidth\labelwidth{[999]}
        \leftmargin\labelwidth
        \advance\leftmargin\labelsep
        \@nmbrlisttrue\def\@listctr{pubctr}
        \setcounter{pubctr}{#1}\addtocounter{pubctr}{-1}}}
\def\@@publist{\list
        {[\arabic{pubctr}]\hfill}{\settowidth\labelwidth{[999]}
        \leftmargin\labelwidth
        \advance\leftmargin\labelsep
        \@nmbrlisttrue\def\@listctr{pubctr}}}
 \relax
\makeatother
\newskip\humongous \humongous=0pt plus 1000pt minus 1000pt

\newif\ifdtup

\relax

\def\be{\begin{equation}}
\def\ee{\end{equation}}
\def\ba{\begin{eqnarray}}
\def\ea{\end{eqnarray}}

\def\no{\noindent}

\def\IR{\relax{\rm I\kern-.18em R}}

\def\IR{\relax{\rm I\kern-.18em R}}
\def\inv{^{\raise.15ex\hbox{${\scriptscriptstyle -}$}\kern-.05em 1}}

\begin{document}
%\draft

%\renewcommand{\theequation}{\arabic{equation}}
\renewcommand{\theequation}{\thesection.\arabic{equation}}

\newcommand{\beq}{\begin{equation}}
\newcommand{\eeq}[1]{\label{#1}\end{equation}}
\newcommand{\ber}{\begin{eqnarray}}
\newcommand{\eer}[1]{\label{#1}\end{eqnarray}}
\newcommand{\eqn}[1]{(\ref{#1})}
\begin{titlepage}
\begin{center}

\hfill hep--th/0403165\\
\vskip -.1 cm
\hfill March 2004\\

\vskip .6in

{\large \bf On the tensionless limit of gauged WZW models}

\vskip 0.6in

{\bf Ioannis Bakas} {\em and} {\bf Christos Sourdis}\footnote{Present address:
Department of Physics, Graduate School of Science, Osaka University, Toyonaka,
Osaka 560-0043, Japan;
e-mail: sourdis@het.phys.sci.osaka-u.ac.jp}
\vskip 0.2in
{\em Department of Physics, University of Patras \\
GR-26500 Patras, Greece\\
\footnotesize{\tt bakas@ajax.physics.upatras.gr \\ sourdis@pythagoras.physics.upatras.gr}}\\

\end{center}

\vskip .8in

\centerline{\bf Abstract}
\no
The tensionless limit of gauged WZW models arises when the level of the underlying
Kac-Moody algebra assumes its critical value, equal to the dual Coxeter number,
in which case the central charge
of the Virasoro algebra becomes infinite. We examine this limit from the
world-sheet and target space viewpoint and show that gravity decouples naturally
from the spectrum. Using the two-dimensional black-hole coset $SL(2,R)_k/U(1)$
as illustrative example, we find for $k=2$ that the world-sheet symmetry is
described by a truncated version of $W_{\infty}$ generated by chiral fields
with integer spin $s \geq 3$, whereas the Virasoro algebra becomes abelian
and it can be consistently factored out. The geometry of target space looks like
an infinitely curved hyperboloid, which invalidates the effective field theory
description and conformal invariance can no longer be used
to yield reliable space-time interpretation. We also compare our results with
the null gauging of WZW models, which correspond to infinite boost in target
space and they describe the Liouville mode that decouples in the tensionless limit.
A formal BRST analysis of the world-sheet symmetry suggests that the central
charge of all higher spin generators should be fixed to a critical value, which is
not seen by the contracted Virasoro symmetry. Generalizations to higher dimensional
coset models are also briefly discussed in the tensionless limit, where similar
observations are made.

\no

\vfill
\end{titlepage}
\eject

\def\baselinestretch{1.2}
\baselineskip 16 pt
\noindent

\section{Introduction}
\setcounter{equation}{0}

The tensionless limit of string theory is a very fascinating but
largely unexplored subject that was first introduced classically
in flat space by letting all points of the string move at the
speed of light, thus leading to the notion of null strings,
\cite{schild}. Aspects of their quantization were subsequently
studied, \cite{ulf1, ulf2}, and it was also found that the concept
of critical dimension does not arise in this case, as the Virasoro
algebra contracts to an abelian structure and the corresponding
BRST operator squares to zero, \cite{ouvry, lizzi} for all
dimensions of space-time; see also \cite{bonelli} for a more
recent discussion of the subject. It is a first indication that
the notion of space-time undergoes a drastic modification as on
passes from tensile to tensionless strings. This limit also arises
naturally in various attempts to formulate a sensible expansion of
string propagation in highly curved backgrounds, since strings
appear to behave as tensionless classical objects in the vicinity
of space-time singularities, \cite{vega}. Other classical
tensionless string models were introduced recently in terms of
geometric actions that are alternative to Polyakov's action, and
their quantization was investigated in connection with higher spin
fields, \cite{savvidy}. On the other hand, there is another
interesting approach to the tensionless limit of string theory,
which arises directly at the quantum level, and it was brought to
light by studying the high energy behavior of string scattering
amplitudes at the Planck scale, \cite{venezia, gross1}. Although
it is not known whether all these theories are equivalent to each
other, or whether they represent different corners of a more
general (yet unknown) framework, there is a common element that
makes tensionless strings special, namely that the Planck mass
becomes zero and all states turn massless as $\alpha^{\prime}
\rightarrow \infty$.

It has been suggested that the tensionless limit represents the
unbroken phase of string theory, where all states appear on equal
footing and they give rise to a huge symmetry group, which
subsequently breaks and gives masses to the string states at lower
energy scales, \cite{gross2}; see also \cite{ed1} for a more
recent discussion in terms of higher spin symmetries. As such, it
could be used to reveal the fundamental symmetry principles of
strings at the Planck scale, and there are also indications that
the theory might be topological in vein, \cite{ed2}, which may
render the classical notion of space-time obsolete in this case or
replace it with another structure. An interesting framework in
which the behavior of tensionless strings may be studied in detail
is provided by the AdS/CFT correspondence when the gauge theory
side becomes weakly coupled (see for instance, \cite{ed1, mani,
sezgin, karch0, karch} and references therein). In any case, since
very little is still known about the tensionless limit of string
theory, and the symmetry breaking patterns of its structure, it is
natural to expect that any further progress in this direction will
be beneficial for the future development of the whole subject.

It is precisely this problem that we are going to address in the
present work by considering the tensionless limit of some exactly
solvable gauged WZW models at the quantum level. In these models,
the tensionless limit arises group theoretically in the
ultra-quantum region, which is well defined and tractable. The
models makes good sense from the world-sheet point of view,
although they invalidate all conventional effective field theory
descriptions based on the $\alpha^{\prime}$-expansion. It will be
shown that gravity decouples naturally from their spectrum in the
form of a Liouville field with infinite background charge, but
there is still a lot of structure that remains and can be treated
in exact terms. Another advantage of these models is the
non-trivial nature of their classical geometry, which receives
substantial $\alpha^{\prime}$ corrections within the usual
perturbative expansion of the renormalization group equations, and
they offer concrete examples for comparing the tensionless limits
before or after quantization. However, we are still unable to
provide a systematic reformulation of the complete theory in terms
of $1/\alpha^{\prime}$ expansion in target space, since this
approach requires the introduction of new concepts and variables
that are purely stringy in nature, without having an analogue in
the language of conventional effective field theories. Finally, it
should be emphasized that the tensionless WZW models do not
necessarily arise as limiting cases within critical string theory,
but this is not a drawback because the critical dimension is not a
useful concept in the tensionless limit.

The supergravity description of string theory, which corresponds
to the opposite (large tension) limit when $\alpha^{\prime}
\rightarrow 0$, provides a consistent truncation of string
dynamics at low energy scales, where only the genuine massless
modes participate, including the graviton and the dilaton, and the
effective action consists of the usual Einstein terms, plus higher
order curvature terms in the $\alpha^{\prime}$-expansion. The
higher order terms are in principle calculable, but their
determination is quite cumbersome unless new symmetry rules are
invented using a more fundamental formulation of string dynamics
that includes all $\alpha^{\prime}$ corrections. This is precisely
a place where the unbroken symmetry of tensionless strings, when
appropriately described, may shed new light into the structure of
all such higher order curvature terms. We note that the situation
is reminiscent of the non-commutative structures that arise in the
deformation approach to quantization: the non-commutative product
admits a power series expansion in Planck's constant with higher
derivative terms that obey consistency requirements, order by
order, following from associativity. The computation of all
deformation terms is made systematic once the notion of classical
geometry is abandoned and one introduces operators acting on the
Hilbert space, as the relevant concepts in the quantum theory, and
declare that the non-commutative product of functions is
isomorphic to the product of quantum operators. Furthermore, the
use of operators provides the only way to treat systematically the
ultra-quantum limit of non-commutative geometry, when Planck's
constant tends to infinity, in which case there is no point to use
power series expansions around the underlying classical concepts.

The tensionless limit we are considering in this paper is taken
directly at the quantum level, which is most natural as large
tension is related to large values of Planck's constant. Recall
that $\alpha^{\prime}$ appears as a loop counting parameter in the
perturbative renormalization group analysis of the world-sheet
sigma model, and the tensionless limit is ultra-quantum in nature
with Planck mass equal to zero. Then, it is natural to expect that
the reformulation of string theory in this case will lead to the
introduction of new concepts that will also be valuable for finite
values of $\alpha^{\prime}$, in analogy with the operator approach
to non-commutative geometry. It should also be added in this
context that several problems of non-commutative field theories
admit a simple formulation in the infinite non-commutativity
limit, \cite{shiraz}, whereas for finite values of the deformation
parameter the treatment becomes more intricate. Thus, tensionless
strings seem to offer the simplest framework in which new ideas
can be brought to light. The WZW models provide a concrete
framework in the attempt to link conformal field theories with
non-commutative structures, following the general program outlined
in reference \cite{jurg}, since the notion of classical geometry
undergoes quantum deformations when the level of the underlying
Kac-Moody algebras assume finite values, which are far away from
the classical large $k$ limit. Then, the tensionless limit that
exists for non-compact models when the level $k$ assumes its
critical value, corresponds to infinitely large non-commutativity
and the correspondence between strings and non-commutative
structures becomes more pronounced.

The effective field theory description of tensionless strings
seems to require the introduction of all massless states on equal
footing, but such an enlarged action and its symmetries principles
are not known for all string states at this moment. We have no
idea about the target space framework that replaces Einstein
gravity and its couplings to other fields when $\alpha^{\prime}
\rightarrow \infty$. However, we can develop an alternative route
based on the world-sheet symmetries of the underlying
two-dimensional quantum field theories that describe building
blocks of string theory vacua, which also make good sense in the
tensionless limit, as for any other value of the string tension.
Typical examples are provided by gauged WZW models based on
non-compact groups, such as the two-dimensional black-hole coset
$SL(2,R)_k/U(1)$ and higher dimensional generalizations thereof,
which are well defined for all values of the central charge of the
underlying Kac-Moody algebra that ranges from the dual Coxeter
number to infinity.

The tensionless limit is reached when $k$ approaches the dual Coxeter
number, and it is well defined in the framework of two-dimensional
quantum field theories. However, this limit is singular in the class of
conformal field theories because the central charge of the Virasoro algebra
becomes infinite and a rescaling of the Virasoro generators is required.
In turn, this leads to a contraction of the conformal symmetry that amounts
to decoupling gravity from the spectrum, but otherwise there is a large
world-sheet symmetry that remains associated to higher spin currents. A
primary aim of the present work is to expose the rich algebraic structures
of these models, which arise as enhanced world-sheet symmetries when $k$
assumes critical values. Put differently, the underlying Kac-Moody algebras
have a large number of null states when their level becomes critical, and
they are responsible for the exact treatment of WZW models in the
tensionless limit. It should be noted, however, that this approach only
applies to non-compact groups, since the compact models can never become
critical by unitarity that restricts the allowed values of $k$.

Our work could be regarded as generalization of some original
ideas introduced in reference \cite{ulf3}, where non-compact WZW
coset were put forward as models for tensionless strings. In this
paper we carry out this program in great detail, using the target
space and the world-sheet description of these models, and find
that gravity decouples naturally from their spectrum. Using the
two-dimensional black-hole coset $SL(2, R)_k/U(1)$, as
illustrative example, we find the exact metric becomes singular at
$k=2$, which is consistent with the naive expectation that strings
behave as tensionless objects in very strong gravitational fields.
However, more careful analysis of the quantum tensionless limit
reveals that gravity plays no role in this case, as it decouples
in the form of a Liouville field with infinite background charge,
and there is no remnant of space-time geometry. These models
resemble ``little string theories" which arise by taking the
string coupling to zero in some configuration of Neveu-Schwarz
five-branes and/or singularities, while keeping the string scale
constant; for a review, see for instance \cite{lst}, and
references therein. Although different, they both define
non-trivial theories which are decoupled from gravity.

The class of models we are considering here turn out to exhibit a rich symmetry
structure in the tensionless limit, which is associated to higher spin
fields and it is systematically described by
a truncated version of the $W_{\infty}$ algebra on the world-sheet of the
resulting two-dimensional quantum theory. Quite remarkably, this algebra
is linear and can be written in closed form using a bilinear realization of its
generators in terms of a complex fermion. It should be noted, however, that
it differs from the usual realizations of $W_{\infty}$-type algebras, as there
is no stress-energy tensor among its generators. It is conceivable that the
tensionless limit of WZW models might also have a topological meaning that characterizes
their behavior after the decoupling of gravity. Although this point
of view will not be developed here, it may be closely related to the topological
phase of tensionless strings in flat space that was advocated before,
\cite{ed2}.

The remaining sections of the paper are organized as follows. In
section 2, we briefly review the world-sheet and space-time
description of gauged WZW models and discuss the exact form of the
metric for all physical values of the level of the underlying
Kac-Moody algebra. Special emphasis is placed on the (Euclidean)
two-dimensional black-hole coset $SL(2,R)_k/U(1)$, which is shown
to exhibit tensionless limit for $k=2$. In section 3, we discuss
the spectrum of the $SL(2, R)_k/U(1)$ coset model, which consists
of two parts coming from a Liouville field and a compactified
boson, respectively, and describe the decoupling of the Liouville
mode when $k=2$. In section 4, we compare our results with the
null gauging of WZW models, which correspond to an infinite boost
in target space, and they describe the Liouville mode that
decouples in the tensionless limit. In section 5, we employ the
theory of non-compact parafermions to construct extended
world-sheet symmetries of the coset $SL(2,R)_k/U(1)$, which
include the Virasoro algebra for $2 < k < \infty$. The resulting
algebraic structure is a non-linear deformation of $W_{\infty}$,
denoted by ${\hat{W}}_{\infty}(k)$, which linearizes in the large
tension limit, $k \rightarrow \infty$. It is also shown that when
$k$ assumes its critical value, $k=2$, ${\hat{W}}_{\infty}(k)$
also becomes linear and coincides with a truncated version of
$W_{\infty}$ generated by all integer higher spin fields with $s
\geq 3$. In this case, the Virasoro algebra abelianizes by
suitable rescaling of the generators, and it can be consistently
factored out as it only depends on the Liouville mode that
decouples in the tensionless limit. The precise identifications
are made in section 6, where the $W_{1 +\infty}$ algebra and its
higher spin truncations are discussed in all generality; we also
comment on their free field realizations as they arise from coset
models. In section 7, we perform a formal BRST analysis of
${\hat{W}}_{\infty}(2)$ as (fundamental) world-sheet symmetry of
tensionless gauged WZW models and show that the central charge of
all higher spin generators should be fixed to a critical value,
which is not seen by the contracted Virasoro symmetry. It turns
out that this symmetry is not anomalous free for the black-hole
coset at $k=2$, but it needs two copies for consistent
implementation. In section 8, we outline generalizations of the
basic framework to higher dimensional coset models when the level
of the underlying Kac-Moody algebras reach their critical values.
Finally, in section 9, we present the conclusions and outline some
directions for future work.

\section{Gauged Wess-Zumino-Witten models}
\setcounter{equation}{0}

In this section we recall the classical description of gauged WZW models and
summarize the group theoretical methods that allow to determine their exact
geometry to all orders of $\alpha^{\prime} \sim 1/k$. It is then possible to take the
ultra-quantum limit by letting $k$ assume its critical value, in order to establish the
exact form of the very strong gravitational field which is responsible for the
tensionless behavior of these models. The geometry becomes singular at $k=g^{\vee}$,
as expected on general grounds, and
conformal invariance can no longer be used within any perturbative renormalization
group scheme to yield reliable space-time interpretation. The main focus is
placed here on the Euclidean
black-hole coset $SL(2, R)_k/U(1)$, which
provides the simplest gauged WZW model based on non-compact groups.
Further examples with higher dimensional coset models will be included in
section 8, where similar observations are also made.

\subsection{Preliminaries}

The WZW models, and their gauged versions, constitute exact conformal
field theories, which are constructed group theoretically, and they can be used as
building blocks for the description of string theory vacua.
These models are based on the Kac-Moody symmetry of a group $G$ that
is generated by the singular part of the operator product expansion of the currents
\be
J_A(z) J_B(w) = f_{AB}^C {J_C(w) \over z-w} + \eta_{AB} {k \over 2(z-w)^2} ~,
\ee
where $f_{AB}^C$ are the structure constants, $\eta_{AB}$ is the standard metric,
and $k$ is the level of the current algebra (see, for instance, \cite{marty, gko, kac}).
We will consider sigma models of the form $G/H$ by gauging appropriately chosen
subgroups of $G$ with special emphasis on non-compact groups for which the level can vary
continuously from the dual Coxeter number $g^{\vee}$ to infinity, i.e., $k\geq g^{\vee}$,
in order to have unitarity. The dual Coxeter number is given by the value of the
quadratic Casimir operator of $G$ in the adjoint representation, so that
\be
g^{\vee} \eta^{AB} = f^{ACD} {f^B}_{CD} ~.
\ee

The non-compact cosets are most appropriate for constructing exact
tensionless models by taking the limit $k \rightarrow g^{\vee}$,
whereas their classical geometry corresponds to the limit $k
\rightarrow \infty$. Also, in this context, one may extrapolate
continuously between the two limits, since the gauged WZW models
are well defined two-dimensional quantum field theories for all
such values of $k$. Compact groups do not allow for this
possibility because their level is quantized and it can never
become critical, i.e., equal to the dual Coxeter number, while
maintaining unitarity; this can be also formally seen by changing
the sign of $k$ in order to pass to the compact group, in which
case the corresponding level is positive and it can never become
equal to $-g^{\vee}$.

The conformal symmetry of WZW models $G_k$ is realized by the Sugawara construction
of their stress-energy tensor,
\be
T(z) = {1 \over k-g^{\vee}} \eta^{AB}J_A J_B (z) ~,
\ee
where normal ordering of the Kac-Moody currents is implicitly assumed. Then, the operator
product expansion
\be
T(z) T(w) = {\partial T(w) \over z-w} + 2{T(w) \over (z-w)^2} + {c \over 2(z-w)^4}
\ee
generates the Virasoro algebra with central charge
\be
c_G = {({\rm dim}G)k \over k-g^{\vee}} ~.
\ee
The stress-energy tensor of the gauged WZW models $G/H$ is simply provided by
the formula, \cite{marty, gko},
\be
T_{G/H} (z) = T_G(z) - T_{H}(z)
\ee
and the corresponding central charge equals to the difference of the two individual terms,
\be
c_{G/H} = c_G - c_H = {({\rm dim}G)k \over k-g^{\vee}} -
{({\rm dim} H)k \over k-h^{\vee}} ~,
\ee
where $g^{\vee}$ and $h^{\vee}$ are the dual Coxeter numbers of $G$ and $H$, respectively.
Thus, we observe that $c_{G/H} \rightarrow {\rm dim}(G/H)$ when $k \rightarrow \infty$,
whereas $c_{G/H} \rightarrow \infty$ when $k \rightarrow g^{\vee}$.
This is precisely the value of interest in the tensionless limit,
since $k-g^{\vee} \sim 1/\alpha^{\prime}$. Better understanding of
this relation will be achieved later using the effective action of the coset conformal
field theories.

The fact that the central charge of the Virasoro algebra becomes infinite
implies that the conformal field theory of the coset model
is singular at $k = g^{\vee}$.
Appropriate rescaling of the Virasoro generators is then required in
order to make the coefficient of the central terms finite, in which case the conformal
algebra contracts to an abelian structure\footnote{This contraction is similar to the
familiar Inon\"u-Wigner contraction of Lie algebras; for example, the $SU(2)$ algebra
contracts to the Heisenberg-Weyl algebra in the infinite spin limit, since a sphere with
infinite radius looks like a two-dimensional plane.}. This issue will be discussed later
in great detail. We will find that the tensionless limit makes perfect sense as
two-dimensional quantum field theory, although it is singular as conformal
field theory.

\subsection{Classical considerations}

The starting point is the ordinary WZW model for a Lie group $G$, which
is defined by the action, \cite{wzw},
\be
S_{\rm WZW} = {k \over 4\pi} \int_{\Sigma}
d^2z {\rm Tr} \left(\partial g^{-1} \bar{\partial} g
\right) + {ik \over 24\pi} \int_{B} d^3 x {\rm Tr} \left(
g^{-1} d g \right)^3 ~.
\ee
Here, $(z, \bar{z})$ are complex coordinates
on the two-dimensional world-sheet
$\Sigma$, whereas the second term is topological and it is defined on a three-dimensional
manifold $B$ whose boundary is $\Sigma$; for all practical purposes $\Sigma$ is taken
to be a sphere and $B$ is a three-dimensional ball. This action has global $G \times G$
symmetry corresponding to $g \rightarrow a g b^{-1}$ with both $a,b \in G$. One may
also consider a variant of the WZW models by gauging a subgroup of the global
symmetry group, but this is not always possible unless the subgroup obeys a certain anomaly
cancellation condition. In the following we consider the gauging of anomaly free
subgroups for the case of non-compact simple Lie groups $G$.

The gauging of the WZW model with respect to a
subgroup $H \subset G$ is implemented by introducing gauge fields $A$ and $\bar{A}$ with
values in the Lie algebra of $H$, and the action is taken to be \cite{wzw},
\cite{kristof, park}
\be
S(g; A, \bar{A}) = S_{\rm WZW} -
{k \over 2\pi} \int d^2z {\rm Tr} \left(A\bar{\partial} g g^{-1} -
\bar{A} g^{-1} \partial g - A g \bar{A} g^{-1} + A \bar{A} \right) .
\ee
$S(g; A, \bar{A})$ is invariant under the local gauge transformations
\be
A \rightarrow h^{-1} (\partial + A) h ~, ~~~~~
\bar{A} \rightarrow h^{-1} (\bar{\partial} + \bar{A}) h ~,
\ee
with $g$ also transforming in a vector-like way, as
\be
g \rightarrow h g h^{-1} ~; ~~~~~~ h \in H~. \label{freed}
\ee
The classical equations of motion that follow by variation with respect to all
fields are
\ba
\delta A & : & ~~~~~~~ \bar{D} g g^{-1} \mid_H = 0 ~, \\
\delta \bar{A} & : & ~~~~~~~ g^{-1} D g \mid_H = 0 ~, \label{katsik1}\\
\delta g & : & ~~~~~~~ \bar{D} \left(g^{-1} D g \right) + F_{z\bar{z}} = 0 ~,
\label{katsik2}
\ea
where $F_{z\bar{z}} = \partial \bar{A} - \bar{\partial} A + [A, \bar{A}]$ is the
field strength and $D$, $\bar{D}$ are the corresponding covariant derivatives.
Then, imposing the condition \eqn{katsik1} on equation \eqn{katsik2},
one arrives at the zero curvature
condition $F_{z\bar{z}} = 0$, and
\be
\bar{D} \left( g^{-1} D g \right) \mid_{G/H} = 0 ~. \label{GHpar}
\ee

We may use appropriate parametrization of the group elements $g(z, \bar{z})$
to fix the gauge freedom \eqn{freed}
and integrate over the gauge fields $A$, $\bar{A}$ in order
to obtain the effective action of the gauged WZW model for the
coset $G/H$. This procedure can be equivalently implemented at the classical level by
first solving for the gauge fields, which act as Lagrange multipliers, and then
substitute the resulting expressions in terms of $g(z, \bar{z})$ back
into the action $S(g; A, \bar{A})$, \cite{kristof, park, kurkut1}.
In particular, choosing a unitary gauge in the fundamental representation of $G$,
we may first fix ${\rm dim}H$ variables
among the total number of ${\rm dim}G$ parameters of the group elements $g$, and
denote by $X^{\mu}$ the remaining ${\rm dim}(G/H)$ target space variables. Then, the
gauge fields can be eliminated from the action $S(g; A, \bar{A})$, using their equations
of motion
\ba
A^a & = & +i \left(C^T - I \right)_{ab}^{-1} L_{\mu}^b \partial X^{\mu} ~, \nonumber\\
\bar{A}^a & = & -i \left(C - I \right)_{ab}^{-1} R_{\mu}^b \bar{\partial}
X^{\mu} ~.
\ea
The indices of the Lie algebra $G$ split as $A=(a, \alpha)$ with
$a \in H$ and $\alpha \in G/H$ and denote the generators of $H$ by $T^a$. Also,
we adopt the following short-hand notations
\be
L_{\mu}^a = -i{\rm Tr} \left(T^a g^{-1} \partial_{\mu} g \right) ~, ~~~~~
R_{\mu}^a = -i{\rm Tr} \left(T^a \partial_{\mu} g g^{-1} \right) ~, ~~~~~
C^{ab} = {\rm Tr} \left(T^a g T^b g^{-1} \right) ~.
\ee
Finally, the sigma model action of the gauged WZW model is written in terms of
these variables as follows,
\be
S = S_{\rm WZW}(g) - {k \over 2\pi} \int d^2z R_{\mu}^a
\left(C^T - I \right)_{ab}^{-1} L_{\nu}^b \partial X^{\mu} \bar{\partial}
X^{\nu} ~.
\ee

In any case, the target space metric depends
only on the gauge invariant parameters that are left to parametrize $g$ after
exploiting the gauge freedom \eqn{freed}; in general, there is also an anti-symmetric
tensor field that originates from the topological term of the action $S_{\rm WZW}(g)$.
The only extra ingredient that requires special attention in the quantum theory is the
introduction of a target space dilaton field due to finite corrections
coming from the integration over the gauge fields. The dilaton should be added
in the effective action, in the usual way, in order to maintain conformal
invariance of the model to lowest order in $\alpha^{\prime} \sim 1/k$,
as $k \rightarrow \infty$. Higher order corrections modify the form of the
background fields and they also provide the exact relation between
the level $k$ and the tension parameter of these models.

The two commuting copies of the Kac-Moody algebra that correspond to the
chiral sectors of the WZW model $G_k$ have remnants in the gauged WZW coset,
and they are associated to the (so called) parafermion currents, \cite{zamo},
\cite{kurkut1, kurkut2}.
They are classically defined by
first parametrizing the gauge fields as pure gauge, i.e., $A=- \partial h h^{-1}$ and
$\bar{A} = - \bar{\partial} \bar{h} \bar{h}^{-1}$ in terms of appropriately chosen
group elements $h$, $\bar{h} \in H$.
Then, introducing the gauge invariant
element $f= h^{-1} g h$ and using the zero curvature condition $F_{z\bar{z}} = 0$,
the classical equations of motion \eqn{GHpar} are written
as chiral conservation laws, $\bar{\partial} \Psi = 0$, where the field
\be
\Psi (z) = {ik \over \pi} f^{-1} \partial f (z)
\ee
defines the classical parafermion current with values in the coset
space $G/H$.
Likewise, the anti-holomorphic parafermion current $\bar{\Psi}$ is defined using
the group elements $\bar{h}$.

It is important to realize in this context that
the parafermion currents are non-local fields, since they have Wilson lines attached
to them which arise by solving $h$ and $\bar{h}$ in terms of $A$ and $\bar{A}$,
respectively, using path-ordered exponentials. Their classical Poisson algebra
is computed using the coset valued matrix elements $\Psi_{\alpha}$, and it
substitutes for the current symmetry algebra of the $G_k$ WZW models. In the
quantum theory, this algebra corresponds to the singular part of the operator
product expansion of the parafermion currents, \cite{zamo},
although non-singular terms are
also important in order to define $W$-algebra generators of the extended conformal
symmetries of these cosets. Later, we will pursue this method directly in the quantum theory
for arbitrary values of the level $k$, and examine the algebraic structures that
result on the world-sheet when $k$ comes close to its critical
value, $g^{\vee}$. In this framework, we will be able to determine the exact properties
of the two-dimensional quantum field theories that correspond to the tensionless
limit of the non-compact WZW models.

We also note for completeness that apart from vector gauging, it is also possible
to perform axial gauging when $H$ is an anomaly free
{\em abelian} subgroup. In this case, the starting point is provided by the action
\be
S^{\prime}(g; A, \bar{A}) = S_{\rm WZW} -
{k \over 2\pi} \int d^2z {\rm Tr} \left(A\bar{\partial} g g^{-1} +
\bar{A} g^{-1} \partial g + A g \bar{A} g^{-1} + A \bar{A} \right) ,
\ee
which is invariant under the axial-like local gauge transformations
\be
g \rightarrow h g h ~; ~~~~~~ h \in H ~, \label{freed2}
\ee
whereas $A$ and $\bar{A}$ transform as in the vector gauging. The rest
proceeds as before, but the computation yields different background fields in target space
which is $T$-dual to the geometry of vector gauging, \cite{elias}.

The simplest example is provided by the choice $G= SL(2, R)$ with
$H$ being an abelian subgroup, \cite{ed3}. One possibility
corresponds to the non-compact abelian subgroup generated by the
third Pauli matrix $\sigma_3$ in the fundamental representation of
$SL(2, R)$. Then, the coset model is $SO(2,1)_k/SO(1,1) \simeq
SL(2, R)_k/R$ with Lorentzian signature and it describes the
classical geometry of a two-dimensional black-hole in the large
$k$ limit. The other possibility corresponds to gauging the
compact abelian subgroup generated by $i\sigma_2$ in terms of the
second Pauli matrix. It leads to the coset model $SO(2,1)_k/SO(2)
\simeq SL(2,R)_k/U(1)$ that describes the geometry of a Euclidean
black-hole. In both cases the gauging can be implemented either
vectorially or axially, since $H$ is abelian, and the resulting
backgrounds are related to each other by $T$-duality. Also, the
Lorentzian and Euclidean black-holes are naturally related to each
other by analytic continuation of their target space coordinates.

The derivation of the explicit expressions is quite standard and
we refer the reader to the original works for further details,
\cite{ed3}, \cite{bars, wadia}. Besides, in the next subsection,
the exact metric and dilaton fields are presented systematically
to all orders in $\alpha^{\prime} \sim 1/k$, and the classical
geometry of the cosets follows when $k \rightarrow \infty$. Higher
dimensional cosets will be discussed briefly in section 8, but
their algebraic and geometric structures become quickly rather
involved, and hence difficult to treat in all generality. In any
case, the two-dimensional black-hole coset provides a good
laboratory for understanding the behavior of gauged WZW models at
critical level. Most of our analysis will be subsequently confined
to applications to the Euclidean black-hole coset.

\subsection{Exact metric and the tensionless limit}

Next, we consider the quantum modifications to the classical geometry,
which are induced by adding
$\alpha^{\prime} \sim 1/k$ corrections to the target space fields
using the perturbative beta function equations. Fortunately, the quantum
corrections can be explicitly worked out in all WZW models by appealing
to different (but equivalent)
methods for the construction of the exact metric and other
background fields to all orders in $\alpha^{\prime}$. Then, one may formally take
the tensionless limit of the exact formulae in order to get a feeling of the
resulting geometry in the ultra-quantum regime. This method also
provides the exact relation between $k$ and $\alpha^{\prime}$
beyond the leading order approximation. In all cases
the geometry becomes singular in the limit $\alpha^{\prime} \rightarrow \infty$, which
agrees with the naive expectation that strings behave as tensionless objects in
strong gravitational fields near the singularities, \cite{vega}.
However, as we will see later, a careful analysis of the quantum theory
shows that gravity decouples in the tensionless limit of WZW models and
there is no remnant of the target space geometry: the theory is non-trivial but
non-geometric.

The simplest way to derive the exact form of the metric is provided by the
Hamiltonian approach, which asserts that the Laplace operator in target
space is given by $L_0 + \bar{L}_0$ in terms of the left and right-moving
Virasoro zero modes. In particular, using the effective action for the
tachyon field $T$, it follows that
\be
L_0 T = \left({\Delta_G^L \over k-g^{\vee}} -
{\Delta_H^L \over k -h^{\vee}} \right)
T ~,
\ee
where $\Delta_G^L$ and $\Delta_H^L$ are the quadratic Casimir operators of the
groups $G$ and $H$, respectively. This equation follows from the
Sugawara construction of the coset model $G/H$, and there is also a similar
action for the operator $\bar{L}_0$ in terms of the operators
$\Delta_G^R$ and $\Delta_H^R$. It can be shown in all generality that
$\Delta_G^L = \Delta_G^R$, whereas
$(\Delta_H^L - \Delta_H^R) T = 0$ is only valid on-shell when acting on the tachyon
field. This is also consistent with the gauge invariance condition
$(J_H^L + J_H^R) T = 0$ which is imposed on the tachyon field by the vector
gauging of the models; for the axial gauging this condition is replaced by
$(J_H^L - J_H^R) T = 0$. Then, the target space metric and dilaton
fields are chosen so that the exact Hamiltonian acts in the
following way,
\be
(L_0 + \bar{L}_0) T = - {e^{-\Phi} \over \sqrt{G}} \partial_i
\left( e^{\Phi} \sqrt{G} G^{ij}
\partial_j T \right) . \label{lapl}
\ee
An invariant expression is $\sqrt{G} {\rm exp} \Phi$, which is independent of $k$.

This identification was first suggested in reference \cite{aben},
and it was subsequently applied
by a number of authors to a variety of conformal field theory models,
\cite{dvv, sfetsos1, tsey1}. The resulting expressions for the background
fields have been tested extensively by comparison to the perturbative expansion
of the beta functions equations to higher orders in $\alpha^{\prime}$.
There are also independent derivations based on the effective action, which can
be made systematic by the exact solvability of these models; for an excellent
exposition of all different approaches see, for instance, \cite{tsey1}. The
quantum analysis also leads to the identification, in appropriate
units\footnote{The coordinates $X^{\mu}$ of non-linear sigma models are
dimensionful but they can be rescaled by the characteristic radius of the manifold
$R$ to dimensionless fields. The WZW action is written in terms of the
group elements $g$ so that the level $k= R^2/\alpha^{\prime}$ is the dimensionless
analogue of the string tension. Classically, the tensionless limit arises when
$k \rightarrow 0$, but quantum mechanically one has to include the shift
by $g^{\vee}$.}
\be
\alpha^{\prime} = {1 \over k-g^{\vee}} ~,
\label{katsik5}
\ee
which follows by replacing $k$ with $k-g^{\vee}$ in all cases.

The tensionless limit is reached when
$k = g^{\vee}$, according to \eqn{katsik5}, but conformal invariance can no longer be
used to yield reliable space-time interpretation of these models. This method suggests
the way to take the tensionless limit in the framework
of the effective field theory, but the target space geometry becomes highly singular.
The appearance of singularities implies that the sigma model description
breaks down in this case as all other massless states
should be included on equal footing, if possible. Consequently, the
tensionless limit of the theory cannot be addressed systematically in the present
framework, which is inadequate as it stands.
It is only considered here to establish the form of the
singularities, which break the validity of the sigma model approach to string theory,
and compare with other methods.

Next, we focus on the exact form of the metric and dilaton fields
for the simplest Euclidean black-hole coset and postpone
generalizations to higher rank spaces until section 8. Since
$SL(2, R)_k/U(1)$ can be constructed by gauging the $U(1)$
subgroup in two different ways, the above Hamiltonian procedure
should be applied separately to the geometry of the momentum and
winding modes. In either case, the same qualitative picture
results when $k=2$, namely that the geometry looks like an
infinitely curved hyperboloid written in different coordinate
patches that depend on the gauging. The intermediate results are
quite standard by now, but they are summarized below following
\cite{dvv}, in order to examine the special limit $k=2$, which is
of interest here. Also, they will be used to discuss the spectrum
in section 3 that relies on the same Hamiltonian method.

Using the standard parametrization of the $SL(2, R)$ group
elements
\be
g = {\rm exp}\left({i \over 2} \theta_{\rm L}
\sigma_2 \right) {\rm exp} \left({1 \over 2} r \sigma_1 \right)
{\rm exp} \left( {i \over 2} \theta_{\rm R} \sigma_2 \right)
\label{groupel}
\ee
in terms of Pauli matrices, we may work out
the Laplacian of the exact sigma model. First, consider the
operator $L_0 = L_0^{SL(2,R)} - L_0^{U(1)}$, where the individual
terms are given by the Sugawara construction in terms of the
Fourier modes, \ba L_0^{SL(2, R)} & = & -{1 \over k-2} \left(
{\cal C}_2 + \sum_{n=1}^{\infty} \left(J_{-n}^+ J_n^- + J_{-n}^-
J_n^+ + 2J_{-n}^3 J_n^3 \right) \right) , \nonumber\\
L_0^{U(1)} & = & - {1 \over k} \left(\left(J_0^3\right)^2 + 2
\sum_{n=1}^{\infty} J_{-n}^3 J_n^3 \right) ~.
\ea
Here, ${\cal C}_2$ denotes the Casimir operator given by the quadratic
expression of zero modes,
\be
{\cal C}_2 = {1 \over 2}
\left(J_0^+ J_0^- + J_0^- J_0^+ \right)
+ \left(J_0^3\right)^2 \label{casim}
\ee
There are also similar expressions for the operators
$\bar{L}_0^{SL(2, R)}$ and $\bar{L}_0^{U(1)}$
in terms of the corresponding anti-holomorphic currents.

Note that only the zero modes $J_0^{\pm}$ and $J_0^3$ contribute
to the action of the operators on the tachyon field, since the
action of the positive modes on highest weight states give zero.
Therefore, it is sufficient to use the differential form of the
global $SL(2, R)$ generators in order to represent the relevant
part of the operator $L_0$. Since
\be
J_0^{\pm} = e^{\mp i
\theta_{\rm L}} \left({\partial \over \partial r} \pm {i \over
{\rm sinh} r} \left({\partial \over \partial \theta_{\rm R}} -
{\rm cosh}r {\partial \over \partial \theta_{\rm L}} \right)
\right) , ~~~~~ J_0^3 = i {\partial \over \partial \theta_{\rm L}}
~,
\ee
the action of $L_0$ and $\bar{L}_0$ on the tachyon field
$T$ is given by the differential operators
\be
L_0 = -{\Delta_0
\over k-2} - {1 \over k} {\partial^2 \over \partial \theta_{\rm
L}^2} ~, ~~~~~~ \bar{L}_0 = -{\Delta_0 \over k-2} - {1 \over k}
{\partial^2 \over \partial \theta_{\rm R}^2} ~, \label{laple} \ee
respectively, where $\Delta_0$ is \be \Delta_0 = {\partial^2 \over
\partial r^2} + {\rm coth} r {\partial \over \partial r} + {1
\over {\rm sinh}^2 r} \left({\partial^2 \over \partial \theta_{\rm
L}^2} + {\partial^2 \over \partial \theta_{\rm R}^2} - 2 {\rm
cosh} r {\partial^2 \over \partial \theta_{\rm L} \partial
\theta_{\rm R}} \right) .
\ee
Clearly, $\Delta_0$ is invariant
under the interchange $L \leftrightarrow R$, as expected on
general grounds. These expressions are also particularly useful
for computing the spectrum of conformal dimensions in the
black-hole coset. Furthermore, the condition $(L_0 - \bar{L}_0)T
=0$ implies the separation of variables $T = T(r, \theta) +
\tilde{T}(r, \tilde{\theta})$ where
\be
\theta = {1 \over 2}
(\theta_{\rm L} - \theta_{\rm R}) ~, ~~~~~~ \tilde{\theta} = {1
\over 2} (\theta_{\rm L} + \theta_{\rm R}) ~, \label{thethe}
\ee
and it leads to different effective descriptions of the geometry
of momentum and winding modes.

Using the variable $\theta$, which arises in the study of the momentum modes,
the action of the Laplacian on $T(r, \theta)$ is given by the
differential operator
\be
L_0 = -{1 \over k-2} \left({\partial^2 \over \partial r^2} +
{\rm coth} r {\partial \over \partial r} + \left(
{\rm coth}^2 {r \over 2} - {2 \over k} \right) {\partial^2 \over
\partial \theta^2} \right) .
\ee
The exact expressions for the effective metric and dilaton fields
follow by comparison with the Laplace operator \eqn{lapl}, and
they assume the special form
\be
ds^2 = {1 \over 2} (k-2) \left(dr^2 + \beta^2 (r) d \theta^2 \right) ,
~~~~~
\Phi = {\rm log} \left( {{\rm sinh} r \over \beta (r)} \right)
\label{data}
\ee
for all $k \geq 2$, where $\beta (r)$ is given by the function
\be
{4 \over \beta^2 (r)} = {\rm coth}^2 {r \over 2} - {2 \over k} ~.
\ee

The results simplify considerably when $k$ approaches
$2$ or infinity, leading to the following geometric structures:
First, as $k \rightarrow \infty$,  the geometry in the
gravitational regime looks like
\be
ds^2  \simeq  {1 \over 2} k \left(dr^2 +
4 {\rm tanh}^2 {r \over 2} d\theta^2 \right) , ~~~~
\Phi  \simeq  2 {\rm log} \left({\rm cosh} {r \over 2} \right) ,
\ee
and describes the familiar semi-infinite cigar when the $\theta$ coordinate
is chosen to be periodic modulo $2\pi$. As such, it satisfies the conditions
for conformal invariance to lowest order in $\alpha^{\prime}$; they read as
$R_{\mu \nu} = \nabla_{\mu} \nabla_{\nu} \Phi$ fixing the normalization of the
dilaton field. Second,
continuing the validity of the exact solution close to the critical level of
the underlying $SL(2,R)_k$ algebra, we find
\be
ds^2  \simeq  {1 \over 2} (k-2) \left(dr^2 +
4 {\rm sinh}^2 {r \over 2} d\theta^2 \right) , ~~~~
\Phi \simeq {\rm log} \left({\rm cosh} {r \over 2} \right) , \label{hyper1}
\ee
which describes the geometry of an infinitely curved hyperboloid in appropriate
coordinates. In this case, there is also a dilaton field that accompanies the
geometry, which is everywhere regular in space and the string coupling
${\rm exp} (-\Phi)$ never becomes infinite. It provides a ``target space"
realization of the tensionless $SL(2,R)_2/U(1)$ coset model
and establishes the singular nature of its effective geometry in the ultra-quantum
regime. However, conformal invariance can not be reliably used in this case because the
beta function equations are
only valid perturbatively in $\alpha^{\prime} \simeq 1/k$
when $k$ is large, and the result should be interpreted with care.

Likewise, the geometry of the winding modes follows by duality transformation
of the $SL(2,R)$ currents,
$J^a \rightarrow J^a$ and ${\bar{J}}^a \rightarrow - {\bar{J}}^a$,
which relate the axial with the vector gauging. The duality
interchanges the angular variables
$\theta \leftrightarrow \tilde{\theta}$ and
the Laplacian on $\tilde{T}(r, \tilde{\theta})$ is represented by
the differential operator
\be
L_0 = -{1 \over k-2} \left({\partial^2 \over \partial r^2} +
{\rm coth} r {\partial \over \partial r} + \left(
{\rm tanh}^2 {r \over 2} - {2 \over k} \right) {\partial^2 \over
\partial {\tilde{\theta}}^2} \right) .
\ee
The relevant expressions for the effective metric and dilaton fields
are also put in the form \eqn{data}, using another function
\be
{4 \over \beta^2 (r)} = {\rm tanh}^2 {r \over 2} - {2 \over k} ~.
\ee
The two cases are formally related to each other by the simple
transformation rule $r \rightarrow r + i\pi / 2$ and $\theta \rightarrow \tilde{\theta}$.
However, the metric and dilaton fields are now singular at
$r = r_0 = 2 {\rm arctanh}\sqrt{2/k}$.
This critical value defines the range of validity of the metric, since the signature
changes beyond it: for $r>r_0$
the signature is $++$, whereas for $r<r_0$ the signature changes to $+-$.

As before, there are two distinct limits corresponding to values of $k$ close to 2
or infinity. In the gravitational regime, $k \rightarrow \infty$, the background fields
are
\be
ds^2 \simeq {1 \over 2} k \left(dr^2 +
4 {\rm coth}^2 {r \over 2} d {\tilde{\theta}}^2 \right) , ~~~~
\Phi \simeq 2 {\rm log} \left({\rm sinh} {r \over 2} \right) ,
\ee
satisfying the conditions for conformal invariance to lowest order in $\alpha^{\prime}$.
This geometry is a trumpet with curvature singularity at the origin of the
coordinate system, and it is dual to the semi-infinite cigar.
On the other hand, when $k \rightarrow 2$, it follows that
\be
ds^2 \simeq {1 \over 2} (k-2) \left(dr^2 -
4 {\rm cosh}^2 {r \over 2} d {\tilde{\theta}}^2 \right) , ~~~~
\Phi \simeq {\rm log} \left({\rm sinh} {r \over 2} \right) , \label{hyper2}
\ee
which provides the ``target space" description of the tensionless
$SL(2, R)_k/U(1)$ model in terms
of an infinitely curved $AdS_2$ space with signature $+-$. Note that the change
of signature occurs because the critical value $r_0$ is pushed to infinity when
$k \rightarrow 2$. Thus, either we are prepared to accept a change of signature
in space, in which case the
string coupling ${\rm exp}(-\Phi)$ becomes infinite at the origin, $r=0$,
but it is regular everywhere else, or else the available space covered by the
coordinate system is completely eaten up to maintain Euclidean signature.
In either case, the singular character of the exact metric indicates, as before, that
the effective theory breaks down in the tensionless limit and conformal invariance
is not particularly useful for exploring the geometry of target space beyond
perturbation theory.

The Lorentzian version of the coset arises by analytic continuation
of the angular variable, so that $r$ remains a spatial coordinate, and it corresponds
to the gauged WZW model $SL(2, R)_k/R$. Then, the tensionless
limit of the two-dimensional geometry is $AdS_2$ with zero radius, using the axial
gauging. On the other hand, vector gauging leads to a confusion of signature, as
before, and it gives rise to an infinitely curved hyperboloid with Euclidean signature
or else all space is eaten up by the quantum corrections.
This coincidence offers a concrete framework
for exploring the relation between the quantum tensionless limit of the
black-hole coset at $k=2$ and the quantization
of classical tensionless strings on $AdS_2$, following the light-cone methods
of references \cite{karch0, karch} (but see also \cite{bonelli}).
It should be further mentioned that
the higher dimensional WZW models $SO(d-1, 2)/SO(d-1, 1)$ with Lorentzian signature
do not represent $AdS_d$ space with zero radius when their effective action is taken to
critical level. As we will see later in section 8, they rather describe non-symmetric
deformations of $AdS_d$ for $d>2$, in the presence of non-trivial anti-symmetric
tensor field, but this may not be so important in the zero
radius limit. These models also exhibit a confusion of signature or else
there is a truncation
of the available space, as for the vector gauged $SL(2, R)_k/U(1)$ coset.
In any case, the theory of non-compact cosets provides exact
models for tensionless strings, which could be used further to understand
the subtle issues of the AdS/CFT correspondence in the zero radius limit.

Summarizing the results of this exposition, we conclude that the
metric sector of gauged WZW models becomes highly
singular at critical $k$ and string propagation behaves as
tensionless theory. Since the effective field theory description breaks down,
other methods should be employed for the exact quantum mechanical treatment of WZW models
in this case. These will be provided later using world-sheet techniques which
make perfect sense in the tensionless limit. In order to motivate some of the
constructions, we will first discuss the behavior of the spectrum close to
critical level and observe a decoupling of gravity from the remaining
fields of the quantum theory.

\section{Spectrum of the $SL(2, R)_k/U(1)$ coset}
\setcounter{equation}{0}

We briefly review the construction of the full
spectrum of conformal dimensions for the $SL(2, R)_k/U(1)$ model,
following \cite{dvv, malda}.
The results are analogous to the spectrum of the compact coset
$SU(2)_N/U(1)$, but with some additional elements for the non-compact group.
Then, we discuss the rescaling which is
necessary to make sense of the special limit $k=2$, and observe that gravity
decouples from the spectrum in the form of a Liouville field with infinite background
charge. The same picture will arise later using world-sheet methods, as the
Virasoro algebra decouples from all remaining symmetries of the coset model.

\subsection{The spectrum for $k>2$}

The spectrum of primary fields of the $SL(2,R)_k/U(1)$ model can
be determined from the states of the $SL(2,R)_k$ conformal field
theory by imposing restrictions on the left and right moving
$J^3$-oscillators,
\be
J_n^3 |{\rm state}> = 0 = \bar{J}_n^3 |{\rm
state}> ~~~~~~~ {\rm for} ~~ n>0 ~, \ee which are supplemented by
the following conditions on the zero modes, \be J_0^3 -
\bar{J}_0^3 = m ~, ~~~~~~~ J_0^3 + \bar{J}_0^3 = nk
\label{omegalr}
\ee
for all integers $m$, $n$. Using the
parametrization \eqn{groupel} of the $SL(2, R)$ group elements,
the condition \eqn{omegalr} translates into the restriction
\be
\omega_{\rm L} = {1 \over 2} (m+nk) ~, ~~~~~~~ \omega_{\rm R} =
-{1 \over 2} (m-nk)
\ee
for the lattice of the corresponding
$U(1)$ quantum numbers. Then, the full spectrum of conformal
dimensions of the coset model can be obtained by diagonalizing the
operators $L_0 = L_0^{SL(2,R)} - L_0^{U(1)}$ and $\bar{L}_0 =
\bar{L}_0^{SL(2, R)} - \bar{L}_0^{U(1)}$.

The action of the $SL(2, R)_k$ currents is computed by taking into account only the
contribution of the zero modes $J_0^{\pm}$ and $J_0^3$, whereas the action of positive
modes on highest weight states gives zero. Thus, it is sufficient to use
the global $SL(2, R)$ generators in order to represent the relevant
part of the operators $L_0$ and $\bar{L}_0$, as in equation \eqn{laple}
in terms of the Casimir ${\cal C}_2$.
Using the defining relations of highest weight states
for both left and right Virasoro movers,
\be
L_0 |l, \omega_{\rm L}> =
h_{l}^{\omega_{\rm L}} |l, \omega_{\rm L}> ~, ~~~~~~~
\bar{L}_0 |l, \omega_{\rm R}> =
\bar{h}_{l}^{\omega_{\rm R}} |l, \omega_{\rm R}> ~,
\ee
the corresponding conformal weights are
\ba
h_l^{\omega_{\rm L}} &=& - {l(l+1) \over k-2} + {1 \over k} \omega_{\rm L}^2 =
-{l(l+1) \over k-2} + {(m+nk)^2 \over 4k} ~, \nonumber\\
\bar{h}_l^{\omega_{\rm R}} &=& - {l(l+1) \over k-2} + {1 \over k} \omega_{\rm R}^2 =
-{l(l+1) \over k-2} + {(m-nk)^2 \over 4k} ~. \label{spect}
\ea
Here, $l$ is the $SL(2,R)$ isospin which is determined by the quadratic
Casimir operator \eqn{casim}
with eigenvalues $c_2 = l(l+1)$.

The allowed values of $l$ follow from the classification of the unitary irreducible
representations of the Lie algebra $SL(2, R)$, supplemented by some additional
restrictions that depend on the central charge $k$ of the Kac-Moody algebra.
Recall at this point that the representations of the global
$SL(2, R)$ algebra fall into three general series according to the allowed
values of the isospin $l$ and the ``magnetic" quantum number $\omega$ that label the
eigenvalues of ${\cal C}_2$ and the $U(1)$ generators, respectively:

\noindent
(a) {\em Principal continuous series}, which have $l=is - 1/2$ with real values of $s$
and $\omega$.

\noindent
(b) {\em Complementary continuous series}, which have real $l \in [-1, -1/2]$ and $\omega$
is also real.

\noindent
(c) {\em Principal discrete series}, which have real $l < -1/2$ and $|\omega| + l$ is
non-negative integer.

\noindent
The discrete series come in two different types, either highest or lowest weight, and
there is also the trivial representation with $l=0$ and $\omega =0$ that should be
added for completeness. Note that the quadratic Casimir is always real with $-l(l+1)$
being positive for the continuous series representations.

The only relevant representations turn out to be the (a) or (c)
series leading to normalizable vertex operators. For the discrete
series, both $|\omega_{\rm L, R}| + l$ should be non-negative
integers, which in turn put restrictions on the allowed range of
$l$ depending on the level $k$. Taking also into account recent
analysis based on spectral flows, \cite{malda, hanan}, one finds
that the allowed range of values for the discrete series is
\be
-{1 \over 2}(k-1) < l < -{1 \over 2} \label{unibodr}
\ee
in order
to have unitarity. There is no such restriction on the principal
continuous series. Thus, we obtain a complete description of the
spectrum \eqn{spect} in either case. It is also important to
stress at this point that there are no discrete representations
appearing at $k=2$, using the unitarity bound \eqn{unibodr}.

As $k \rightarrow 2$,
only the principal continuous series representations become relevant with
\be
h_l^{\omega_{\rm L}} =  {s^2 + 1/4 \over k-2} +
{(m+nk)^2 \over 4k} ~, ~~~~~~
\bar{h}_l^{\omega_{\rm R}} =  {s^2 + 1/4 \over k-2} +
{(m-nk)^2 \over 4k} ~, \label{spect2}
\ee
but the first term blows up in the limit. Its contribution will be attributed to a
Liouville field that arises in the free field representation of the coset model and
produces the same spectrum as above.
Also, in section 4, the same Liouville theory will
control the effective description of the $SL(2, R)_k/U(1)$ coset
model, which is obtained by introducing a boost with very high Lorentz factor in
the Lie algebra.
This Liouville theory describes the radial coordinate of the coset model and has to
decouple at critical level; otherwise, the dimensions \eqn{spect2} will contain an
infinite part, which is natural to expect at very high energy scales but not in
tensionless models. Put differently, unless the gravitational effects of Liouville theory
can be consistently removed, the high energy limit cannot be considered as being
tensionless. This crucial point will be clarified further in the following for
the gauged and the ungauged WZW models.
It is also interesting to recall that the spectral flow of the continuous
representations correspond to long string states in the $SL(2,R)_k$ WZW
theory, \cite{malda, seiwit}. On the other hand, the discrete representations and their
spectral flow correspond to short strings, but they do not arise at $k=2$. Since
the radial coordinate of long strings is effectively described by a Liouville
theory with background charge $Q \sim 1/\sqrt{k-2}$, as follows from the
exact analysis of the problem in the context of the $D1/D5$ brane system, \cite{seiwit},
one is lead to suspect that their radial dynamics decouples in the tensionless
limit.

\subsection{The decoupling of Liouville field at $k=2$}

Next, we examine the reductions that arise at critical level of the $SL(2, R)_k$
current algebra. The conformal dimensions \eqn{spect2} consist of two terms that
behave differently when $k \rightarrow 2$; as they stand, the
first blows up like $1/(k-2)$
and the second remains finite. Likewise, the central charge of the Virasoro
algebra blows up like $1/(k-2)$. Thus, a rescaling is required in order to
make sense of these infinities in a systematic way.

It is instructive for this purpose to describe the spectrum of conformal dimensions
\eqn{spect} for arbitrary level $k$ using another conformal field theory
\be
S = {1 \over 2\pi} \int d^2 z \left(\partial \phi_1 \bar{\partial} \phi_1 + \partial
\phi_2 \bar{\partial} \phi_2 + {Q \over 4} \sqrt{g} R \phi_1 \right) \label{Qtheo}
\ee
that contains two free scalar fields, one of them with background charge $Q$ and the other
compactified on a circle with radius $R$. The field $\phi_2$ has stress-energy tensor
\be
T_2 (z) = -{1 \over 2} (\partial \phi_2)^2 (z)
\ee
with Virasoro central charge $c=1$ and the full mass spectrum consists of states with
dimensions
\be
M^2 = {1 \over 2} \left({m \over 2R} + n R \right)^2 ~,
\ee
where $m$ labels the Kaluza-Klein modes and $n$ the winding modes, which are both integer.

On the other hand, the stress-energy tensor of the field $\phi_1$ is improved due to the
background charge, and it is
\be
T_1(z) = - {1 \over 2}
(\partial \phi_1)^2 (z) + {Q \over 2} \partial^2 \phi_1 (z) ~. \label{canonst}
\ee
The corresponding primary fields are vertex operators of the general form
$V(z) = {\rm exp}(q\phi_1(z))$
with conformal dimensions equal to
\be
h = -{1 \over 2} q(q+Q)
\ee
with respect to $T_1(z)$. Choosing appropriate values for the momenta and the
background charge of the Liouville field $\phi_1$, and fixing the periodicity
of the field $\phi_2$ as follows,
\be
q= l \sqrt{2 \over k-2} ~, ~~~~~ Q = \sqrt{2 \over k-2} ~, ~~~~~
R= \sqrt{k} R_0 ~,
\ee
where $R_0 = 1/\sqrt{2}$ is the ``self-dual" radius, we find that the net
spectrum is the same as in equation \eqn{spect} above.
Similarly, identifications are worked out in the anti-holomorphic sector of the model.

The total stress-energy tensor
of the conformal field theory \eqn{Qtheo} is
\be
T(z) = T_1(z) + T_2(z) = -{1 \over 2} (\partial \phi_1)^2(z)
- {1 \over 2}
(\partial \phi_2)^2 (z) + {1 \over \sqrt{2(k-2)}} \partial^2 \phi_1 (z) ~ \label{pstrem}
\ee
with Virasoro central charge
\be
c=2{k+1 \over k-2} ~,
\ee
which is the same as for the coset
model $SL(2,R)_k/U(1)$. Therefore, the fields $\phi_1$ and $\phi_2$ can be used
to provide a free field realization of the gauged WZW model, although the
conformal field theory \eqn{Qtheo} is not
meant to be equivalent to $SL(2, R)_k/U(1)$. Also, as we will see later in section 5,
the expression \eqn{pstrem} provides the stress-energy tensor of the
black-hole coset in terms of free fields that arise in the parafermionic construction
of the underlying $SL(2, R)_k$ current algebra, and it comes as no surprise that the
two spectra coincide.

This construction suggests the rescaling of the Virasoro generators
which is required before taking the limit of the theory at critical level. The infinite
contribution of the background charge is removed by considering
\be
\tilde{T}(z) = \lim_{k \rightarrow 2} \left( \sqrt{2(k-2)} T(z) \right) =
\partial^2 \phi_1 (z) ~, \label{mamiri}
\ee
which satisfies the simplified operator product expansion
\be
\tilde{T}(z) \tilde{T} (w) \sim {1 \over (z-w)^4} ~,
\label{mamiri0}
\ee
up to an overall (irrelevant) constant.
The rescaling removes the infinity from the central charge of the Virasoro algebra,
but the price to pay is the contraction of the Virasoro algebra to an abelian structure
generated by the derivative of the $U(1)$ current $\partial \phi_1$. Thus,
$SL(2,R)_2/U(1)$ is a singular conformal field theory, but otherwise it makes
perfect sense as a quantum theory.

The rescaling \eqn{mamiri} forces the constituent fields that appear in the free
field realization of the $SL(2, R)_k/U(1)$ coset model to decouple at
critical level, since only $\phi_1$ is contributing to $\tilde{T}(z)$.
The decoupling of the Liouville field $\phi_1$ is also
reflected in the rescaled form of the conformal dimensions that follow by multiplying
\eqn{spect2} with $\sqrt{k-2}$. The first term still grows large
as $1/\sqrt{k-2}$, whereas the contribution of $\phi_2$
goes to zero as $\sqrt{k-2}$.
Disregarding the very heavy states associated to the Liouville field, we
obtain a tensionless model where gravity plays no role, but there is still
some non-trivial structure associated to the field $\phi_2$ that remains behind.
It is our purpose to investigate some aspects of the residual theory
in a more systematic way.

In addition, as we will see in section 5, the $SL(2, R)_k$ current algebra
admits a free field realization in terms of three scalar fields, using the
parafermionic construction of the $SL(2,R)_k/U(1)$ coset plus one extra boson.
A degeneration takes place at $k=2$, since one of the two bosons that
parametrize the parafermions of the coset decouples naturally from the rest.
In this case, $SL(2, R)_2$
makes perfect sense without need for rescaling of the affine currents,
but the Sugawara construction
is singular; it provides extra reason to believe
that gravity plays no role in the tensionless limit. As for the remaining field,
it assumes a non-geometric role in the exact description of the coset model.
This unexpected reduction should also be held responsible for the
null states that arise in the representation theory of the $SL(2, R)_k$
algebra at $k=2$.

Finally, we also note for completeness that there is a rescaling of the Liouville
field $\phi_1$,
\be
\tilde{\phi}(z) = \sqrt{2(k-2)} ~ \phi_1 (z) ~,
\ee
which can be consistently implemented in $SL(2, R)_k$ and $SL(2, R)_k/U(1)$ and
leads to the following expression for the rescaled Virasoro operator
\be
\tilde{T}^{\prime} (z) = \lim_{k \rightarrow 2} \left( 2(k-2) T(z) \right) =
-{1 \over 2} (\partial \tilde{\phi})^2 (z) + \partial^2 \tilde{\phi} (z)
\label{gaudinh}
\ee
at critical level. In this case, $\tilde{\phi}$ can be viewed as a null boson with
\be
<\tilde{\phi}(z) \tilde{\phi}(w)> = 0 ~,
\ee
and as a result
\be
\tilde{T}^{\prime} (z) \tilde{T}^{\prime} (w) \sim 0
\ee
without having singular terms. Then, $\tilde{T}^{\prime} (z)$
is strictly abelian, unlike \eqn{mamiri0},
and it provides the central elements in the enveloping algebra of
$SL(2, R)_k$, which are
non-trivial only for $k=2$. This rescaling is consistent but contrary to the
previous case the resulting null field $\tilde{\phi}$
does not decouple from the operators of the WZW model at critical level. The
systematic study of this limit also proves interesting in many
respects as it connects quite naturally the Casimir \eqn{gaudinh} with the
Hamiltonian of completely integrable quantum spin chains, \cite{reshet}. These
issues will not be discussed here but in forthcoming publications.

\subsection{Kac-Kazhdan determinant formula}

We complete our general discussion by including the Kac-Kazhdan determinant formula
for the highest weight representations of the $SL(2, R)_k$ current algebra and
the corresponding determinant of the inner product of states of the coset module
$SL(2, R)_k/U(1)$, \cite{kazh}; see also \cite{peskin} for a more comprehensive
discussion of the subject. It offers a complementary understanding of the
huge degeneracy that is expected to arise at critical level for the algebra as for the
coset space model.

Let us consider the holomorphic sector of the current algebra and denote by $D_N$
the determinant of inner products of all states in the highest weight module
which are lying at a given $L_0^{SL(2, R)}$ level $N$ with $J_0^3$ charge equal to
$m$; the anti-holomorphic sector will not be discussed, but it can be treated in
a similar way. Clearly, the individual inner products of states depend on $m$ and
$k$, as well as the value of the Casimir operator ${\cal C}_2$.
It is convenient to define the operators
\be
{\cal J}_{+n} = {\cal C}_2 + (m+n-1)(m+n) ~, ~~~~~
{\cal J}_{-n} = {\cal C}_2 + (m-n+1)(m-n)
\ee
for all $n>0$. Then, for general values of $k$, the determinant assumes the form
\ba
D_N & = & (-1)^{r_3(N)} C_N (k-2)^{r_3(N)} \prod_{n=1}^{N} \left({\cal J}_{-n}
{\cal J}_{+n} \right)^{P_3(N, n)} \times \nonumber\\
& & ~~~~ \prod_{r, s =1}^N \left({\cal C}_2 + {1 \over 4}\left(r(k-2) + s+1\right)
\left(r(k-2) +s -1\right) \right)^{p_3(N-rs)} ~,
\ea
where $C_N$ is a positive numerical constant and $r$, $s$ are constrained so that
$rs \leq N$. The exponents are all positive but their exact form is not very important
in the present work. Note that the determinant vanishes at $k=2$, as noted before,
due to the appearance of many null states.

The determinant formula for the representations of the coset model can also be found
using the factorization
\be
D_N = \prod_{q=0}^{N} D_N^{(q)} ~,
\ee
where $D_N^{(q)}$ denotes the determinant of inner products of states at level $N$ of the
$SL(2, R)$ representation which are at level $q$ of the $U(1)$ current algebra.
According to this rearrangement, $D_N^{(0)}$ provides the determinant of inner products
of states in the coset space module, which is of interest here, and it assumes the
following form
\ba
D_N^{(0)} & = & C_N^{\prime} \left(1 - {2 \over k} \right)^{r_2(N)}
\prod_{n=1}^{N} \left({\cal J}_{-n}
{\cal J}_{+n} \right)^{P_2(N, n)} \times \nonumber\\
& & ~~~~ \prod_{r, s =1}^N \left(C_2 + {1 \over 4}\left(r(k-2) +
s+1\right) \left(r(k-2) +s -1\right) \right)^{p_2(N-rs)} ~,
\ea
where $C_N^{\prime}$ is another positive numerical constant, $rs
\leq N$, and the exponents are appropriately chosen as before.
Note that the factors appearing in the first line are all positive
when $k>2$ and ${\cal C}_2$ is taken in the continuous series
representations, whereas the factors in the second line are
positive when $k>2$ and ${\cal C}_2 >0$. Thus, the determinant of
the coset module is always positive for the continuous series
representations provided that $k>2$, which is required for
unitarity. In the tensionless limit we only have to consider
continuous series representations, since the unitarity bound
\eqn{unibodr} squeezes all discrete representations, but the
determinant also vanishes in this case.

In summary, for unitary highest weight representations all states have strictly
positive norm and there are no non-trivial null states when $k>2$. When the
level of the $SL(2, R)_k$ algebra becomes critical many null states make their
appearance and lead to huge degeneracies of the spectrum.

\section{Liouville field and null gauging}
\setcounter{equation}{0}

The background of a two-dimensional black-hole
provides a classical solution of the same theory that describes
a Liouville field coupled to $c=1$ matter. The $c=1$ matrix model superficially has
a one-dimensional target space, as it appears in its initial formulation,
but in fact it is more naturally understood in terms of
a two-dimensional space, \cite{jpol}, where the extra dimension is provided by the
Liouville mode. In view of this correspondence, the radial variable $r$ of the coset model
can be interpreted as Liouville field, which is always space-like in either
Euclidean or Lorentzian version of the model. The relation becomes very clear in the
weak coupling region $r \rightarrow \infty$, where the string coupling
${\rm exp}({-\Phi})$ tends to zero and the dilaton grows linearly as
$\Phi \sim r$. Thus, in this region, the geometry of the black-hole is asymptotic
to the two-dimensional geometry of the $c=1$ matrix model, but their equivalence
is not valid everywhere.

The precise relation between the two models is better understood
by revisiting the interpretation of a (non-critical) string theory
in $d-1$ dimensions as a string theory in $d$ dimensions with the
extra dimension being provided by the Liouville
field\footnote{This relation is usually stated for critical
strings in $d$ dimensions, although here we want eventually to
apply it to tensionless models where there is no notion of
criticality; likewise, the conformal symmetry degenerates in the
tensionless limit and cannot be employed in the argument. These
issues introduce complications that invalidate the effective field
theory description and they will not be addressed properly in the
present context.}, as in the $c=1$ matrix model. Following
\cite{ed3}, we note that the reverse map exists only if the
gradient of the dilaton field of a $d$-dimensional model has the
same space-time character with the Liouville coordinate; in this
case, the extra coordinate can be gauged away, using conformal
transformations, and one arrives at a lower dimensional model, as
it can be easily seen from the modified transformation law of the
target space variables $\delta X^{\mu} \sim G^{\mu \nu}
\nabla_{\nu} \Phi$ to lowest order in $\alpha^{\prime}$. However,
if the gradient of the dilaton changes character or if it becomes
singular at certain points, the reverse map will break down. This
is precisely the situation we encounter for the two-dimensional
coset model and as a result {\em the black-hole cannot be regarded
as a theory of $c=1$ matter coupled to two-dimensional (Liouville)
gravity}. This, in turn, suggests that the theory which remains
after the decoupling of gravity in the tensionless limit of the
coset model is not $c=1$ matter in isolation; instead it is a
variant of it or an exotic phase thereof, which is still poorly
understood. We will briefly return to it in section 6, where the
symmetries of the model will be represented as fermion bilinears
and compared to the usual $W_{1+\infty}$ symmetry of the ordinary
$c=1$ matrix model.

Next, in order to understand the role of the Liouville field in the
tensionless limit, we consider the null gauging of WZW models. Equivalently,
the same theory is obtained by making an infinite boost in the Lie algebra of $SL(2,R)$.
In this case, we will find that the target space
geometry exhibits a drastic reduction to only one dimension, which is always
space-like and corresponds to the radial coordinate $r$. Then, since we are
committed to interpreting the Liouville field as a spatial coordinate of the
black-hole geometry and not as a (Euclidean) time variable, it follows that
the null gauging of $SL(2,R)$ captures the gravitational sector of the model
that decouples when $k=2$.

We proceed with some background material on the null gauging of WZW models
for non-compact groups, putting the emphasis on $SL(2, R)_k$; examples with higher rank
groups will be presented later in section 8.
It is well known that the group $SL(2,R)$ has three different conjugacy classes
of subgroups that are isomorphic either to the group of rigid rotations in two
dimensions, $SO(2)$, or the Lorentz group $SO(1,1)$, or the
isotropy group of light-like vectors, $E(1)$, which in this case coincides with the
Euclidean group in one dimension. Their generators are $i\sigma_2$, $\sigma_3$ and
$\sigma^+ = \sigma_3 + i\sigma_2$, respectively, but in the latter case the
generator is nilpotent since $(\sigma^+)^2= 0$.
The standard gauging of the $SL(2, R)$ WZW model is taken either with respect
to the group $SO(2)$ or
$SO(1,1)$, thus giving rise to the Euclidean or Lorentzian two-dimensional
black-hole cosets. The null gauging was also considered
in the literature before, \cite{iran1, iran2, iran3},
by forming the $SL(2, R)_k/E(1)$ WZW model, and a
striking result was found, at least at it appears at first sight, namely that the
classical target space geometry degenerates to one dimension.
The result follows easily by applying
the usual prescription for axial or vector gauging of the
corresponding subgroup $E(1)$.

Let us consider the action of WZW models and discuss first their axial gauging.
Using the standard parametrization of the $SL(2, R)$ group elements $g$ in the
fundamental representation,
\be
g = \left(\begin{array}{ccc}
a & & u \\
  & &   \\
-v& & b
\end{array} \right) ~; ~~~~~ ab+uv = 1 ~,
\ee
we fix the gauge by choosing $a+b=0$. The parameters
\be
\chi = u-v ~, ~~~~~~ w = a-b-u-v
\ee
remain invariant under axial transformations and provide a gauge invariant
parametrization of the coset space $SL(2, R)/E(1)$.
Then, the action of the gauged WZW model assumes the form
\be
S(g; A, \bar{A}) = {k \over 2 \pi}
\int d^2 z \left( {\partial w \bar{\partial} w \over w^2} -
{1 \over w^2} \mid A + {1 \over 2w^2} (\chi \partial w - w \partial \chi) \mid^2
\right) ,
\ee
which after the integration of the gauge fields yields the following metric and dilaton
fields
\be
ds^2 = k {\partial w \bar{\partial} w \over 2w^2} ~, ~~~~~~ \Phi = 2 {\rm log} w
\ee
in the large $k$ limit. Finally, introducing a scalar field $\phi$, so that
$w = \pm {\rm exp} \phi$, depending on the sign of $w$, one arrives at the
one-dimensional theory in target space
\be
S= {k \over 2 \pi} \int d^2 z \left(\partial \phi \bar{\partial} \phi \right)  +
{1 \over 4 \pi} \int d^2 z \sqrt{g} R \phi ~, \label{redeq}
\ee
which includes the contribution of a linear dilaton and describes the action of
a free scalar field with background charge. This is the theory of a Liouville
field with zero cosmological constant, whereas the other field $\chi$ decouples
from the geometry and it does not appear in the action.

We also note for completeness that the vector
gauging proceeds in a similar way by making the gauge choice $u-v=0$ and then
using the two gauge invariant parameters $w$ and $a+b$ to describe the coset.
By eliminating the gauge fields, the same reduction occurs in target space, as
before, and the classical geometry of the vector gauged $SL(2, R)_k/E(1)$ model
is described again by the effective action \eqn{redeq}. Thus, the
end result is insensitive to the gauging prescription, which appears to be self-dual
with respect to $T$-duality transformations, and
there is a drastic dimensional reduction in target space.

The same result admits an interesting group theoretic description
by boosting the subgroup $H$ that appears in ordinary gauged
models to very large Lorentz factor, as an alternative to null
gauging. For the simplest class of $SL(2, R)_k$ models, let us
consider the transformation
\ba
\sigma_1(\beta) & = & e^{-\beta
\sigma_1} \sigma_1 e^{\beta \sigma_1} =
\sigma_1 ~, \nonumber\\
i\sigma_2(\beta) & = & e^{-\beta \sigma_1} i\sigma_2 e^{\beta \sigma_1} =
({\rm sinh} 2\beta ) \sigma_3 + ({\rm cosh} 2\beta ) i\sigma_2 ~,
\nonumber\\
\sigma_3(\beta) & = & e^{-\beta \sigma_1} \sigma_3 e^{\beta \sigma_1} =
({\rm cosh} 2\beta ) \sigma_3 + ({\rm sinh} 2\beta ) i\sigma_2 ~,
\label{bosta}
\ea
which introduces a boost with parameter $\beta$,
\be
{\rm tanh} 2 \beta = {v \over c} ~.
\ee
It describes a Lorentz transformation in the Lie algebra of $SL(2,R) \simeq SO(2, 1)$ when
there are two frames moving perpendicular to the spatial direction $\sigma_1$ with relative
velocity $v$. As $\beta$ ranges from
$0$ to infinity, $i\sigma_2(\beta)$ interpolates smoothly between
$i\sigma_2(0) = i\sigma_2$ and $i\sigma_2(\infty)$, which becomes proportional
to the nilpotent element $\sigma^+$. Likewise, $\sigma_3(\beta)$ interpolates
between $\sigma_3$ and $\sigma^+$ as $\beta$ ranges from $0$ to infinity.
Thus, one may use the group
generated by $i\sigma_2(\beta)$ to gauge the $SL(2, R)_k$ WZW model
and obtain the geometry of $SL(2, R)_k/E(1)$ in the infinite boost limit
of the usual black-hole coset model. This prescription
refers to the Euclidean black-hole coset by boosting the compact subgroup
generated by $i\sigma_2$, whereas the Lorentzian model
can be treated similarly provided that one boosts the generator of the non-compact
subgroup $\sigma_3(\beta)$.
In either case, one
arrives at the same description of the coset model $SL(2, R)_k/E(1)$ in terms
of the Liouville field \eqn{redeq} which is always
{\em space-like} and independent of
the axial or vector gauging.

The gauging of the action can be
worked out in detail for all values of $\beta$ in order to obtain a systematic
expansion of the target space geometry in powers of
${\rm exp}(-4 \beta)$ for the boosted abelian subgroup.  More precisely, using
\be
i\sigma_2(\beta) = {1 \over 2} e^{2 \beta} \left(\sigma^+ - e^{-4\beta}
\sigma^- \right) ,
\ee
the action for the boosted model is written in the form
\ba
& & S(g; A, \bar{A})  =  S_{\rm WZW} -
{k \over 2 \pi} \int d^2 z {\rm Tr} \left(A\sigma^+ \bar{\partial} g g^{-1}
\pm \bar{A} \sigma^+ g^{-1} \partial g \pm A \bar{A} \sigma^+ g \sigma^+ g^{-1}
\right)  \nonumber\\
& & ~~~~~~ + {k \over 2 \pi} e^{-4\beta}  \int d^2 z {\rm Tr}
\left(A\sigma^- \bar{\partial} g g^{-1}
\pm \bar{A} \sigma^- g^{-1} \partial g
\pm A \bar{A} (\sigma^- g \sigma^+ g^{-1}
+ \sigma^+ g \sigma^- g^{-1} \pm 2) \right)  \nonumber\\
& & ~~~~~~ \mp {k \over 2 \pi} e^{-8 \beta} \int d^2 z {\rm Tr} \left(A \bar{A}
\sigma^- g \sigma^- g^{-1} \right) ~,
\ea
using the nilpotent elements
$\sigma^+ = \sigma_3 + i\sigma_2$ and $\sigma^- = \sigma_3 - i\sigma_2$.
Here, the $\pm$ signs refer to the axial and vector gauging, respectively, which
are both treated together. Also, $A$ and $\bar{A}$ are rescaled appropriately to
absorb the factor ${\rm exp}2\beta$, which appears in the definition of the boosted
generator and becomes infinite when $\beta \rightarrow \infty$.
Fixing the gauge and performing the integration over the fields
$A$ and $\bar{A}$, the effective action turns out to be \eqn{redeq} plus subleading
terms of order ${\cal O}({\rm exp}(-4 \beta))$ that account for the coupling of
the other field $\chi$. Further
details can be found in the literature, \cite{iran1, iran2, iran3}.

The action that results in this case describes the classical geometry of the
coset $SL(2, R)_k/E(1)$ when $k$ tends to infinity. Quantum corrections
will also change $k$ to $k-2$, as it is customary in WZW models, but there
are no other modifications since the geometry of target space is one-dimensional.
Rescaling $\phi$ with $\sqrt{(k-2)/2}$, so that the corrected kinetic term of
Liouville theory \eqn{redeq} becomes canonically normalized, the linear dilaton term
gives rise to the stress-energy tensor \eqn{canonst} with background charge
$Q = \sqrt{2/(k-2)}$ that becomes infinite as $k$ approaches 2.
Thus, null gauging selects only the Liouville
sector of the black-hole coset and provides the effective description of the radial
coordinate $r$ associated to the generator $\sigma_1$. On the other hand, the
tensionless limit of the ordinary gauged WZW model exhibits a non-trivial structure
after the decoupling of Liouville field, but its form appears to be
non-geometric. The algebraic structure of the residual $SL(2,R)_2/U(1)$ model will be
examined later using the world-sheet formulation of two-dimensional cosets.

We conclude this section with a few remarks that will further clarify the meaning of
the infinite boost in the Lie algebra of non-compact WZW models from different points
of view. We first consider the ungauged non-compact WZW model and explain how
null strings arise by accelerating its semi-classical solutions to very high Lorentz
factor. Then, the quantization of classical null strings can be attempted as usual,
\cite{ulf1, ulf2}, but their theory cannot capture the properties of the $SL(2, R)_2$ model
which is taken directly at the quantum level. We will also consider the gauged WZW
model $SL(2, R)_k/U(1)$ and compare different limits that describe high energy strings.

Recall that the $SL(2,R)_k$ WZW model admits short and long
string solutions using the spectral flow of different geodesics in $AdS_3$,
\cite{malda, seiwit}. Short strings correspond to {\em time-like geodesics} which
are described by group elements
\be
g= U \left(\begin{array}{ccc}
{\rm cos} (\alpha \tau) & & {\rm sin}(\alpha \tau) \\
  &  &  \\
-{\rm sin}(\alpha \tau) & & {\rm cos}(\alpha \tau)
\end{array} \right) V
\ee
where $U$, $V$ belong in $SL(2,R)$ and $\tau$ is the world-sheet time coordinate; if
$U=V=1$, the solution represents a particle sitting at the center of $AdS_3$. Their
monodromy matrix is
\be
M=\left(\begin{array}{ccc}
{\rm cos} (\alpha \pi) & & {\rm sin}(\alpha \pi) \\
  &  &  \\
-{\rm sin}(\alpha \pi) & & {\rm cos}(\alpha \pi)
\end{array} \right) \in SO(2)
\label{mono1}
\ee
and belongs to the elliptic conjugacy class of $SL(2, R)$ generated by $i\sigma_2$.

On the other hand, long strings correspond to {\em space-like geodesics} which
are described by group elements
\be
g = U \left(\begin{array}{ccc}
e^{\alpha \tau} & & 0\\
  &  &  \\
0 &   & e^{-\alpha \tau}
\end{array} \right) V
\ee
where $U$, $V$ belong in $SL(2,R)$;
if $U=V=1$, the solution is a straight line cutting the space-like section
$t \equiv \tilde{\theta} =0$ of
$AdS_3$ diagonally. In this case, their monodromy matrix is
\be
M=\left(\begin{array}{ccc}
e^{\alpha \pi} & & 0\\
  &  &  \\
0 &   & e^{-\alpha \pi}
\end{array} \right) \in SO(1,1)
\label{mono2}
\ee
and belongs to the hyperbolic conjugacy class of $SL(2,R)$ generated by $\sigma_3$.

Since both elements $i\sigma_2$ and $\sigma_3$ are transformed to $\sigma^+$ by
performing a boost \eqn{bosta} with very high Lorentz factor, we find that the monodromy
matrices \eqn{mono1} and \eqn{mono2}
correspond to the parabolic conjugacy class $E(1)$ of $SL(2,R)$ which arises by
contraction in either case.
As a result, there is no distinction between short and long strings since
they both correspond to {\em null geodesics} in $AdS_3$ when $\beta \rightarrow \infty$.
Thus, null strings make their appearance as spectral flows of geodesics in the
$SL(2, R)$ model in the limit that all string solutions are speeding with infinitely
large Lorentz factor\footnote{As such they should be compared to the ultra-relativistic limit
of rotating strings in $AdS_3 \times S^3$, which become effectively tensionless in the
limit of large angular momentum in $S^3$, \cite{mateos}, and have applications to
weakly coupled Yang-Mills theory within the AdS/CFT correspondence.}. It should be noted
in this context that the spectral flow stretches a geodesic solution
in the time direction (parametrized by $\tilde{\theta}$) by adding $w\tau$ and rotates
it around the center $r=0$ of $AdS_3$ by adding $w\sigma$ to the angular coordinate
$\theta$. The resulting classical solutions depend on the spatial world-sheet coordinate
$\sigma$ and describe circular strings that wind $w$ times around the center of the space.
For null strings, however, the world-sheet is degenerate and every point moves independently
along a null geodesic, \cite{schild, vega}, as the speed of light on the world-sheet
becomes effectively zero. This will be possible only if the corresponding
semi-classical solutions have $w=0$ so that there is no $\sigma$ dependence. Put differently,
the spectral flow is not a symmetry of the $SL(2,R)_k$ model when the generators
of the Lie algebra \eqn{bosta} are infinitely boosted, which also explains why the
Hilbert space of WZW model does not contain states other than the long and short
strings associated to the spectral flow of the continuous and discrete representations.
In any case, we will not
attempt to discuss the quantization of classical null strings, as the approach we are
following here is taken directly at the quantum level of WZW models and the results are
expected to be different: the current algebra $SL(2,R)_2$
still exhibits the symmetry of spectral flow in the quantum tensionless limit
and the world-sheet never becomes degenerate
in our approach.

Next, passing to the coset model $SL(2,R)_k/U(1)$, we observe that
only the spatial direction $\sigma_1$ survives when $\beta
\rightarrow \infty$, since it is perpendicular to the direction of
the boost and it remains unaffected. Thus, after all, it is not
surprising that the null gauging yields a Liouville field for the
radial coordinate $r$, which parametrizes the gravitational sector
of the model. This reduction is quite different from the
simplifications that arise in the vicinity of cosmological
singularities, where the spatial points become causally
disconnected and gravity reduces to one-dimensional system that
depends only on time. Such a limit was studied extensively in
four-dimensional gravity, following the original work of Belinski,
Khalatnikov and Lifshitz (BKL), \cite{bkl1}, but it also received
attention in modern day string theory and in the small tension
expansion of M-theory, \cite{bkl2}; alternatively, it can be
viewed as a strong coupling limit of gravity so that the Planck
mass becomes zero. In the BKL case, the bulk tensionless limit of
the gravitational theories is defined in an ultra-relativistic way
by letting the velocity of light in target space become zero.
Then, time derivatives of fields dominate over their spatial
gradients and they subsequently lead to an ultra-local reduction.
As a result, the target space is parametrized by only one
coordinate, which is clearly time-like for space-times with
Lorentzian signature; the same limit can also arise in spaces with
Euclidean signature, upon analytic continuation, but the variable
that remains cannot be interpreted as the radial coordinate $r$.

It is instructive to study the difference between the two reductions in the context
of high energy strings. Recall that there are two different ways to obtain a string
of very high energy: either accelerate an ordinary microscopic string to very high
Lorentz factor or stretch it to become very large of macroscopic size.
However, not both of them correspond to the tensionless limit. The first case
yields ultra-relativistic strings which are always tensionless
as in the BKL limit. The tensionless limit also describes high
energy strings which propagate in spaces with compact spatial directions because their
size is bounded from above; as a result, the kinetic energy dominates the contribution
of their tension and the strings become effectively tensionless in the high energy limit.
However, for string propagation in non-compact spaces, the
contribution of their tension can also grow large without bound and turn to
macroscopic objects at very high energies. Clearly, in this case, the high energy limit
is not ultra-relativistic and the strings do not behave as tensionless objects.

With this in mind, we may now look at string configurations in the
semi-classical geometry of a two-dimensional black-hole (or
$AdS_3$ for the same matter). The radial variable $r$ parametrizes
the non-compact spatial direction and high energy strings can also
stretch to macroscopic size. There is no other way to describe
tensionless high energy strings in these models but decouple the
radial degree of freedom from the remaining variables. This
provides a classical prescription that prevents the appearance of
macroscopic high energy strings, but need not be always
appropriate to use for addressing questions in quantum WZW models
at critical level. However, it helps to explain the physical
difference between the two dimensional reductions -- BKL versus
null gauging -- and provides an intuitive (yet classical) way to
think about the decoupling of gravity in the tensionless limit of
WZW models. Of course, this decoupling also arises in the
tensionless limit of the quantum theory, since otherwise the
conformal dimensions \eqn{spect2} receive contributions from the
Liouville field that become infinite at $k=2$.

\section{World-sheet symmetries of the $SL(2, R)_k/U(1)$ coset}
\setcounter{equation}{0}

In this section we outline the construction of an infinite
dimensional chiral algebra, which acts as extended world-sheet
symmetry of the gauged WZW model $SL(2, R)_k/U(1)$ for all $k \geq
2$. The results we present here are based on earlier work using
the concept of non-compact parafermions and their free field
realization in terms of two scalar fields with background charge.
We arrive at a rather complicated algebraic structure that
simplifies considerably in two special limits, as $k\rightarrow
\infty$ and $k \rightarrow 2$. We first review the essential
details of our working framework, using the parafermion currents
to define the $W$-algebra of coset models, and then apply the
results to the tensionless limit where it is found that the
world-sheet symmetry is a higher spin truncation of the algebra
$W_{\infty}$ with spin $s \geq 3$. In this case, the Virasoro
algebra decouples consistently from all commutation relations, and
contracts to an abelian structure.  In the other limit, $k
\rightarrow \infty$, the symmetry algebra of the model is
$W_{\infty}$, which is generated by all integer spins $s \geq 2$.
Thus, the interpolation between the two algebraic structures may
be used further to provide a systematic algebraic framework for
studying the role of $1/\alpha^{\prime}$ corrections in gauged WZW
models.

\subsection{Parafermions and W-algebras}

The two-dimensional conformal field theories of gauged WZW models contain
a collection of chirally conserved currents $\psi_l (z)$, and their Hermitean
conjugates $\psi_l^{\dagger}(z) = \psi_{-l} (z)$, which are
semi-local fields that interpolate between bosons and fermions.
In particular, the parafermion
currents $\psi_l(z)$, and their anti-holomorphic partners $\bar{\psi}_l(\bar{z})$,
have fractional conformal dimensions which are determined by the mutual locality
exponent with respect to the monodromy properties of their correlation functions.
Parafermion algebras were
first introduced in the simplest family of $SU(2)_N/U(1)$ models,
\cite{zamo}, which have
a $Z_N$ symmetry that accounts for their charges; here $l= 0, 1, \cdots , N-1$
and $\psi_0(z) = \psi_0^{\dagger}(z) = 1$.
In this case, the central charge of the Virasoro algebra is $c_{\psi}=2(N-1)/(N+2)$
and ranges from $1/2$ to $2$, as $N$ assumes all integer values $2, 3, \cdots$
which are allowed
by unitarity. Their introduction proves advantageous for the systematic
description of all primary fields on group manifolds and their cosets, such as
$SU(2)_N$ and $SU(2)_N/U(1)$,
since the two classes of conformal field theories are related to each other by
subtracting and then adding back free bosons. At the same time, parafermions
can also be used to construct unitary representations of the ${\cal N} = 2$
superconformal algebra with central charge $c= 3N/(N+2) = c_{\psi} + 1$.

The basic structure of the parafermion algebra is described by the operator
product expansions, \cite{zamo},
\ba
\psi_{l_1} (z) \psi_{l_2} (w) & = & C_{l_1 , l_2} (z-w)^{\Delta_{l_1 + l_2} -
\Delta_{l_1} - \Delta_{l_2}} \left(\psi_{l_1 + l_2} (w) + {\cal O}(z-w)
\right) , \label{opepaa}\\
\psi_{l_1} (z) \psi_{l_2}^{\dagger} (w) & = & C_{l_1, -l_2 -k}
(z-w)^{-2 \Delta_{l_2}} \left(\psi_{l_1 - l_2} (w) + {\cal O} (z-w) \right) ,
\label{opepab}\\
\psi_{l} (z) \psi_l^{\dagger} (w) & = & (z-w)^{-2 \Delta_l}
\left(1 + {2 \Delta_l \over c_{\psi}} (z-w)^2 T_{\psi} (w) +
{\cal O} (z-w)^3 \right) \label{opepa}
\ea
where $C_{l_1, l_2}$ are appropriately chosen structure constants determined
by associativity. Also,
$\Delta_l$ is the conformal dimension of $\psi_l(z)$ and $\psi_l^{\dagger}(z)$,
which equals to $l(N-l)/N$ in the simplest case of the $SU(2)_N/U(1)$
WZW model. It is also implicitly assumed that
$l_1 > l_2$ in the operator product expansion \eqn{opepab}.
The operator product expansion \eqn{opepa}, $\psi_l(z) \psi_l^{\dagger}(w)$, gives
rise to the stress-energy tensor $T_{\psi}$ with central charge $c_{\psi}$,
as well as to a collection of
other chiral currents with integer spin that appear to higher orders in the
power series expansion. These currents, in turn, form the extended conformal
algebra of the model, which is known as $W$-algebra; for example, the $W$-algebra
of the coset $SU(2)_N/U(1)$ is generated by chiral fields with integer spin
$2, 3, \cdots N$ and it is denoted by $W_N$. Further generalizations to
higher dimensional coset models are also known, but their structure is more
intricate for non-abelian gauging. The commutation relations of $W$-algebras are
in general non-linear. For a collection of papers on the subject, see, for
instance, \cite{walg}.

Here, we are mainly concerned with the generalization of parafermions to non-compact
groups, such as $SL(2, R)_k$, and the construction of the $W$-algebra of the
corresponding coset model
$SL(2, R)_k/U(1)$ for all $k \geq 2$; special emphasis will be placed later in
the particular limit $k=2$. Recall that
non-compact parafermions were initially introduced as a tool to break
the $c=3$ barrier of the ${\cal N}=2$ superconformal algebra and construct
unitary representations with $c>3$ from the $SL(2, R)_k$ algebra
by subtracting and then adding back a free boson, \cite{peskin, lyken, oscar}.
We may formally pass from the
compact to the non-compact coset by letting $N=-k$ with $k$ ranging
continuously from
$2$ to infinity. In this case, the number of independent parafermion fields
$\psi_l (z)$ is infinite,
as there is no $Z_N$ symmetry to truncate the number of their components to
$N-1$. Also, the operator product expansions \eqn{opepaa}--\eqn{opepa} can be
extended to the non-compact model in a straightforward way, setting
\be
\Delta_l = {l (k+l) \over k}
\ee
and
\be
c_{\psi} = 2 {k+1 \over k-2} ~, \label{cent} ~.
\ee
These expressions follow from the corresponding values of the $SU(2)_N/U(1)$
coset model by changing $N$ to $-k$, and likewise for
the structure constants of the parafermion
algebra, which are determined to be
\be
C_{l_1, l_2} = \left({\Gamma (k+l_1 + l_2) \Gamma(k) \Gamma (l_1 + l_2 + 1)
\over \Gamma (l_1 + 1) \Gamma (l_2 + 1) \Gamma (k+ l_1) \Gamma (k+ l_2)}
\right)^{1/2} ~.
\ee
We also note that the operator
product expansion of non-compact parafermion currents $\psi_{l_1}(z) \psi_{l_2}(w)$,
which is shown in equation \eqn{opepaa}, contains no singular terms, since
the exponents $\Delta_{l_1 + l_2} - \Delta_{l_1} - \Delta_{l_2} =
2l_1 l_2 / k$ are always positive; thus, the operators $\psi_l (z)$
commute among themselves for all positive values of
$l=0, 1, 2, 3, \cdots $. Non-trivial
commutation relations arise only among $\psi_l(z)$ and their Hermitean conjugate
partners.

Clearly, the infinitely generated algebra of non-compact parafermions allows
for $c_{\psi} \geq 2$, and as a result the tensionless limit
$c_{\psi} \rightarrow \infty$ is reached by letting $k \rightarrow 2$; this
possibility does not arise for the compact coset model, since the central charge of
its Virasoro algebra can never exceed the $c_{\psi} = 2$ barrier. The non-compact
parafermion currents appear to have integer or half-integer dimensions in the
tensionless limit, which are given by the special values
\be
\Delta_l(k=2) = {1 \over 2}l(l+2) ~; ~~~~~~ l=0, 1, 2, 3, \cdots ~.
\ee
In reality, however, these conformal dimensions are all zero since the
Virasoro algebra with infinite central charge contracts to an abelian structure
by appropriate rescaling of the stress-energy tensor; it is a simple consequence
of the singular nature of the conformal field theory $SL(2, R)_k/U(1)$ at $k=2$.
In any case, experience with operator algebras suggests that the corresponding
$W$-algebra should be linear.
In fact, we will be able to determine its exact algebraic structure and
show that it can be identified with a consistent higher spin truncation of
$W_{\infty}$, whereas the Virasoro algebra becomes abelian and it
decouples naturally from the spectrum of the world-sheet currents. This decoupling
can already be seen in equation \eqn{opepa}, since the coefficient of the term
$T_{\psi}$ becomes zero as $c_{\psi} \rightarrow \infty$; the same result
holds after rescaling $T_{\psi}$ by $\sqrt{k-2}$ in order to make sense
of the tensionless limit as singular conformal field theory.

It is convenient to introduce free field realizations of the operators that
arise in two-dimensional conformal field theories in order to simplify calculations
with abstract operator algebras and compute correlation functions. Thus, for the
$SL(2, R)_k/U(1)$ model, we introduce two free scalar fields
$\{\phi_i (z); i=1, 2\}$
and represent the parafermion currents $\psi_1(z)$ and
$\psi_{-1}(z)$ as follows, \cite{peskin, lyken},
\be
\psi_{\pm 1} (z) = {1 \over \sqrt{2k}} \left(\mp \sqrt{k-2} \partial \phi_1 (z)
+ i \sqrt{k} \partial \phi_2 (z) \right) {\rm exp} \left(\pm i \sqrt{{2 \over k}}
\phi_2 (z) \right) \label{ffrep}
\ee
for all $k \geq 2$. The fields $\phi_i(z)$ are both space-like with two point
functions
\be
<\phi_i (z) \phi_j (w)> = - \delta_{ij} {\rm log} (z-w) ~.
\ee
The expressions \eqn{ffrep} follow from the realization of
$SU(2)_N/U(1)$ parafermion currents in terms of two free scalar fields, using the
formal continuation $N \rightarrow -k$. The higher parafermion currents $\psi_l (z)$
also admit free field realizations, but their exact description will not be needed
in the present work.

The parafermion algebra can be converted into the $SL(2, R)_k$ current algebra by
introducing operators
\be
J^{\pm} (z) = \sqrt{k} \psi_{\pm 1} (z) {\rm exp} \left(\pm \sqrt{{2 \over k}}
\chi (z) \right) , ~~~~~ J^3 (z) = - \sqrt{{k \over 2}} \partial \chi (z) ~,
\ee
which dress the parafermions with the addition of an extra free scalar field
$\chi (z)$; likewise, there is a
parafermionic construction of the ${\cal N} = 2$ superconformal algebra, \cite{peskin}.
We have, in particular, the operator product expansions\footnote{There are two different
ways in the literature to define the currents $J^{\pm}$ by considering $J^1 \pm i J^2$ or
$iJ^1 \mp J^2$. Their hermiticity properties are different and likewise the hermiticity
properties of the corresponding parafermions $\psi_{\pm 1}$ are also different.}
\ba
J^+ (z) J^- (w) & = & {k \over (z-w)^2} -2 {J^3 (w) \over z-w} ,
\nonumber\\
J^3 (z) J^{\pm} (w) & = & \pm {J^{\pm} (w) \over z-w} ,
\nonumber\\
J^3 (z) J^3 (w) & = & -{k/2 \over (z-w)^2} ,
\ea
which provide a free field realization of the $SL(2, R)_k$ current
algebra in terms of three scalar fields $\phi_1 (z)$, $\phi_2 (z)$ and
$\chi (z)$. In this case, the stress-energy tensor of the
coset model $SL(2, R)_k/U(1)$,
which arises in the operator product expansion
$\psi_1(z) \psi_1^{\dagger}(w)$,
is realized in terms of two free scalar fields, as
\be
W_2 (z) \equiv T_{\psi} (z) = -{1 \over 2} (\partial \phi_1)^2 -{1 \over 2}
(\partial \phi_2)^2 + {1 \over \sqrt{2(k-2)}} \partial^2 \phi_1 ~, \label{parste}
\ee
where one of them, denoted by $\phi_1 (z)$, appears with background charge and
accounts for the value \eqn{cent} of the central charge of the Virasoro
algebra for all $k$. Note that the expression \eqn{parste} coincides with
the stress-energy tensor \eqn{pstrem} that was introduced earlier in our
discussion of the spectrum.

Following the bootstrap method of conformal field theory, we define the
{\em primary} higher spin generators $W_s(z)$ of the extended conformal
operator algebra of the model $SL(2, R)_k/U(1)$, using the expansion, \cite{hat},
\ba
& & \psi_1 (z+ \epsilon) \psi_{-1} (z) = \epsilon^{-2{k+1 \over k}}
\left(1 + {k-2 \over k} \left(\epsilon^2 + {1 \over 2} \epsilon^3 \partial
+ {3 \over 20} \epsilon^4 \partial^2 + {1 \over 30} \epsilon^5 \partial^3
\right) W_2 (z) \right. \nonumber\\
& & ~~~ \left. - {1 \over 4} \left(\epsilon^3 + {1 \over 2} \epsilon^4 \partial +
{1 \over 7} \epsilon^5 \partial^2 \right) W_3 (z) + {(6k+5)(k-2)^2 \over
2k^2 (16k-17)} \left(\epsilon^4 + {1 \over 2} \epsilon^5 \partial \right)
:W_2^2:(z) \right. \nonumber\\
& & ~~~ \left. + {1 \over 32} \left(\epsilon^4 + {1 \over 2} \epsilon^5 \partial
\right) W_4 (z)
- {(10k + 7) (k-2) \over 4k (64k - 107)} \epsilon^5 :W_2 W_3:(z) \right.
\nonumber\\
& & ~~~ \left. - {1 \over 3 \cdot 2^7} \epsilon^5 W_5 (z)
+ {\cal O} (\epsilon^6) \right) \label{master}.
\ea
Here, normal ordered products are defined, as usual, by subtracting the
singular terms plus the finite terms that are total derivatives of lower
dimension operators appearing in the operator product expansion. Then,
using the free field realization of the parafermion currents $\psi_{\pm 1} (z)$,
we can obtain explicit expressions for the higher spin generators in terms of two
free fields for all values of $k$. Of course, the case $k=2$, which is relevant
in the tensionless limit, is special because the parafermion currents \eqn{ffrep}
are realized in terms of one scalar field only, $\phi_2$, and likewise
for the resulting higher spin generators $W_s(z)$; this case will be treated
separately later.

We may extract the higher spin generators and compute their operator
product expansions in order to identify the structure of the resulting $W$-algebra
for all values of $k$. We will follow the general construction presented in
reference \cite{hat}. The spin 3 operator turns out to be
\ba
W_3(z) & = & 2i \sqrt{{2 \over k}} \left({3k-4 \over 3k}(\partial \phi_2)^3 +
{1 \over 6} \partial^3 \phi_2 + {k-2 \over k} (\partial \phi_1)^2 \partial
\phi_2 \right. \nonumber\\
& & ~~~~~~ \left. + {k-2 \over k} \sqrt{{k-2 \over 2}} \partial^2 \phi_1
\partial \phi_2
- \sqrt{{k-2 \over 2}} \partial \phi_1 \partial^2 \phi_2 \right) ,
\label{spf1}
\ea
which yields the following operator product expansion with itself
\ba
& & W_3(z+ \epsilon) W_3(z) = {16 \over 3}{(k+1)(k+2)(3k-4) \over k^3}{1 \over \epsilon^6}
+ 16 {(k+2)(k-2)(3k-4) \over k^3} \cdot \nonumber\\
& & ~~~~~~ \cdot \left({1 \over \epsilon^4} +
{1 \over 2}{\partial \over \epsilon^3} + {3 \over 20}
{\partial^2 \over \epsilon^2} + {1 \over 30} {\partial^3 \over \epsilon} \right)
W_2(z) + 2{2k-3 \over k} \left({1 \over \epsilon^2} + {1 \over 2}{\partial \over
\epsilon}\right) W_4(z) \nonumber\\
&& ~~~~~~ + 2^7 {(k+2)(3k-4)(k-2)^2 \over k^3 (16k-17)}\left({1 \over \epsilon^2}
+ {1 \over 2}{\partial \over \epsilon}\right) :W_2^2:(z) ~.
\ea

Here, $W_4(z)$ is the primary spin 4 operator which is defined by the parafermionic
operator product expansion \eqn{master}, and it turns out to be
\ba
W_4(z) & = & -{4(3k-4) \over k^2 (16k - 17)} \left((k-12)(2k-1)(\partial \phi_2)^4
-2(2k^2 + 2k +3) (\partial^2 \phi_2)^2 \right) \nonumber\\
& & -{4 (k-2) \over k^2 (16k-17)} \left(8 (k^2 -k+1)\left(\partial \phi_2 \partial^3 \phi_2
+ \partial \phi_1 \partial^3 \phi_1 \right) + (k-2)(6k+5) (\partial \phi_1)^4 \right.
\nonumber\\
& & ~~~~~~ \left. + 6(2k^2 -13k+8) (\partial \phi_1)^2 (\partial \phi_2)^2 -2 (6k^2 -12k +1)
(\partial^2 \phi_1)^2 \right) \nonumber\\
& & + {8 \sqrt{2(k-2)} \over 3k^2 (16k-17)} \left(3(k-2)(6k+5) (\partial^2 \phi_1)
(\partial \phi_1)^2 - 6k(16k-17) \partial \phi_1 \partial \phi_2 \partial^2 \phi_2
\right. \nonumber\\
& & ~~~~~~ + \left. 3(2k-3)(19k-8) \partial^2 \phi_1 (\partial \phi_2)^2 +
(k^2 -k+1) \partial^4 \phi_1 \right) .
\label{spf2}
\ea
Higher spin generators and their operator product expansions can be constructed
recursively in a similar fashion, using the free field
realization of the parafermions $\psi_{\pm 1}(z)$, but their structure becomes
quickly rather involved for all $2<k<\infty$.

For generic values of the Kac-Moody level $k$, it is very
difficult to iterate the algorithm and extract the structure of
the underlying $W$-algebra in closed form together with the free
field realization of its higher spin generators. However, the
boundary values $k=2$ and $k = \infty$ are rather special, since
many simplifications occur and the whole procedure becomes
tractable. As we will see later, there are special {\em
quasi-primary} bases of the $W$-algebra that remove all non-linear
terms in a systematic way when $k$ reaches its boundary values.
This result is also closely related to the local character of the
parafermion currents at $k=2$ and $k=\infty$, since $\psi_{\pm
1}(z)$ are represented as derivatives of fermions and a boson,
respectively. The details will be made available shortly.

Before we proceed further, let us recall a
number of qualitative results which are known for arbitrary $k$ by
performing sample calculations with generators up to spin 5, \cite{hat}.
First, the extended
conformal symmetry of the $SL(2, R)_k/U(1)$ coset model appears to be
infinitely generated\footnote{It is natural to
expect that the extended conformal symmetries of
non-compact coset models are infinitely generated as they correspond to
irrational conformal field theories. In these cases there are no finitely generated
$W$-algebras whose representations can be used to organize the operator content
in terms of a finite number of primary field blocks,
unlike the simpler case of compact coset models.
Likewise, the family of non-compact parafermions $\psi_{\pm l}(z)$ is infinitely
large, in contrast to the compact space parafermions, which form a
finite family. It should be noted, however, that the $W$-algebra that arises here
has null fields of spin $s \geq 6$, and as a result it appears to be non-freely
generated for all values of $k$, \cite{feher}. This occurrence is closely connected
to the existence of non-trivial unitary quasi-free representations of
$W_{\infty}$-type algebras, but we will not consider their implications
any further.}
by chiral operators $W_s(z)$ with all integer spin
$s \geq 2$. Second, for generic values of $k$, the resulting $W$-algebra is non-linear
as there is no field redefinition that can bring it into a linear form. Thus,
it is a non-linear deformation of the $W_{\infty}$ algebra, with coefficients
that depend on $k$; for this reason, the world-sheet symmetry of the
black-hole coset is denoted by $\hat{W}_{\infty}(k)$ for generic $k$.
Third, we may formally extend the validity of the algebra
$\hat{W}_{\infty}(k)$ to all real values $-\infty < k < +\infty$ and set $k=-N$,
where $N$ is any positive integer greater or equal than 2. Then, it appears
that all higher spin generators with $s>N$ become null, as it can be seen
by considering the two-point functions $<W_s(z+ \epsilon) W_s(z)>$. In this
case, all generators with $s>N$ can
be consistently set equal to zero, thus rendering $\hat{W}_{\infty}(k=-N)$
finitely generated and isomorphic to the algebra $W_N$. This result is also
consistent with the formal relation between the two coset models $SL(2, R)_k/U(1)$
and $SU(2)_N/U(1)$ for $k=-N$. Thus, the algebra $\hat{W}_{\infty}(k)$ is
rather universal, as it can be viewed as a continuous generalization
of the $W_N$ algebras for all {\em real} values of the level $k$.

Within this
general framework we also encounter the world-sheet symmetry
of the tensionless non-compact coset model $SL(2, R)_2/U(1)$, which is
clearly identified with $\hat{W}_{\infty}(k=2)$.
The detailed description of its algebraic
structure is one of the primary goals of the present work. At the same time,
the non-linear algebra $\hat{W}_{\infty}(k)$, which appears for generic values
of $k$, provides a concrete framework for exploring the symmetries of the model
at finite tension, and may offer an understanding of the exact nature of
$1/{\alpha}^{\prime}$ corrections from the world-sheet viewpoint; it is an
important problem with far reaching consequences that should be investigated in
the future. Finally, we note for completeness that much larger
symmetries may arise in the tensionless limit with $\hat{W}_{\infty}(k=2)$
being the smallest subalgebra of a much larger world-sheet symmetry group;
for example, we may include all higher parafermion currents
$\psi_{\pm l}(z)$ and investigate whether they all form an enlarged symmetry group
together with the higher spin generators $W_s(z)$ at $k=2$. Such generalizations
are also lying beyond the scope of the present work.

The complete structure of the $\hat{W}_{\infty}(k)$
algebra can be determined alternatively,
without relying on the specific realization of the parafermion currents in terms of
free fields,
using the correlation functions of the elementary fields $\psi_{\pm 1}(z)$.
Recall at this point the recursive relations among the parafermionic correlation
functions,
\ba
& & <\psi_1 (z_1) \cdots \psi_1 (z_n) \psi_1^{\dagger} (w_1) \cdots
\psi_1^{\dagger} (w_n)> = \prod_{i=2}^n (z_1 - z_i)^{2/k}
\prod_{j=1}^n {1 \over (z_1 - w_j)^{2/k}} \nonumber\\
& & ~~~~~~ \cdot \sum_{a=1}^n \left({1 \over (z_1 - w_a)^2} -
{2 \over k(z_1 - w_a)} \left(\sum_{l=2}^n {1 \over w_a - z_l} -
\sum_{m \neq a} {1 \over w_a - w_m}\right) \right) \nonumber\\
& & ~~~~~~ \cdot \prod_{q=2}^n {1 \over (z_q - w_a)^{2/k}}
\prod_{p=1}^{a-1} (w_p - w_a)^{2/k} \prod_{r=a+1}^n (w_a - w_r)^{2/k}
\nonumber\\
& & ~~~~~~ \cdot <\psi_1(z_2) \cdots \psi_1 (z_n) \psi_1^{\dagger} (w_1)
\cdots \hat{\psi}_1^{\dagger} (w_a) \cdots \psi_1^{\dagger}(w_n)> ~,
\ea
which determine all such $2n$-correlation functions in terms of
lower $2(n-2)$-correlators. They follow from the corresponding relations
for the compact coset parafermions, \cite{zamo}, by setting $k=-N$.
Then, since $<\psi_1(z) \psi_1^{\dagger}(w)>
= 1/(z-w)^{2(k+1)/k}$, as follows from equation \eqn{opepa},
we easily obtain
\be
<\psi_1(z_1) \psi_1^{\dagger} (z_2) \psi_1 (z_3) \psi_1^{\dagger} (z_4)>
= \left({z_{13} z_{24} \over z_{12} z_{14} z_{34} z_{23}} \right)^{2/k}
\left({1 \over z_{12}^2 z_{34}^2}
\left(1 - {2 \over k} {z_{12} z_{34}
\over z_{23} z_{24}} \right) + (z_2 \leftrightarrow z_4) \right) ,
\label{4ptf}
\ee
where $z_{ij} = z_i - z_j$. Likewise, we may obtain explicit expressions
for the six-point correlation functions and so on.

Using the parafermionic four-point correlation function \eqn{4ptf}, as well as
the operator product expansion \eqn{master} that contains the chiral fields $W_s(z)$
in power series, we may obtain the two-point correlation functions among the
$W$-algebra generators
\ba
<W_2 (z) W_2 (0)> &=& {k+1 \over k-2} {1 \over z^4} ~, \nonumber\\
<W_3 (z) W_3 (0) > &=& {16 \over 3} {(k+1) (k+2) (3k-4) \over k^3}
{1 \over z^6} ~,  \nonumber\\
<W_4 (z) W_4 (0)> &=& {2^{10} (k+1) (k+2) (k+3) (2k-1) (3k-4) \over k^4 (16k -17)}
{1 \over z^8} ~, \\
<W_5 (z) W_5 (0)> &=& {9 \cdot 2^{15} (k+1) (k+2) (k+3) (k+4)
(2k-1) (5k-8) \over 5k^5 (64k-107)} {1 \over z^{10}} ~, \nonumber
\ea
and so on for all other higher spin fields. They account for the central terms
appearing in the commutation relations of the non-linear algebra
$\hat{W}_{\infty}(k)$. Continuing further, we may compute the structure constants
of the algebra that appear as coefficients in the singular terms of the operator
product expansion $W_s(z + \epsilon) W_{s^{\prime}}(z)$ for all generators.
This is achieved by considering the
parafermionic six-point function
$<\psi_1(z_1) \psi_1^{\dagger} (z_2) \psi_1 (z_3) \psi_1^{\dagger} (z_4) \psi_1 (z_5)
\psi_1^{\dagger}(z_6)>$, which is naturally related to the three-point functions of
$W$-generators when combined with the expansion \eqn{master}. Unfortunately, although
this procedure is straightforward, it is rather cumbersome to implement in all
generality in order to extract the complete structure of $\hat{W}_{\infty}(k)$
in closed form for arbitrary values of the level $k$. Thus, either way, the exact
structure of the non-linear algebra $\hat{W}_{\infty}(k)$ remains out of reach
in all generality.

The remaining part of this section is devoted to the derivation of the linear
structures that arise in the two special limits $k \rightarrow \infty$ and
$k \rightarrow 2$. The first is already known in the literature and it will be
only briefly discussed for completeness. The second is new and it appears here
for the first time. Both limiting values will be given a systematic description
in section 6 in the framework of $W_{\infty}$-type
algebras, \cite{bak1, bak2, prs1, prs2, bak3}.

\subsection{W-algebra at $k= \infty$}

The algebra $\hat{W}_{\infty} (k)$ linearizes in the limit $k
\rightarrow \infty$, as it can be easily seen by introducing an
appropriate {\em quasi-primary} basis for its higher spin
generators. Note that in the large $k$ limit the basic parafermion
fields become local and they simplify to the $U(1)$ currents
\be
\psi_{1}(z) = i \partial \phi (z) ~, ~~~~~~ \psi_{-1}(z) = i
\partial \bar{\phi}(z) ~, \ee where $\phi(z)$ is a complex free
boson \be \phi(z) = {1 \over \sqrt{2}} \partial (\phi_2 + i
\phi_1) ~.
\ee
Then, their operator product expansion can be
written in the form
\be
\psi_1(z+ \epsilon) \psi_{-1} (z) = { 1
\over \epsilon^2} \left( 1 + \sum_{s=2}^{\infty} {(-1)^s (2s-1)!
\epsilon^s \over 2^{2(s-2)}(s-1)! (s-2)!} \sum_{n=0}^{\infty}
{(s+n-1)! \epsilon^n \over n! (2s+n-1)!}
\partial^n \tilde{W}_s (z) \right) \label{muster}
\ee
where $\tilde{W}_s(z)$ are appropriately chosen {\em quasi-primary} generators
that absorb all non-linear terms of the operator product expansion \eqn{master}
as $k \rightarrow \infty$.

The $W$-generators that result in this case are given by simple bilinear
expressions in free field realization
\ba
& & \tilde{W}_2(z) = W_2(z) = - \partial \phi \partial \bar{\phi} ~, \nonumber\\
& & \tilde{W}_3(z) = W_3(z) = -2 \left(\partial \phi \partial^2 \bar{\phi} -
\partial^2 \phi \partial \bar{\phi} \right) , \nonumber\\
& & \tilde{W}_4(z) = W_4(z) + 6 :W_2^2:(z) =
-{16 \over 5} \left(\partial \phi \partial^3 \bar{\phi}
-3 \partial^2 \phi \partial^2 \bar{\phi} + \partial^3 \phi \partial
\bar{\phi} \right)
\ea
and so on. $\tilde{W}_2(z)$ is the stress-energy tensor of the model with central
charge equal to its classical value $c=2$,
whereas all other operators $\tilde{W}_s(z)$ are higher
spin currents with $s=3, 4, \cdots$.

More generally, it is known that the complete system of {\em quasi-primary}
operators is given in the large $k$ limit by the general expression, \cite{bak2},
\be
\tilde{W}_s(z) = {2^{s-3} s! \over (2s-3)!!(s-1)} \sum_{k=1}^{s-1} (-1)^k
{s-1 \choose k}{s-1 \choose k-1} \partial^k \phi \partial^{s-k} \bar{\phi}
\label{ffrep1}
\ee
for all values of spin $s \geq 2$. Obviously,
the operator product expansion among these operators,
$\tilde{W}_s(z) \tilde{W}_{s^{\prime}}(w)$,
leads to a linear algebra thanks to
the bilinear form of the generators \eqn{ffrep1}.
The commutation relations can also be written
in Fourier modes using the standard prescription
\be
[\tilde{W}_n^s , \tilde{W}_m^{s^{\prime}}] = \oint_{C_0}
{dw \over 2 \pi i} w^{m+s^{\prime} -1}
\oint_{C_w} {dz \over 2 \pi i} z^{n+s-1}
\tilde{W}_s (z) \tilde{W}_{s^{\prime}} (w) ~,
\ee
where $C_w$ is a contour around $w$ and $C_0$ a contour around 0.
The algebra that results in this case is denoted by $W_{\infty}$ and
its structure will be described in the next section together with
several other technical details that are also relevant for the tensionless
limit of the coset model.

\subsection{W-algebra at $k=2$}

Next, we examine the structure of $\hat{W}_{\infty}(k)$ when the level of the
$SL(2, R)_k$ algebra assumes its critical value, $k=2$.
In this case, the background charge of the $\phi_1$ boson
becomes infinite and appropriate rescaling of the stress-energy tensor is
required in order to make the parafermionic Virasoro algebra well defined.
We first rescale the operator $W_2(z)$ shown in equation \eqn{parste}
by $\sqrt{k-2}$ and then
take the limit $k \rightarrow 2$, which leads to the identification
$\tilde{W}_2(z) \sim
\partial^2 \phi_1$, as in equation \eqn{mamiri} before.
The rescaling amounts to an abelian contraction of the Virasoro algebra, since
$\tilde{W}_2(z)$ is the derivative of a $U(1)$ current with
$\tilde{W}_2(z + \epsilon) \tilde{W}_2(z) \sim 1 / \epsilon^4$.
Introducing Fourier modes
\be
\tilde{L}_n =
{1 \over 2\pi i} \oint_0 dz \tilde{W}_2(z) z^{n+1} \cdot
\left\{ \begin{array}{ll} 1/n ~, ~~~~~
& {\rm for} ~~ n \neq 0 ~, \\
1 ~, ~~~~~
& {\rm for} ~~ n=0 ~, \end{array} \right.
\ee
which are also conveniently rescaled by $n$, we arrive at the $U(1)$ current algebra
\be
[\tilde{L}_n , \tilde{L}_m] = n \delta_{n+m, 0}
\ee
that replaces the Virasoro algebra of the coset model at $k=2$. It is realized by
the field $\phi_1$ alone.

On the other hand, the parafermion currents simplify at $k=2$ and they assume the
following form
\be
\psi_{\pm 1} (z) = {i \over \sqrt{2}} \partial \phi_2 (z)
e^{\pm i \phi_2 (z)}  ~.
\ee
These currents depend only on the scalar field $\phi_2 (z)$ and they are
well defined without the need for rescaling. Then, their operator product expansion
gives rise to $W$-generators, as usual, but they do not contain the
stress-energy tensor in the spectrum. The decoupling arises from the independence
of $\psi_{\pm 1}(z)$ from $\phi_1(z)$, whereas $\tilde{W}_2(z)$ depends only on
it. It can also be seen directly from the operator
product expansion of the parafermions \eqn{master}, where the rescaling of $W_2(z)$
by $\sqrt{k-2}$ eliminates all $W_2$-dependent terms when $k \rightarrow 2$.
Then, in this limit, we claim that the operator product expansion \eqn{master} can
be written in a simpler form, which is analogous to the expansion \eqn{muster},
\be
\psi_1(z+ \epsilon) \psi_{-1} (z) = { 1 \over \epsilon^3} \left(
1 + \sum_{s=3}^{\infty} {(-1)^s (2s-1)! \epsilon^s
\over 2^{2(s-2)}(s-1)! (s-2)!}
\sum_{n=0}^{\infty} {(s+n-1)! \epsilon^n \over n! (2s+n-1)!}
\partial^n \tilde{W}_s (z) \right) , \label{muster2}
\ee
using a new system of appropriately chosen generators
$\tilde{W}_s(z)$ for all $s \geq 3$. This basis is constructed by
absorbing all composite $W$-operators that arise in the operator
product expansion \eqn{master}, as for the large $k$ limit.

It is important to realize in this context that the parafermion currents
$\psi_{\pm 1}(z)$ become total derivatives of a more elementary system of
free fermions with components $(\Psi , \bar{\Psi})$, which are defined to be
\be
\Psi(z) = e^{-i \phi_2(z)} ~, ~~~~~~ \bar{\Psi}(z) = e^{i \phi_2(z)}
\label{bosoni}
\ee
with two-point function
\be
<\Psi(z) \bar{\Psi} (w)> = {1 \over z-w} = <\bar{\Psi}(z)
\Psi (w)> ~. \label{bega}
\ee
In reality they form the components of a fermionic ghost and conjugate ghost
system with dimension $1/2$.
Then, for $k=2$, the operator product expansion of the conjugate fields
\be
\psi_1(z) = {1 \over \sqrt{2}} \partial \bar{\Psi}(z) ~, ~~~~~~
\psi_{-1}(z) = -{1 \over \sqrt{2}} \partial \Psi (z)
\ee
can only give rise to fermion bilinears,
in close analogy with the boson bilinears that arise in the operator product
expansion of the parafermions for $k \rightarrow \infty$. As a result,
it is also expected here that the operator product expansion of the
new currents $\tilde{W}_s(z)$ will lead to an infinite
dimensional algebra with linear commutation relations. Its structure
will be subsequently determined by extracting the exact form of $\tilde{W}_s(z)$
as fermion bilinears and computing their operator product expansions
using the two-point function \eqn{bega}. Equivalently, we may first bosonize
the complex fermions using a free scalar boson $\phi_2$, as given
by the defining relations \eqn{bosoni}, and express all formulae in
terms of $\phi_2$.

We summarize below the result of the computations that were performed in this
case. We have the following
realization of the $W$-generators as fermion bilinears
\be
\tilde{W}_s(z) = {2^{s-4} (s+1)! \over (2s-3)!! (s-1)} \sum_{k=0}^{s-3}
(-1)^k {s-1 \choose k} {s-1 \choose k+2} \partial^{k+1} \bar{\Psi}
\partial^{s-k-2} \Psi (z)   \label{ffrep2}
\ee
for all $s \geq 3$, which are obtained
by taking into account the normalizations appearing in equation \eqn{muster2} above.
The lowest lying field is $\tilde{W}_3(z) = 2 \partial \bar{\Psi} \partial \Psi (z)$,
which equals to the normal ordered product $-4:\psi_1 \psi_{-1}:(z)$ as required by
the power series expansion \eqn{muster2}. Next, we have $\tilde{W}_4(z) =
8(\partial \bar{\Psi} \partial^2 \Psi - \partial^2 \bar{\Psi} \partial \Psi)$ and
so on for all other higher currents that can be readily found from equation
\eqn{ffrep2}.
The result is analogous to the free field realization \eqn{ffrep1} that was
encountered in the limit $k \rightarrow \infty$, but it is not the same.

The currents can be equivalently written using the free field $\phi_2(z)$.
Their bosonization leads to the following expressions for the $W$-generators,
\ba
\tilde{W}^{3}(z) & = &{i \over 3} \left( 2(\partial \phi_2)^3 + \partial^3 \phi_2
\right) , \nonumber\\
\tilde{W}^{4}(z) & = & 4 \left( (\partial \phi_2)^4 +(\partial^{2} \phi_2)^2
\right) , \nonumber\\
\tilde{W}^{5}(z) & = & {8i \over 35} \left(
-84 (\partial \phi_2)^5 + 90 (\partial \phi_2)^2 \partial^3 \phi_2 - 240
\partial \phi_2
(\partial^{2}\phi_2)^2 + \partial^5 \phi_2 \right) , \nonumber\\
\tilde{W}^{6}(z) & = & {64 \over 3} \left(
-4 (\partial \phi_2)^6 + 12(\partial \phi_2)^3 \partial^3\phi_2 -
24 (\partial \phi_2)^2
(\partial^2 \phi_2)^2 - 2(\partial^3 \phi_2)^2 + 2(\partial^2 \phi_2)
\partial^4 \phi_2 \right), \nonumber\\
\tilde{W}^7(z) & = & {128 i \over 693} \left(1980 (\partial \phi_2)^7 - 11970
(\partial \phi_2)^4 \partial^3 \phi_2 + 21420 (\partial \phi_2)^3
(\partial^2 \phi_2)^2 +
420 (\partial \phi_2)^2 \partial^5 \phi_2 \right. \nonumber\\
& & \left. +
6300 \partial \phi_2 (\partial^3 \phi_2)^2 - 6720 \partial \phi_2
\partial^2 \phi_2
\partial^4 \phi_2 + 630 (\partial^2 \phi_2)^2 \partial^3 \phi_2 +
\partial^7 \phi_2
\right) , \ea and so on. The first representatives of this list
should be compared with the corresponding expressions for the
higher spin fields derived for general values of $k$ after taking
the limit of the currents \eqn{spf1} and \eqn{spf2} at critical
level, and they turn out to be the same. The higher spin fields
$\tilde{W}_s(z)$ provide the right basis for the algebra by
absorbing the non-linear terms that arise at higher spins. It is
rather difficult to extract the free boson realization of all
currents in closed form, but this is not really a handicap of our
general construction as we already have them expressed as fermion
bilinears for all $s \geq 3$.

It is now straightforward procedure to verify that the operator
products of these currents take the form
\ba
& & \tilde{W}^3 (z + \epsilon) \tilde{W}^3(z) =
{16 \over \epsilon^6} + \left({1 \over \epsilon^2}
+ {1 \over 2} {\partial \over \epsilon} \right) \tilde{W}^4(z),
\nonumber\\
& & \tilde{W}^3(z + \epsilon) \tilde{W}^4(z) =
96 \left({1 \over \epsilon^4} + {1 \over 3}
{\partial \over \epsilon^3} + {1 \over 14}{\partial^2 \over \epsilon^2}
+ {1 \over 84}{\partial^3 \over \epsilon} \right) \tilde{W}^3 (z) + {5 \over 3}
\left({1 \over \epsilon^2} + {2 \over 5} {\partial \over \epsilon}
\right) \tilde{W}^5 (z), \nonumber\\
& & \tilde{W}^3(z + \epsilon) \tilde{W}^5(z) =
{1440 \over 7} \left({1 \over \epsilon^4} +
{1 \over 4}{\partial \over \epsilon^3} + {1 \over 24}{\partial^2 \over
\epsilon^2} + {1 \over 180} {\partial^3 \over \epsilon} \right) \tilde{W}^4(z) +
{9 \over 4} \left({1 \over \epsilon^2} + {1 \over 3}{\partial \over \epsilon}
\right) \tilde{W}^6(z), \nonumber\\
& & \tilde{W}^4(z + \epsilon) \tilde{W}^4(z) = {1536 \over \epsilon^8} + 8 \left(
{36 \over \epsilon^4} + 18 {\partial \over \epsilon^3} + 5 {\partial^2
\over \epsilon^2} +{\partial^3 \over \epsilon} \right) \tilde{W}^4(z) +
3 \left({1 \over \epsilon^2} + {1 \over 2}{\partial \over \epsilon} \right)
\tilde{W}^6(z),
\nonumber\\
& & \tilde{W}^{4}(z+ \epsilon) \tilde{W}^5(z) = {138240 \over 7}
\left({1 \over \epsilon^6} + {1 \over 3}
{\partial \over \epsilon^5} + {1 \over 14}{\partial^2 \over \epsilon^4} +
{1 \over 84}{\partial^3 \over \epsilon^3} + {5 \over 3024} {\partial^4 \over
\epsilon^2} + {1 \over 5040} {\partial^5 \over \epsilon} \right)
\tilde{W}^3(z)
\nonumber\\
& & ~~~~~~~~~~~~~~~~~~~~~~ + {4080 \over 7} \left({1 \over \epsilon^4} +
{2 \over 5}{\partial \over \epsilon^3}
+ {1 \over 11}{\partial^2 \over \epsilon^2} + {1 \over 66}{\partial^3 \over
\epsilon} \right) \tilde{W}^5(z) + {3 \over 5} \left({7 \over \epsilon^2} +
3{\partial \over \epsilon} \right) \tilde{W}^7(z), \nonumber\\
& & \tilde{W}^3(z+ \epsilon) \tilde{W}^6(z) =
{10240 \over \epsilon^6} \tilde{W}^3(z) + {1120 \over 3} \left(
{1 \over \epsilon^4} + {1 \over 5}{\partial \over \epsilon^3} + {3 \over 110}
{\partial^2 \over \epsilon^2} + {1 \over 330}{\partial^3 \over \epsilon}
\right) \tilde{W}^5(z) \nonumber\\
& & ~~~~~~~~~~~~~~~~~~~~~~~ + {2 \over 5}
\left({7 \over \epsilon^2} + 2 {\partial \over
\epsilon} \right) \tilde{W}^7(z), \label{masope}
\ea
and so on for higher spin commutators.

The infinite dimensional algebra $\hat{W}_{\infty}(2)$ that result in tensionless
limit of the non-compact coset $SL(2, R)_k/U(1)$ can be systematically described as a
higher spin truncation of $W_{1 + \infty}$.
As we will see next, its form is rather unique and provides the extended symmetry
algebra of the model in closed form. The same framework also helps to describe the
linear form of $\hat{W}_{\infty}(k)$ that arises in the large $k$ limit
in a unifying way. For all other values, $2 < k < \infty$,
the complete structure of $\hat{W}_{\infty}(k)$ still
remains out of reach because of the non-linear terms in the commutation relations.

\section{$W_{1+\infty}$ and its higher spin truncations}
\setcounter{equation}{0}

We set up the framework by considering
the infinite dimensional algebra of all differential operators
on the circle, namely $\{f(x) D^n ; n = 0, 1, 2, \cdots \}$,
where $D$ denotes the derivative operator with respect to $x \in S^1$. Their
algebra assumes the form
\be
[f(x) D^n , g(x) D^m] = \left(n f(x) g^{\prime}(x) - m f^{\prime} (x)
g(x) \right) D^{n+m-1} + ~ {\rm lower ~ order ~ terms} ~, \label{lead}
\ee
where the subleading terms are lower order differential operators that follow
by making use of Leibnitz's rule
\be
D^n f(x) = \sum_{k=0}^{n} {n \choose k} f^{(k)}(x) D^{n-k} ~.
\ee
The leading order terms in equation \eqn{lead} give rise to the
algebra of area preserving diffeomorphisms on the cylinder
$T^{\star}S^1$, whereas the inclusion of all subleading terms
provides a non-linear deformation of it, which is also known
as Moyal algebra on $T^{\star}S^1$.

It is more convenient in the sequel to introduce a complex
parameter $z= e^{ix}$ and work with Laurent series in $z$, instead
of trigonometric functions of $x$ defined on the circle, and use
$\partial$ to denote the derivative operator with respect to $z$.
Then, the differential operators $\{z^{n+s-1} \partial^{s-1} \}$
with $n \in Z$ and $s \in Z^{+}$ provide a basis for writing the
commutation relations of the resulting infinite dimensional Lie
algebra. It contains the centerless Virasoro algebra
\be
[L_n ,
L_m] = (n-m) L_{n+m} ~, \label{viro}
\ee
which is generated by the
first order differential operators $L_n = -z^{n+1} \partial$ with
$s=2$ and it is associated with the algebra of point canonical
transformation on $T^{\star}S^1$. Then, the complete algebra of
all differential operators can be viewed as a module of the
centerless Virasoro algebra with each $z^{n+s-1} \partial^{s-1}$
term having conformal weight $s$, \cite{bak1}.

The infinite dimensional algebra $W_{1 + \infty}$ is defined to be
the central
extension of the algebra of all differential operators on $S^1$.
The central extensions are described systematically in the
mathematics literature using the logarithm of the derivative operator,
${\rm log} \partial$, whose commutator is defined to be
\be
[{\rm log} \partial , A] = \sum_{k\geq 1} {(-1)^{k-1} \over k}
a^{(k)} (z) \partial^{n-k}
\ee
when acting on a differential operator $A = a(z) \partial ^n$. The
result is a pseudo-differential operator that involves negative powers of the
derivative operator, and therefore it is natural to consider the pairing
\be
{\cal C}(A, B) = \int {\rm res} \left([A, {\rm log} \partial ] \circ B
\right) \label{cocycle}
\ee
among any two differential operators $A$ and $B$, \cite{bak3}, \cite{khesin, radul1}.
The computation is
performed using the calculus of pseudo-differential operators and
{\em res} is the residue function of the resulting operator
$[A, {\rm log} \partial ] \circ B$ given by the coefficient of its
$\partial^{-1}$ term. For example, for the case of first order differential
operators, one finds
\be
{\cal C}(z^{n+1}\partial , z^{m+1}\partial) = {1 \over 6} (n^3 - n)
\int z^{n+m-1} dz  \sim (n^3 - n) \delta_{n+m, 0} ~,
\ee
which coincides with the usual cocycle formula of the Virasoro algebra that
describes central extensions of \eqn{viro}. More generally, it is known that
the 2-cocycle \eqn{cocycle} provides the
unique non-trivial central extension of the
algebra of all differential operators on the circle, \cite{khesin, radul1}. It can be
computed explicitly for all elements of the algebra, since
\be
{\cal C}(f(z) \partial^n , g(z) \partial^m ) = {n! m! \over (n+m+1)!} \int
f^{(m)}(z) g^{(n+1)}(z) dz ~.
\ee

After this brief outline of the exact mathematical structure, it
is convenient to write down the complete system of commutation
relations of the algebra $W_{1+\infty}$ in closed form by
introducing a basis that diagonalizes the resulting central terms.
Using the system of differential operators, \cite{bak3}, \be V_n^s
= -B(s) \sum_{k=1}^s \alpha_k^s {n+s-1 \choose k-1} z^{n+s-k}
\partial^{s-k} ~, \label{maibas}
\ee
which consist of special series of operators of order up to $s-1$ for all
$s=1, 2, 3, \cdots$, with coefficients
\be
B(s) = {2^{s-3} (s-1)! \over (2s-3)!!} ~, ~~~~~~ \alpha_k^s =
{(2s-k-1)! \over [(s-k)!]^2} ~, \label{numcoef}
\ee
we find that the 2-cocycle assumes the form
\be
{\cal C}(V_n^s , V_m^{s^{\prime}}) = - {B^2(s) \over 2s-1}
{(n+s-1)! \over (n-s)!} \delta_{s, s^{\prime}} \delta_{n+m, 0} ~,
\label{central1}
\ee
and it is diagonal in the indices $s$ and $s^{\prime}$.

Furthermore, it can be
shown that the commutation relations of the centrally extended
algebra of differential operators assume the following form, \cite{prs1},
\ba
[V^s_n,V^{s^\prime}_m] & = & \left((s^{\prime}-1)n-(s-1)m
\right) V^{s+s^\prime-2}_{n+m}
+\sum_{r\ge 1}
g^{ss^\prime}_{2r}(n,m; \mu)V^{s+s^\prime-2-2r}_{n+ m} \nonumber\\
&  & + c_s(\mu) n(n^2 -1) (n^2 - 4) \cdots (n^2 - (s-1)^2)
\delta_{s, s^{\prime}} \delta_{n+m, 0} ~, \label{ala1} \
\ea
where the structure constants of the algebra are given by
\be
g_{2r}^{ss^\prime}(n,m; \mu)={\phi^{ss^\prime}_{2r}(\mu) \over
2(2r+1)!} \,N^{ss^\prime}_{2r}(n,m) ~, \label{ala2} \ee with \be
\phi^{ss^\prime}_{2r}(\mu)=\sum_{k= 0}^{r} {(-{1 \over 2} -2\mu)_k
({3 \over 2} +2\mu)_k (-r-{1 \over 2} )_k (-r)_k \over k! (-s +{3
\over 2})_k (-s^{\prime} + {3 \over 2})_k (s+s^{\prime} -2r- {3
\over 2})_k} \label{ala3} \ee and \be N_{2r}^{ss^\prime}(n,
m)=\sum_{k=0}^{2r+1}(-1)^k{2r+1\choose
k}(2s-2r-2)_k[2s^\prime-k-2]_{2r+1-k}[s-1+n]_{2r+1-k}[s^\prime-1+m]_k
~. \label{ala4}
\ee
In the above formulae $(a)_n$ and $[a]_n$
denote the ascending and descending Pochhammer symbols,
respectively,
\be
(a)_n = a(a+1) \cdots (a+n-1) ~, ~~~~~ [a]_n =
a(a-1) \cdots (a-n+1)
\ee
with $(a)_0 = 1 = [a]_0$. Finally, the
coefficients of the central terms are given by the expression
\be
c_s(\mu) = {c \over 2^{1 -2|\mu|}} {2^{2(s-3)} (s+2\mu)! (s-2\mu
-2)! \over (2s-1)!! (2s-3)!!} ~, \label{central}
\ee
where the
overall coefficient $c$ is left arbitrary and coincides with the
value of the central charge of the Virasoro subalgebra generated
by the Fourier modes $V_n^2$.

The commutation relations of the full $W_{1 + \infty}$ algebra
correspond to the choice of the free parameter $\mu = -1/2$, in
which case the $s$-dependent coefficient of the central terms
\eqn{central} coincides with the expression derived from the
2-cocycle formulae \eqn{central1} in the basis \eqn{maibas}. Other
choices of $\mu$ describe various consistent truncations of $W_{1
+ \infty}$ that we will encounter shortly. For $\mu = -1/2$, the
generators $V_n^s$ represent the Fourier modes parametrized by $n
\in Z$ of chiral fields with spin $s$,
\be
V^s (z) =
\sum_{n=-\infty}^{+\infty} V_n^s z^{-n-s} ~,
\ee
in the framework
of two-dimensional conformal field theories. Indeed, the
commutation relations \eqn{ala1} of $W_{1 + \infty}$ can be
converted into operator product expansions
\be
V^s(z)V^s{^\prime}(w) \sim -\sum_{r\ge 0} f^{ss{^\prime}}_{2r}
(\partial_z,
\partial_w; \mu) {V^{s+s{^\prime}-2-2r}(w) \over z-w}
-c_s (\mu)\delta_{s,s^{\prime}} \partial_z^{2s-1} {1 \over z-w}
\label{magie1}
\ee
with
\be
f_{2r}^{ss^{\prime}}(n, m; \mu)={\phi^{ss^{\prime}}_{2r} (\mu) \over 2(2r+1)!}
M^{ss{^\prime}}_{2r}(n, m) ~,
\ee
where
\be
M_{2r}^{ss^\prime}(n, m) =\sum_{k=0}^{2r+1}(-1)^k{2r+1\choose
k}(2s-2-2r)_k[2s^\prime-k-2]_{2r+1-k} n^{2r+1-k} m^k ~.
\label{magie2}
\ee
The operators $M_{2r}^{ss^{\prime}}(\partial_z , \partial_w)$ are obtained by replacing
the powers appearing in $n$ and $m$ by the corresponding
derivative operators with respect to $z$
and $w$, respectively.

Thus, in the framework of two-dimensional conformal field
theories, $W_{1 + \infty}$ can be regarded as an extended
world-sheet symmetry generated by an abelian $U(1)$ current
$V^1(z)$, the stress-energy tensor $V^2(z)$ with central charge
$c$, which is determined by the conformal field theory that
realizes the symmetry, and an infinite collection of higher spin
fields $V^s(z)$ for all integer values $s \geq 3$. For $\mu =
-1/2$, we observe that the operator product expansion $V^s(z)
V^{s^{\prime}}(w)$ contains all fields with spin $s + s^{\prime}
-2r$, starting from $s+s^{\prime}-2$ and terminating at $V^2(w)$
or $V^1(w)$ for $s+s^{\prime}$ even or odd, respectively. In this
case, $W_{1 + \infty}$ is formulated as infinite dimensional
linear algebra by choosing a {\em quasi-primary} field basis with
$<V^s(z) V^{s^{\prime}}(w)> \sim \delta_{s, s^{\prime}} /
(z-w)^{s+ s^{\prime}}$. Note that in a {\em primary} basis, which
requires the use of non-linear field redefinitions, the
commutation relations of the algebra $W_{1 + \infty}$ are
non-linear and, hence, less tractable in closed form.

Higher spin truncations of $W_{1 + \infty}$ cannot be obtained by
simply setting some of the lower spin generators equal to zero, as
this prescription is not consistent with the Lie algebra
commutation relations. Instead, consistent truncations can be made
systematic by twisting the generators
\be
\tilde{V}_n^s = V_n^s +
~{\rm lower ~ spin ~ terms} \label{twist}
\ee
so that the new
elements $\tilde{V}_n^s$ close among themselves for all values of
$s$ bigger or equal than a fixed integer value $M$, which provides
the lower spin bound. The truncation is consistent when the
remaining lower spin generators do not appear in the commutation
relations of the generators with $s \geq M$; of course,
non-trivial commutation relations will arise among the lower and
higher spin generators so that the complete structure of the $W_{1
+ \infty}$ algebra is only described in a different basis. For
example, as it is well known, the infinite dimensional algebra
$W_{\infty}$ generated by all fields with spin $s \geq 2$ follows
from $W_{1+ \infty}$ by appropriate twisting. Likewise, the
twisting procedure can be generalized to construct higher spin
algebras with $s \geq M$ for any choice of the lower cutoff
integer $M$. We will present their construction using appropriate
choice of bases in the algebra of all differential operators, from
which explicit twisting formulae can be obtained for the higher
spin truncations of $W_{1+\infty}$. Such higher spin algebras can
be studied as mathematical curiosities on their own right, but
here we find that they also characterize the chiral algebra of the
gauged WZW models in the tensionless limit. Their precise meaning
will become evident shortly by elaborating on the relevance of the
lower spin generators that decouple entirely from the spectrum. In
this context, the twisting \eqn{twist} is only a method for their
systematic construction from $W_{1 + \infty}$, while their
physical interpretation depends on the circumstances that they
arise in quantum field theory.

It can be shown that the following system of differential
operators of order bigger or equal than $M-1$ and less or equal
than $s-1$,
\be
\tilde{V}_n^s = -{(s-M)! \over (s-1)!} B(s)
\sum_{k=1}^{s- M +1} {(s-k)! \over (s-k-M+1)!} \alpha_k^s {n+s-1
\choose k-1} z^{n+s-k} \partial^{s-k} ~, \label{twista}
\ee
where
$B(s)$ and $\alpha_k^s$ are given by equation \eqn{numcoef}, as
before, form a closed algebra for all integer values of spin $s
\geq M$, \cite{bak3}. The logarithmic 2-cocycle remains diagonal
in this basis and the commutation relations of the corresponding
centrally extended higher spin algebra assume the general form
\eqn{ala1}-\eqn{ala4} with parameter
\be
\mu = {1 \over 2}(M-2) ~.
\ee
Also, for this choice of $\mu$, the central terms are given by
equation \eqn{central} for all $s$, up to an overall numerical
value which is conveniently normalized to $c/2^{1-2|\mu|}$ and it
depends on the model. Note that the operator product expansions
$\tilde{V}^s(z) \tilde{V}^{s^{\prime}}(w)$ that emerge in this
case involve all terms of the form $\tilde{V}^{s+s^{\prime} -2r}$
starting from $\tilde{V}^{s+s^{\prime}-2}(w)$ and terminating at
$\tilde{V}^{M+1}(w)$ or $\tilde{V}^M(w)$ for $s+s^{\prime} +M-1$
even or odd, respectively. The resulting infinite dimensional
algebra will be denoted by $W_{\infty}^{(M)}$ and involves all
generators with $s \geq M$. If $M \geq 3$, the higher spin algebra
is not conformal, as it does not contain the Virasoro algebra, and
the term ``spin" should not be taken at face value; nevertheless,
we will continue to call higher spin algebras all such consistent
truncations of the original $W_{1 + \infty}$ algebra.

It is now straightforward to put the results of the previous section into a
more systematic framework. First, the algebra $\hat{W}_{\infty}(k)$ linearizes
in the large $k$ limit and becomes isomorphic to the algebra $W_{\infty}$
for which $M=2$. In this case the free field realization of the
{\em quasi-primary} operators \eqn{ffrep1} that follow from the parafermionic
operator product expansion satisfy the commutation relations \eqn{ala1} with
$\mu = 0$. This result is already known in the literature,
\cite{bak2}, and it will not be
discussed further. Second, and most important result for the purposes of the
present work is the identification of $\hat{W}_{\infty}(k)$ at critical level
$k=2$ with the higher spin algebra $W_{\infty}^{(M)}$ for $M=3$.
Indeed, it can be verified that the free field realization of the
operators \eqn{ffrep2} satisfy the commutation relations \eqn{ala1} with
$\mu = 1/2$ provided that the following rescaling of generators is also taken
into account,
\be
\tilde{W}_n^s = {s-2 \over 2} \tilde{V}_n^s ~. \label{resca}
\ee
The linear structure of the algebra $\hat{W}_{\infty}(2)$ was already established
in the previous section for all generators, thanks to the bilinear form of the
currents $\tilde{W}_s (z)$ in terms of the complex fermion system
$(\Psi, \bar{\Psi})$. Also, the operator product expansions \eqn{masope} can be
identified with the corresponding commutation relations of $W_{\infty}^{(3)}$ after
the rescaling \eqn{resca}. Then, these sample calculations suggest the
exact equivalence of the algebras
$\hat{W}_{\infty}(2)$ and $W_{\infty}^{(3)}$ for all values of $s \geq 3$. The complete
proof relies on the uniqueness of the algebra $W_{1 + \infty}$,
and its higher spin truncations, following from the uniqueness of the Moyal algebra
as linear deformation of the algebra of area preserving diffeomorphism
of $T^{\star}S^1$, \cite{paul}, and the uniqueness of its central
extensions, \cite{khesin, radul1}.
Besides, this can also be verified
independently by direct computation of the commutation relations among all other
higher spin fields, which thus prove their exact equivalence.

Summarizing, we have shown that {\em the chiral operator algebra
of the gauged WZW model $SL(2, R)_k/U(1)$ at critical level $k=2$
is the higher spin truncation of $W_{1+\infty}$ generated by
chiral fields with $s \geq 3$.} In this case we find that the
fermionic realization \eqn{ffrep2} yields central terms for the
higher spin generators with coefficients $c_s(\mu = 1/2)$ given by
equation \eqn{central} for
\be
c=1 ~.
\ee
The value of the central
charge does not have the usual meaning as in conformal field
theories because the chiral algebra $W_{\infty}^{(3)}$ does not
contain the Virasoro algebra, which is decoupled from the
spectrum. The Virasoro algebra contracts to an abelian structure
which commutes with all other generators $\tilde{W}_s(z)$ after
rescaling of the Virasoro generators by $\sqrt{k-2}$. Thus, in the
present case, one should think of the algebra $W_{\infty}^{(3)}$
as having life on its own, and it cannot be extended to the full
$W_{1+\infty}$ symmetry by (un)twisting the generators. Otherwise,
the quantum field theory of the tensionless coset $SL(2,
R)_2/U(1)$ would remain conformal after the decoupling of the
Liouville field; this possibility is ruled out by the singular
character of the Sugawara construction in terms of $SL(2, R)$
currents as there is no stress-energy tensor in the spectrum at
critical level.

It is also instructive to compare the free field representation \eqn{ffrep2} with the
fermionic representation of $W_{\infty}^{(3)}$ that results by twisting the
higher spin generators of $W_{1+\infty}$. Recall that $W_{1+\infty}$ also admits
a free field realization with fermion bilinears
\be
V_s(z) = 2^{s-3} {(s-1)! \over (2s-3)!!} \sum_{k=0}^{s-1}
(-1)^k {s-1 \choose k}^2 \partial^k \bar{\Psi}
\partial^{s-k-1} \Psi (z) ~,
\label{wadiman}
\ee
which satisfy the commutation relations \eqn{ala1} with Virasoro central
charge $c=1$, \cite{prs2}.
Then, introducing appropriate field redefinitions
$\tilde{V}_n^s = V_n^s + {\rm lower ~ spin ~ terms}$,
we obtain new elements $\tilde{V}_n^s$ that represent the algebra of higher spin
operators \eqn{twista} as fermion bilinears, \cite{bak3},
\be
\tilde{V}_s(z) = 2^{s-3} {(s-M)! (s+M-2)! \over (s-1)! (2s-3)!!} \sum_{k=0}^{s-M}
(-1)^k {s-1 \choose k} {s-1 \choose k+M-1} \partial^k \bar{\Psi}
\partial^{s-k-1} \Psi (z)
\ee
with $s \ge M$ and central terms having $c= (-1)^{M-1} 2^{1-2|\mu|}$,
according to our normalizations.
The central charge is easily computed by noting that the
lowest spin operator $\tilde{V}_M(z)$ and its two-point function are given by
\be
\tilde{V}_M(z) = 2^{2(M-2)} \bar{\Psi} \partial^{M-1} \Psi (z) ~; ~~~~
<\tilde{V}_M(z) \tilde{V}_M(w)> = (-1)^{M-1}{[2^{2(M-2)} (M-1)!]^2 \over (z-w)^{2M}} ~.
\ee
Then, passing to Fourier modes and comparing with the coefficient of the central term
\eqn{central} for $s=M$ and $\mu = (M-2)/2$, we arrive at the value of $c$ given above.

This procedure provides another fermionic realization of the algebra
$W_{\infty}^{(M)}$ for all integer values of the lower spin $M$, which can be
subsequently specialized to $M=3$ and compared with the representation \eqn{ffrep2}.
The two realizations are different from each other since the corresponding
spin 3 operators $\tilde{V}_3(z)$ are $4 \bar{\Psi} \partial^2 \Psi (z)$ and
$4 \partial \bar{\Psi} \partial \Psi (z)$,
respectively. Likewise, their higher spin representatives are also different from
each other, although both of them have $c=1$ when $M=3$. It should be noted, however,
that they differ by total derivative terms of lower spin fields including
$\bar{\Psi} \partial \Psi (z) - (\partial \bar{\Psi}) \Psi (z)$ and
$\bar{\Psi} \Psi (z)$. These two expressions provide the stress-energy tensor and the
$U(1)$ number current of the
conformal field theory of free fermions, but they do not represent physical operators
of the coset model at critical level. Put differently, $SL(2, R)_2/U(1)$ is not meant to
be equivalent to the theory of free fermions, but only the algebra of its operators
is realized in terms of some fermion bilinears; all other fermionic expressions, like
$\bar{\Psi} \Psi$ and $\bar{\Psi} \partial \Psi - (\partial \bar{\Psi}) \Psi$,
do not correspond to operators of the coset model.
Thus, it is not surprising that there is no physical Virasoro generator to append
to the chiral algebra of the coset $SL(2, R)_2/U(1)$ and extend the free field
realization \eqn{ffrep2} of $W_{\infty}^{(3)}$ to the full $W_{1+\infty}$ algebra.
As we have already stressed, the coset model is a
singular conformal field theory in the tensionless
limit.

This comparison may also have important
consequences for the characterization of the $SL(2,R)_2/U(1)$ quantum theory in
connection with the $c=1$ matrix model. It is well known that in the fermionic description
of the matrix model there are infinitely many conserved quantities associated
with conserved currents of higher spin of the form \eqn{wadiman}, which form a $W_{1+\infty}$
algebra with Virasoro central charge $c=1$ (see, for instance, \cite{igori});
they include the fermion number
associated with the spin-1 current $\bar{\Psi} \Psi$ followed by the stress-energy
tensor and a collection of
higher spin generators.
One the other hand, it has already been noted in section 4 that the $SL(2,R)_k/U(1)$
model can not be regarded as a theory of $c=1$ matter coupled to two-dimensional gravity,
since this identification is only valid in the asymptotic (weak coupling) region of the
classical geometry. Therefore, in the tensionless limit, where gravity decouples in the form
of a Liouville field, the remnants cannot be possibly identified with the ordinary
$c=1$ model; instead, the theory that remains should be thought
as a variant or an exotic phase of the $c=1$ matrix model. The fermionic realization of the
corresponding symmetry algebra $W_{\infty}^{(3)}$ suggests that this unknown model
could also be formulated in terms of fermion fields, as in the ordinary case, but without
having lower spin currents among its physical operators. We expect that better understanding
will be achieved in the future by studying the matrix model for the two-dimensional black-hole,
\cite{kuta}, which is based on a conjectured equivalence between the $SL(2,R)_k/U(1)$ coset
and the sine-Liouville model, and the dynamics of vortices.

\section{BRST analysis of the world-sheet symmetry}
\setcounter{equation}{0}

In bosonic string theory, where the Virasoro algebra is the
underlying world-sheet symmetry, nilpotency of the BRST operator
\be
{\cal Q} = \sum_{n=-\infty}^{+\infty} L_n c_n -{1 \over 2}
\sum_{n=-\infty}^{+\infty} \sum_{m=-\infty}^{+\infty} (n-m)
:c_{-n} c_{-m} b_{n+m}: ~, \label{brstvi}
\ee
provides the
critical value of the central charge $c=26$, \cite{Kato}. Here,
$b_n$ and $c_n$ are the Fourier modes of a fermionic ghost system
$(b, c)$ with conformal weights $(2, -1)$, respectively, which are
associated with reparametrization invariance. On the other hand,
when the Virasoro algebra contracts to a $U(1)$ current algebra,
as in the commutation relations $[L_n , L_m] = n\delta_{n+m, 0}$
for the case of tensionless strings, the corresponding BRST
operator squares to zero without the need to impose any
restrictions on the coefficient of the central term, \cite{ouvry,
lizzi}, \cite{bonelli}. Thus, the concept of critical dimension,
which renders the quantum theory of strings free of Weyl
anomalies, appears to be lost when $\alpha^{\prime} \rightarrow
\infty$. It comes as no surprise, since the very notion of
space-time where the evolution of strings takes place is not
relevant in the tensionless limit, and the cancellation of the
Weyl anomaly that otherwise breaks space-time Lorentz invariance
is not an issue anymore. The result is also consistent with the
arbitrariness of the coefficient that remains in the central terms
after rescaling the Virasoro generators to absorb the infinite
value of the central charge when $\alpha^{\prime} \rightarrow
\infty$. It is for this reason that $SL(2, R)_2/U(1)$ can be taken
as model for tensionless strings, although it does not arise as a
limiting case within critical string theory.

The purpose of
this section is to show that for two-dimensional quantum field theories
that possess additional higher spin symmetries, the BRST charge is not
nilpotent unless the coefficient of the corresponding
higher spin central terms is fixed
to a critical value. Thus, although the contracted Virasoro symmetry is not
sufficient to distinguish among different tensionless models,
nilpotency of the BRST charge for the
$W_{\infty}^{(3)}$ world-sheet symmetry may be used, instead, to impose
severe restrictions on tensionless string model building. There is an implicit
assumption that one makes in this case, namely that $W_{\infty}^{(3)}$ serves
as a fundamental world-sheet symmetry in the tensionless limit.
Then, cancellation of the anomalies associated to all higher spin generators
with $s \geq 3$ will provide a substitute of the critical dimension, but there
is no space-time interpretation of its value as gravity
has decoupled.  We will find that the
central terms \eqn{central}, which correspond to the choice $\mu = 1/2$,
\be
c_s(\mu = 1/2) = {2^{2(s-3)} (s-3)! (s+1)! \over (2s-3)!! (2s-1)!!} c
\label{central17}
\ee
cancel by the corresponding ghost contributions
provided that the overall coefficient is fixed to the value
\be
c=2 ~, ~~~ \forall ~ s \geq 3 ~.
\ee

The analysis we perform in this
section can only be viewed at the present time as
curiosity of the symmetries arising in the tensionless limit of gauged WZW
models. It should be used in a more fundamental way in case
the theory of tensionless strings admits a systematic reformulation as
$W$-strings for the non-conformal algebra $W_{\infty}^{(3)}$. Such
point of view will not be confirmed here, but it will be investigated
separately as a viable possibility in future work. Apart from this
issue, it should be mentioned that the BRST analysis also serves
as an additional consistency check of the truncation
procedure leading to the infinite dimensional algebra
$W_{\infty}^{(3)}$ and its non-trivial central extensions.

Recall that the BRST operator for a Lie algebra with generators $T^a$ and
structure constants $f^{ab}{}_c$ is generally given by
\be
{\cal Q} = c_a
T^a - {1 \over 2} f^{ab}{}_c c_a c_b b^c \label{charge}
\ee
where $(c_a,b^a)$ is a pair of Faddeev-Popov ghosts that possess opposite
statistics to the generators $T^a$; as such, they satisfy
$\{c_a,b^b\}=\delta_a^b$ and all other anticommutators vanish. For
finite dimensional Lie algebras with trivial cohomology groups, as for all
simple Lie algebras, the operator ${\cal Q}$ is always
nilpotent and there is no anomaly that needs to be cancelled.
However, for infinite dimensional algebras that admit non-trivial
central extensions, the quantum theory may be anomalous
as the ghost contributions do not always balance the bosonic
part of the central charge in order to have ${\cal Q}^2 = 0$. This is
common to all symmetries arising in two-dimensional
conformal field theory, thus leading to critical values of the
central charges.

The BRST operator of the Virasoro algebra, and all other algebras that
contain higher spin fields among the generators, can be constructed
by first introducing a system of ghost fields
$(b_s(z) , c_s(z))$ for
each current $V_s(z)$ with weights $(s, 1-s)$,
respectively, and
\be
<b_s(z) c_s(w)> = {1 \over z-w} = <c_s(z) b_s(w)> ~.
\ee
Then, the operator ${\cal Q}$ is defined to be the charge
\be
{\cal Q} = {1 \over 2\pi i} \oint dz j(z)
\ee
associated to the BRST current of the
corresponding symmetry algebra
\be
j(z) = \sum_{\{s_i\}} V_{s_i} (z) c_{s_i} (z) + \cdots ~.
\ee
Here, the summation is taken over all generators of the
extended world-sheet symmetry and dots denote appropriately
chosen $ccb$-type terms whose exact form depends on the structure constants
of the underlying Lie algebra; for instance, for the Virasoro algebra
generated by the stress-energy tensor with spin 2, we have
$j(z) = T(z)c_2(z) + c_2 \partial c_2 b_2 (z)$, which yields the BRST
operator \eqn{brstvi} in terms of Fourier modes. One may also work out similar
expressions for $W$-algebras, as we will see later in detail.

The nilpotency of the BRST operator, ${\cal Q}^2 = 0$,
implies that the operator product expansion of the currents
$j(z)j(w)$ vanishes up to total derivative terms. This is precisely
how the Virasoro anomaly cancels in bosonic string
theory to yield $c=26$. For extended conformal symmetries,
the ghost contribution to
the central term of the Virasoro subalgebra comes from all
possible higher spin generators and it turns out to be
\be
c_{\rm gh} (2) = - \sum_{\{s_i\}} (6s_i^2 - 6s_i + 1) ~.
\label{zetaf}
\ee
Each $(b_s, c_s)$ system contributes the characteristic value
$-(6s^2 - 6s + 1)$.
Thus, for finitely generated $W$-algebras,
the total contribution to $c_{\rm gh}(2)$ is finite and its value
depends on the spin content of the additional symmetries. Anomaly
cancellation requires $c/2 + c_{\rm gh} (2) = 0$, which fixes $c$ to a given
critical value. For instance, for
Zamolodchikov's $W_3$ algebra, which is generated by the stress-energy tensor
and an additional chiral field with spin 3, the ghost contribution is
$-13 -37 = -50$
and therefore the critical central charge of $W_3$-strings is $c=100$,
\cite{mieg}.
For infinitely generated algebras, as for $W_{1 + \infty}$ and
its higher spin truncations that were encountered before, the sum that
determines $c_{\rm gh}(2)$ diverges and additional regularizations have
to be taken into account in order to make good sense of it. Also, in
all these cases, one has to make sure that the coefficients of the singular
terms that arise in the operator product expansion
$j(z) j(w)$ vanish all at once to all orders in $z-w$. The
central terms of the higher spin generators receive different
ghost contributions $c_{\rm gh}(s)$ that appear to order $1/(z-w)^{2s}$
for each value of $s$ and their cancellation is a prerequisite for the
consistency of any regularization scheme that makes the BRST operator
nilpotent. This singles out a definite
value for the central charge $c$ for which the $W$-algebra becomes
free of anomalies.

There is a definite regularization scheme that makes $c_{\rm gh}(2)$ finite
for all infinitely generated conformal algebras, which can also be extended to
all higher spin central charges $c_{\rm gh}(s)$ in a consistent way.
This problem was investigated for the first time for the algebra $W_{\infty}$,
where the formal expression \eqn{zetaf} can be made finite using the zeta function
regularization,
\be
c_{\rm gh}(2) = - \sum_{j=0}^{\infty} \left(6(j+1)^2 + 6(j+1) + 1 \right)
= - 6 \zeta(-2, 1) - 6 \zeta(-1, 1) - \zeta (0, 1) ,
\label{bernu1}
\ee
while setting for convenience $j=s-2 \ge 0$. Here, $\zeta(s, a)$ is defined as usual
\be
\zeta(s,a)=\sum_{n\ge 0} (n+a)^{-s} ~,  \label{zeta}
\ee
which converges for $s>1$, but it has a pole at $s=1$. Using
the analytic continuation of \eqn{zeta} to negative integer values
$s=-l$, we define the regularized zeta function in terms
of the Bernoulli polynomials\footnote{We list the first few Bernoulli
polynomials that are needed for the calculations presented here and in the
remaining part of this section: $B_1(x) = x-1/2$, $B_2(x) =x^2 -x +1/6$,
$B_3(x) = x^3 -3x^2/2 + x/2$, $B_4(x) = x^4 -2x^3 + x^2 -1/30$,
$B_5(x) = x^5 -5x^4/2 + 5x^3/3 - x/6$, $B_6(x) = x^6 -3x^5 +5x^4/2 -
x^2/2 +1/42$, $B_7(x) = x^7 -7x^6/2 +7x^5/2 -7x^3/6 +x/6$,
$B_8(x) = x^8 -4x^7 +14x^6/3 - 7x^4/3 + 2x^2/3 -1/30$,
$B_9(x) = x^9 - 9x^8/2 + 6x^7 -21x^5/5 +2x^3 -3x/10$, and so on. In general
they satisfy the relation $B_n^{\prime}(x) =n B_{n-1}(x)$ with $B_n(0)$ equal
to the Bernoulli numbers.}
\be
\zeta (-l, a)_{\rm reg} = -{B_{l+1} (a) \over l+1} ~.
\ee
Then,
we easily find that the regularized value of the ghost contribution to the
Virasoro algebra
is $c_{\rm gh}(2) = 1$, \cite{Yamagishi}.
Hence, the BRST operator of $W_{\infty}$ is nilpotent at
the spin 2 level provided that the bosonic representation of the algebra has
the opposite central charge, $c=-2$.
Furthermore, it can be verified that this choice of
central charge is also critical for the higher spin
generators of the algebra, thus leading to similar anomaly cancellations
between $c_{\rm gh}(s)$ and the ``matter" part of the corresponding
higher order central terms.
Likewise, for the $W_{1+\infty}$ algebra, similar analysis shows that the
regularized value is $c_{\rm gh}(2) = 0$, \cite{Popevf}, as the summation over
$j=s-2$ extends from $-1$ to infinity, and the critical value turns out to be $c=0$
in this case.

We now turn to the non-conformal algebra $W_{\infty}^{(3)}$, which is of interest
here, and introduce a pair of ghost fields $(b_s, c_s)$ for each generator
with $s \geq 3$. Using the general formulation of this algebra in terms of the structure
constants $g_{2r}^{ss^{\prime}}(n,m; \mu)$ with $\mu = 1/2$, we may
write the BRST operator in terms of Fourier modes,
\be
{\cal Q}=\sum_{s\ge3}\tilde{V}^s_mc^m_s-
{1 \over 2}\sum_{{s,s^{\prime}\ge3}}\sum_{r\ge0}
g^{ss^{\prime}}_{2r}(n,m; \mu=1/2):c^{-n}_sc^{-m}_{s^{\prime}}
b^{n+m}_{s+s^{\prime}-2r}: ~,
\ee
where summation over $n$ and $m$ is also implicitly assumed.
Reverting to the coordinate representation as more appropriate for the
operator product expansions, we write the BRST current of the algebra
in the form
\be
j(z)=\sum_{s\ge3}\tilde{V}^s
c_s(z) - \sum_{s,s^{\prime} \geq 3} \sum_{r\ge0} f^{ss^{\prime}}_{2r}
(\partial_{c_s},\partial_{c_{s^{\prime}}};
\mu=1/2):c_s c_{s^{\prime}}
b^{s+s^{\prime} -2r}:(z) ~ ,
\label{devdist}
\ee
where $f_{2r}^{ss^{\prime}}$ are given by equations
\eqn{magie1}-\eqn{magie2}, as before,  and $\partial_{c_s}$ denotes the
derivative operator $\partial/\partial z$
acting on the ghost field $c_s$. The summations also range from
$3$ to $\infty$, as dictated by the operator content of the
algebra $W_{\infty}^{(3)}$.

The operator product expansion of the corresponding BRST currents
$j(z)j(w)$ follows from general considerations and it consists of the series
of terms
\ba
j(z)j(w) & = &
(5!\,c_3+c_{\rm gh}(3)){:c_3(z)c_3(w): \over (z-w)^6}+
(7!\,c_4+c_{\rm gh}(4)){:c_4(z)c_4(w): \over (z-w)^8} \nonumber\\
& & +(9!\,c_5+c_{\rm gh}(5))
{:c_5(z)c_5(w): \over (z-w)^{10}}+ {\cal O} \left((z-w)^{-12} \right) ,
\label{opejj}
\ea
where normal ordering is used to extract all singular terms in
different orders depending on $s$.
Then, taking into account equation \eqn{central17} for the normalization of
the central terms $c_s$ and computing the ghost
contribution to the higher spin terms $c_{\rm gh}(s)$, one may verify
that the numerical coefficients of the singular terms vanish
all at once for the same choice of the central charge $c$.
The nilpotency of the BRST
operator should be implemented consistently level by level in $s$ and the critical
charge should be the same for all spins. This procedure is rather
cumbersome to follow in all generality, since the zeta function regularization
of $c_{\rm gh}(s)$ are not available in closed form
for arbitrary values of $s$. We will present the result of explicit
calculations for $s=3, 4$ and $5$, which yield the same value $c=2$ for the
corresponding critical charge. Higher spin calculations can also be carried out,
and some general arguments about the consistency of the regularization scheme to
all levels will be presented later.

Starting with $c_{\rm gh}(3)$, we note that its computation involves
contributions from double contractions of the terms
$\sum_{s\geq 3} :c_sc_3b_{s+1}:(z):c_{s+1}c_3b_s:(w)$ that appear in the
operator product expansion $j(z)j(w)$,
with derivatives also distributed on the ghost fields according to
equation \eqn{devdist}. After some calculation, they give rise to the
infinite sum
\ba
c_{\rm gh}(3) & = & -8 \sum_{s=3}^{\infty} \phi_2^{3, s+1} (\mu = 1/2)
\left(10s(2s-1)(s-1)^2 + 15 (2s-1) (s-1)^2 + 6 \right. \nonumber\\
& & \left. + 10s(s-1)(2s-1) +30(s-1)^2 + 10(s-1)(2s-1) + 30(s-1)
\right) ,
\ea
where
\be
\phi_2^{3s}(\mu = 1/2) = 1- {15 \over
(2s-1)(2s-3)} ~.
\ee
Setting $j=s-3$, we reorganize the sum and
compute it as follows, using zeta function regularization,
\ba
c_{\rm gh}(3) & = & 16 \sum_{j\ge0}\left(-10(j+3)^4+45(j+3)^2-{77
\over 4}
+{45 \over 8(2j+7)}-{45 \over 8(2j+5)}\right) \nonumber\\
& = & 16 \left( -10\zeta(-4,3)+45\zeta(-2,3)-{77 \over 4}\zeta(0,3)
-{9 \over 8} \right) =-128 ~. \label{bernu2}
\ea
The constant term $-9/8$ that appears in the second line has remained
after the pairwise cancellations between the last two fractional
terms shown in the first line.
Then, since
the coefficient of the singular term $(z-w)^{-6}$ appearing in equation
\eqn{opejj} is $5! c_3 +c_{\rm gh}(3) = 64c + c_{\rm gh}(3)$, we conclude that
$c=2$ at this level, as advertised before. It is the first nilpotency condition
that determines the critical value of the central charge $c$ of $W_{\infty}^{(3)}$.

Next, we present some technical details and outline the way to handle the
computations for arbitrary spin
level in order to achieve the desired cancellation of anomalies when $c=2$.
We set up the general framework for evaluating the ghost contribution
to the spin $s$ central charge, following earlier work on
infinite dimensional algebras of $W_{\infty}$-type, \cite{Popevf}.
We have, in particular,
\be
c_{\rm gh}(s)= - \sum_{r=0}^{s-2}\sum_{s^{\prime} \ge s_0}
{\phi^{s, s^{\prime} +1}_{2r}(\mu)
\phi^{s,s+s^{\prime} -2r-1}_{2s -2r -4}(\mu) \over
4(2r+1)! (2s-2r-3)! [(2s^{\prime} -2r -1)!]^2} X(s, s^{\prime}, r; \mu)
\label{X-f}
\ee
where $s_0 = \max(2\mu+ 1, 2r-s +3 +2\mu)$ and
\ba
& & X(s, s^{\prime}, r;\mu) =
\sum_{k=0}^{2r+1\;}\sum_{k^\prime=0}^{2s-2r-3\;}
\sum_{l=0}^{2r+1-k\;}\sum_{l^\prime=0}^{2s-2r-3-k^\prime}
(2r+1-l+l^{\prime})! (2r+k^{\prime} +1)! (2s^{\prime}-k)! \nonumber\\
& & ~~~~~~~~~~ \times {(2s-2r-3-l^\prime+l)!(2s-2r-3+k)!
(2s+2s^{\prime}-2r-4-k^\prime)! \over
k!\;k^\prime!\;l!\;l^\prime!\;
(2r+1-k-l)!(2s-2r-3-k^\prime-l^\prime)!} ~.
\ea
All these terms
arise by making appropriate double contractions of the ghost terms
that contribute to $c_{\rm gh}(s)$ for any given spin $s$, and
they are valid for all higher spin truncations of $W_{1+\infty}$
characterized by $\mu$. Of course, here, we must set $\mu = 1/2$
and proceed with explicit calculations.

The ghost contribution to the spin 4 central charge is given by
the following three infinite sums
\ba
c_{\rm gh}(4) & = & -32
\sum_{j\ge1} \left( 21(j+{5 \over 2})^6-210(j+{5 \over 2})^4+{4473
\over 8}(j+{5 \over 2})^2 -{3321 \over 8}
-{4725 \over 128(2j+3)} \right. \nonumber\\
& & \left. +{6075 \over 64(2j+5)^2}+{4725 \over 128(2j+7)} \right)
\nonumber\\
& & -64 \sum_{j\ge1} \left(
35(j+{3 \over 2})^6-245(j+{3 \over 2})^4+{3675 \over 8}
(j+{3 \over 2})^2-{495 \over 8}
-{7155 \over 128(2j+1)} \right. \nonumber\\
& & \left. -{6075 \over 128(2j+1)^2}+{7155 \over 128(2j+5)}
- {6075 \over 128(2j+5)^2} \right) \nonumber\\
& & -32 \sum_{j\ge3} \left(
21(j+{1 \over 2})^6-210(j+{1 \over 2})^4+{4473 \over 8}
(j+{1 \over 2})^2-{3321 \over 8}
-{4725 \over 128(2j+3)} \right. \nonumber\\
& & \left. +{6075 \over 64(2j+5)^2}+{4725 \over 128(2j+7)}
\right) ~.
\label{bernu3}
\ea
Each sum has to be
evaluated independently by shifting the summation index
appropriately and making use of the zeta function regularization.
After some calculation the result turns out to be
\ba
c_{\rm gh}(4) & = & -64 \left(21 \zeta(-6, 7/2) -210 \zeta(-4,7/2)
+{4473 \over 8} \zeta(-2, 7/2) \right. \nonumber\\
& & \left. - {3321 \over 8} \zeta(0, 7/2) +
35 \zeta(-6, 5/2) -245 \zeta(-4, 5/2) \right. \nonumber\\
& & \left. + {3675 \over 8} \zeta(-2, 5/2)
-{495 \over 8} \zeta(0, 5/2) -{3177 \over 64} \right) = -16^2 \cdot 12 ~.
\ea
Since the numerical coefficient of the corresponding term in the operator
expansion of the BRST currents is $7! c_4 + c_{\rm gh}(4) =
16^2 \cdot 6c + c_{\rm gh}(4)$, we find once more
that the anomaly cancellation occurs at $c=2$ for spin $4$.

The ghost contribution to the spin 5 anomaly can be evaluated in a similar
way following the general expression \eqn{X-f}, but the individual terms are
quite a lot in number. Here, we only present
the end result of the calculation using zeta function regularization of the
various sums,
\ba
c_{\rm gh}(5) & = & 1024 \left(
-9 \zeta(-8,4)+{615 \over 4}\zeta(-6,4)-{13245 \over 16}\zeta(-4,4)+
{106815 \over 64}\zeta(-2,4) \right. \nonumber\\
& & \left. -{260181 \over 256}\zeta(0,4)
-63 \zeta(-8,3)+{3225 \over 4}\zeta(-6,3)-{399165 \over 112}
\zeta(-4,3) \right. \nonumber\\
& & \left. +{2609175 \over 448}\zeta(-2,3)-{412947 \over 256}\zeta(0,3)-
{13005 \over 256} \right)
=-16^3 \cdot {288 \over 7} ~.
\ea
Then, since the overall coefficient of the corresponding singular term
in the operator product expansion \eqn{opejj} is
$9! c_5 +c_{\rm gh}(5) = 16^3 \cdot 144c/7 + c_{\rm gh}(5)$,
we find again that the algebra is free of
anomalies for $c=2$, as required for consistency at all spin level.

The computation of the
higher ghost central charges becomes extremely more complicated as
the spin level increases, but in all cases we have examined
so far the result is the same and singles out $c=2$ as the critical
value of the non-conformal algebra $W_{\infty}^{(3)}$. Actually, this
is not surprising because the ghost currents satisfy the same operator
algebra, and therefore all $c_{\rm gh}(s)$ are also related to each other
in terms of a single central charge $c$, as in equation \eqn{central17} above.
Thus, if there is cancellation at a given spin level, it will also hold at all levels.
What is rather surprising, however, is the consistency of the regularization
scheme based on zeta functions, which seems to respect the structure of the
symmetry algebra. It is not guaranteed to be so because there is an
intrinsic ambiguity to regularize the divergent sums $\sum_{n \geq 0} (n+a)^{-s}$
when $s=0$; in particular, $\sum_{n\geq 0} 1$ can be chosen to be equal to
$\zeta(0, a) = - B_1(a) = -a+1/2$, while $a$ remains arbitrary. This arbitrariness
may lead to any value of the critical central charge at a given spin level,
which then can create discrepancies by comparing the results at different levels
or require fine tuning level-by-level.

It is remarkable that there is a natural regularization scheme which appears
to be consistent with the nilpotency of the BRST charge
and it resolves the ambiguous choice of the free parameter
$a$ in all terms that involve $\sum_{n \geq 0} 1$. The prescription we
followed simply sets $a$ equal to the value of its companion terms within a
given bracket of divergent sums. For instance, $a=1$ in the bracket of
divergent sums \eqn{bernu1}, whereas $a$ is chosen to be $3$ in the bracket of
divergent sums that appear in \eqn{bernu2}; likewise, $a$ is taken to be $7/2$,
$5/2$ and $7/2$ in the three different brackets of divergent sums shown in equation
\eqn{bernu3}, respectively.
This method works consistently for all $W_{\infty}$-type algebras, but it has
no rigorous foundation to this day. It resembles the description of quantum
corrections in Kaluza-Klein theories, where a vanishing one-loop vacuum energy
in higher dimensions has a lower dimensional interpretation as a divergent sum
that also equals to zero by zeta function regularization. Unfortunately, we
do not know of a higher dimensional theory that collects the infinite collection
of all two-dimensional higher spin currents into a single entity. If this
is properly understood, our results will be put into a firm and better frame.
In any case, such higher dimensional interpretation could also be used to provide
a more fundamental definition of the tensionless limit.

Finally, we return to the tensionless coset model $SL(2, R)_2/U(1)$ that exhibits
the $W_{\infty}^{(3)}$ symmetry with $c=1$. As such, it is not free of anomalies
but it provides a building block for constructing $c=2$ models with nilpotent
BRST charge. If higher spin symmetries play a fundamental role in the ultimate definition
of tensionless strings, the BRST analysis above will serve as a guide for model
building. In that case it will be possible to gauge the entire world-sheet symmetry, as
it is anomalous free for $c=2$, and examine its breaking pattern by including
$1/\alpha^{\prime}$ corrections.

\section{Generalizations to higher dimensional cosets}
\setcounter{equation}{0}

Generalizations to higher dimensional coset models can be described in a similar
fashion by gauging different subgroups $H$ of higher rank non-compact groups $G$.
Here, we will focus attention on the cosets $SO(n,1)_k/H_k$ by gauging
the maximal compact subgroup $H=SO(n)$ in a vectorial way so that the signature of
the classical geometry is always Euclidean; instead, if we consider $H=SO(n-1,1)$
the construction is similar but the signature will be Lorentzian.
Note that the critical level is
$k=2$ when $n=2$ or $3$, whereas for $n \geq 4$ the critical
level turns out to be $k=n-1$.

Let us begin with the simplest non-abelian coset model for $n=3$. Summarizing
the results of earlier work, \cite{sfetsos1}, we present the solution to
the perturbative beta function equations by including all $\alpha^{\prime}$ corrections.
The exact result is obtained by the Hamiltonian method, as in section 2, and
leads to
\be
ds^2 = 2(k-2) \left(dr^2 + G_{\theta \theta} d\theta^2 + G_{\phi \phi} d\phi^2
+ 2G_{\theta \phi} d\theta d\phi \right) ,
\ee
where the metric components depend on the level as follows,
\ba
G_{\theta \theta} & = & \beta^2 (r) \left({\rm tanh}^2 r - {1 \over k-1} {1 \over
{\rm cos}^2 \theta} \right) , \nonumber\\
G_{\phi \phi} & = & \beta^2 (r)
\left({\rm tanh}^2 r ~ {\rm cot}^2 \phi ~ {\rm tan}^2 \theta
+ {{\rm coth}^2 r \over {\rm cos}^2 \theta} - {1 \over k-1} {1 \over {\rm cos}^2
~ \theta {\rm sin}^2 \phi} \right) , \nonumber\\
G_{\theta \phi} & = & \beta^2 (r) {\rm tanh}^2 r ~ {\rm cot}\phi ~ {\rm tan}\theta ~.
\ea
The function $\beta(r)$ turns out to be
\be
\beta^{-2} (r) = 1 - {1 \over k-1} \left({{\rm coth}^2 r \over {\rm cos}^2 \theta}
+ {{\rm tanh}^2 r \over {\rm sin}^2 \phi} \left(1 + {\rm cos}^2 \phi ~ {\rm tan}^2 \theta
\right) \right) + {1 \over (k-1)^2} {1 \over {\rm cos}^2 \theta ~ {\rm sin}^2 \phi}
\ee
and the dilaton equals to
\be
\Phi = {\rm log} \left({{\rm sinh}^2 2r ~ {\rm sin}^2 \phi ~ {\rm cos}^2 \theta \over
\beta(r)} \right)
\ee
so that $\sqrt{G} {\rm exp} \Phi$ is independent of $k$, as required on general grounds.
There is also an anti-symmetric tensor field
\be
B_{\phi \theta} = {\beta^2 (r) \over 2(k-1)} {\rm tanh}^2 r ~ {\rm cot} \phi ~
{\rm tan} \theta ~,
\ee
which is present for all $k < \infty$.

As $k \rightarrow \infty$, $\beta(r) \rightarrow 1$ and the classical geometry of the coset
$SO(3, 1)_k/SO(3)_k$ simply reads
\ba
ds^2 & = & dr^2 + {\rm tanh}^2 r \left(d\theta + {\rm cot}\phi ~ {\rm tan} \theta ~
d\phi \right)^2 + {{\rm coth}^2 r \over {\rm cos}^2 \theta} d\phi^2 ~, \nonumber\\
\Phi & = & {\rm log} \left({\rm sinh}^2 2r ~ {\rm sin}^2 \phi ~ {\rm cos}^2 \theta
\right) ,
\ea
whereas the anti-symmetric tensor field vanishes in this case. The signature of target
space is clearly $+++$ in the entire region covered by the coordinates
$(r, \theta, \phi)$. The geometry also exhibits curvature singularities in places
where the string coupling ${\rm exp}(-\Phi)$ blows up. On the other hand, extending
the validity of the exact quantum geometry to all $k \geq 2$, it is easy to
observe that the signature changes in places where $\beta^2(r)$ is not positive definite.
For instance, if $r$ and $\phi$ are restricted in a region where
${\coth}^2 r ~ {\rm sin}^2 \phi \geq 1$, whereas $\theta$ remains arbitrary,
the components $G_{\theta \theta}$ and $G_{\phi \phi}$ will have opposite signs
as $k \rightarrow 2$. Thus, we encounter a situation which is similar to the vector
gauged $SL(2)_k/U(1)$ coset model at critical level, as noted in section 2.
The range of the coordinate system should be restricted to regions of Euclidean
signature, or else there is a change of signature due to quantum corrections.
Furthermore, the metric is multiplied with $k-2$ and it becomes singular
at critical level. This behavior is expected on general grounds and it provides
another example of tensionless models.

We may also consider the
non-abelian coset model $SO(3, 1)/E(2)$, \cite{iran3}, which is obtained by gauging
the non-semisimple subgroup of the Lorentz group
in four-dimensional Minkowski space. Recall that the group
$SO(3,1)$ has six generators $J_i$ and $K_i$ with $i=1,2,3$, which correspond to the
rigid rotations $SO(3)$ and the three boosts, respectively.
On the other hand, the Euclidean
group $E(2)$ is generated by the following combinations of Lorentz generators,
\be
E_1 = J_1 - K_2 ~, ~~~~~ E_2 = J_2 + K_1 ~, ~~~~~ E_3 = J_3 ~,
\ee
which give rise to a semi-direct product group structure.
The generators of translations $R^2$ in $E(2)$ are null in the four-dimensional fundamental
representation of the Lorentz group, since
\be
{\rm Tr} E_1^2 = {\rm Tr} E_2^2 = E_1^3 = E_2^3 = 0 ~,
\ee
whereas $E_3$ generates the subgroup of rotations $SO(2)$.
Then, the vector gauging of the $SO(3,1)_k/E(2)$ WZW model can be carried out as usual
using gauge connections $A$, $\bar{A}$ with values in $E(2)$. Parametrizing the elements
$g \in SO(3,1) \simeq SL(2, C)$ as $4 \times 4$ matrices and making appropriate
gauge choices, one may eliminate the gauge fields and derive the effective space-time
action of the coset model. It turns out that the result is described by the theory
of a single free boson with background charge, as in equation \eqn{redeq}, thus showing
that a drastic reduction of dimensions takes place in target space, as in section 4.
Therefore, null gauging describes the Liouville field that will decouple at critical level
when $k$ is shifted to $k-2$ upon quantization, as before.

Furthermore, the null gauging can be equivalently described by boosting the
$SO(3)$ subgroup of $SO(3,1)$ and then take the infinite boost limit.
More precisely,
the boosted version of the non-abelian WZW coset $SO(3,1)/SO(3)$ is defined by
introducing new generators
\be
J_1(\beta) = e^{-\beta K_3} J_1 e^{\beta K_3} ~, ~~~~~
J_2(\beta) = e^{-\beta K_3} J_2 e^{\beta K_3} ~, ~~~~~
J_3(\beta) = J_3 ~, \label{boosta9}
\ee
which amount to considering $SO(3)_{\beta}$ with boost parameter
$\beta$. Of course, there is an
isomorphism $SO(3)_{\beta} \simeq SO(3)$ for all values of the deformation parameter
$\beta$ with the exception of the limiting value
$\beta \rightarrow \infty$, in which case a contraction
takes place to $E(2)$. The effective action of the vector gauged WZW model
$SO(3,1)/SO(3)_{\beta}$
can be computed in powers of ${\rm exp}(-2 \beta)$ in the large level and
large boost limit. Using appropriate redefinition of variables, as those shown
in reference \cite{iran3}, the result is the Liouville action \eqn{redeq}, plus
subleading terms of order ${\cal O}({\rm exp}(-2 \beta))$, which contain
two more additional fields that interact exponentially.

The transformation \eqn{boosta9} can be extended to the three remaining
generators by defining
\be
K_i(\beta) = e^{-\beta K_3} K_i e^{\beta K_3} ~.
\ee
In the limit $\beta \rightarrow \infty$, there is a contraction of $SO(3, 1)$
using the boost which is perpendicular to the spatial direction
associated to $K_3$.
Clearly, $K_3 (\beta) = K_3$ remains inert and it corresponds to the
radial coordinate that parametrizes the Liouville action. Thus, the infinite
boost limit selects the non-compact spatial direction $r$ which decouples from
the rest, as in the simplest $SL(2, R)$ case discussed in section 4.
The null gauging of WZW models was also considered in reference \cite{ctiradk}
but in a different way; in that case, the resulting background exhibits a
more general Toda-like structure and there is no dimensional reduction to only
one Liouville field.

The discussion can be repeated for all $SO(n,1)_k/SO(n)_k$ cosets
with higher values of $n \geq 4$. The exact form of the metric,
dilaton and anti-symmetric tensor fields can be worked out in
close form, case by case, following \cite{sfetsos1}. The resulting
expressions are quite lengthy and they are not included here, but
it turns out that they all yield tensionless models in the limit
$k \rightarrow n-1$. The decoupling of gravity manifests in the
form of a Liouville field, which is also described by gauging the
Euclidean group $E(n-1)$; equivalently, it follows by contracting
the maximal compact subgroup of $SO(n, 1)$ to $E(n-1)$ in the
infinite boost limit. This seems to be a universal result for all
maximally gauged WZW models based on $SO(n,1)_k$ current algebras,
\cite{iran3}. In all cases, one is left with a non-geometric
theory of $n-1$ bosons after the decoupling of gravity at critical
level. Since the effective description of these remnants is not
known, one hopes to gain insight into their quantum theory by
appealing to world-sheet methods, as for the simplest $SO(2,
1)/SO(2) \simeq SL(2, R)/SO(2)$ coset.

It is rather unfortunate
that the theory of non-abelian parafermions is still poorly understood beyond the
semiclassical limit $k \rightarrow \infty$, \cite{kurkut2}, and there are
no exact formulae available for them in the quantum regime. However, it is
conceivable that the exact representation of the basic parafermion currents
$\Psi^{\alpha}(z) \in SO(n, 1)/SO(n)$ can be obtained by other means when
the level assumes its critical value, depending on $n$. Experience with the
$n=2$ case suggests that the parafermions should be expressed in terms of
derivatives of (multi-component) fermions at critical level,
and their operator product expansion
should yield the corresponding $W$-generators as fermion bilinears. These
fermions could also be expressed as vertex operators of $n-1$ real scalars, via
bosonization, with the introduction of appropriate two-cocycle
(twist) factors. Furthermore,
the parafermions could be dressed by $SO(n)$ currents to provide a free field
realization of the $SO(n,1)_k$ current algebra at critical level. This is
reminiscent of the fermionic realization of current algebras $\hat{G}_k$ at special
values of $k$, using fermions in various representations of $G$ (see, for instance,
\cite{gko} and references therein); for example,
taking fermions in the adjoint
of a compact group $G$ yields representations of $\hat{G}_k$ with $k=g^{\vee}$
as fermion bilinears, whereas for orthogonal groups representations of level $1$
are constructed choosing fermions in the fundamental representation.
It should be emphasized, however, that the parafermionic representation is
different but it has not been spelled out in all generality for all
$k$ -- in particular for $k=g^{\vee}$, which is of interest here.
If this issue is resolved, we will be able to provide explicit construction
of the $W$-algebra that underlies all non-abelian coset models at critical level.
It might turn out that the relevant world-sheet symmetries are matrix generalizations
of $W_{\infty}$, and its higher spin truncations, which have been studied
elsewhere in different context, \cite{bak3, bak4}.

Finally, it will be interesting to consider more general coset models $G/H$,
where $H$ is not necessarily restricted to the maximal compact subgroup of $G$, and
extend the analysis of their quantum properties at critical level.

\section{Conclusions and discussion}
\setcounter{equation}{0}

We studied a certain class of tensionless models by considering gauged
WZW theories for non-compact groups $G_k$ at critical level, $k = g^{\vee}$, equal to the
dual Coxeter number of $G$. The WZW
models are unitary exact conformal field theories for all values of
$k > g^{\vee}$, and $k-g^{\vee}$ is naturally
identified with their tension parameter. The behavior of these theories is somewhat
singular at critical values of
$k$ because the central charge of the Virasoro algebra becomes
infinite, and it is appropriate to rescale the Virasoro generators in order to
make it finite. Then, in this limit, the Virasoro algebra contracts to an abelian
structure and conformal invariance is lost, thus leading to decoupling of gravity
from the spectrum. Apart from this incident, the resulting gauged WZW models make
perfect sense as two-dimensional quantum field theories, and
they exhibit infinite dimensional world-sheet symmetries generated by higher spin
fields. For $SL(2, R)_2/U(1)$ the complete structure of the underlying $W$-algebra was
determined and found to correspond to a higher spin truncation of $W_{\infty}$
by excluding the Virasoro generators.

Gauged WZW models also provide a testing ground for comparing different tensionless
limits that often appear in the literature. A natural question that arises in this
context is whether the tensionless limit of the quantum theory is equivalent to
the quantization of the classical tensionless strings on a given background.
Spaces that exhibit no $\alpha^{\prime}$ corrections, such as flat space or
pp-waves, provide exact
solutions to the beta-function equations to all orders in perturbation
theory, and they are expected to yield the same tensionless theory
in either case. However, for spaces that the
$\alpha^{\prime}$ corrections are substantial, as for the gauged WZW models, the
two methods need not be equivalent. The simplest example of this kind is
the coset $SL(2, R)_k/U(1)$ at $k=2$, which
defines the quantum tensionless limit of the two-dimensional black-hole
model as singular conformal field theory. In this case, the decoupling of
gravity manifests as decoupling of the Liouville field, and the operators of the
coset model, which include the basic parafermion currents, are faithfully realized
without it, in terms of the remaining boson. The same phenomenon can be seen
directly at the level of the $SL(2, R)_k$ current algebra, which is realized
in terms of two bosons at $k=2$, rather than using three bosons as for all
other values of $k>2$. Thus, there is no real remnant of the target space geometry
at $k=2$, although the space looks formally like an infinitely curved hyperboloid,
which is obtained by
including all $\alpha^{\prime}$ corrections in the perturbative beta function
equations and then letting $k-2 \sim 1/\alpha^{\prime} \rightarrow 0$.
In the same context, the null gauged WZW model $SL(2, R)_k/E(1)$
yields a Liouville field with infinite background charge at $k=2$, which
describes the gravitational sector that decouples in the tensionless limit.

The results of this study may be of more general value and provide
a way to resolve singularities that appear in the classical theory
of gravitation. In fact, the classical gravitational field becomes
very strong close to space-time singularities, where string
propagation behaves as tensionless theory. These singularities may
then be resolved within the complete quantum theory of strings,
when appropriately defined by including the effect of higher spin
massive states, by simply demanding the decoupling of gravity from
all other string states. This novel possibility seems to arise
quite naturally in the quantum tensionless limit of WZW models,
and it is very different in nature from any other classical
considerations of the problem. Thus, it appears that quantum
strings in strong gravitational fields will experience no gravity
after all. The result may be taken as indication that the
tensionless limit of quantum conformal field theories has a
topological character. However, it is not clear at this moment how
to utilize the infinitely many generators of the world-sheet
symmetries of the gauged WZW models in order to give a more
precise topological characterization of the resulting
two-dimensional, but non-conformal, quantum theories. There might
be similarities of these models with the theory of topological
orbifolds, which were introduced in the past in an attempt to gain
intuition about the unbroken phase of string theory in flat space
when $\alpha^{\prime} \rightarrow \infty$, \cite{ed2}. Work is in
progress towards this direction.

Apart from these general issues, there are a few open questions that could
be addressed directly in the simplest case of WZW models based on the
$SL(2, R)_k$ current algebra. There is an
enhancement of symmetries that follows from the behavior of the Kac-Kazhdan
determinant formula of the
$SL(2, R)_k$ current
algebra. The determinant vanishes at $k=2$ and it implies the existence of
several null states in the Verma module of the current algebra, \cite{kazh, peskin};
the same conclusion also applies to the
field theory of the coset model. It will be interesting to use them systematically
in order to provide a complete description of the corresponding WZW models at
critical level and clarify their interpretation. Also, one may take advantage of the
non-linear algebra $\hat{W}_{\infty}(k)$ that exists for all values
of $k$ in the coset model and extrapolates between the two linear algebras $W_{\infty}^{(3)}$
and $W_{\infty}$ at $k=2$ and $\infty$, respectively, \cite{hat}.
This algebra could be used as guide to
understand the $1/\alpha^{\prime}$ corrections away from the critical value
of $k$ using world-sheet methods. It may also offer a concrete framework to
understand the lifting of
degeneracies away from $k=2$ and explain the mechanism that introduces the coupling
to gravity. Needless to say, this is a rather important aspect of our
working framework, which should be
investigated separately in the future in great detail.

Generalizations to higher dimensional coset models $G/H$ based on higher rank
non-compact groups $G$ are also interesting to consider, but they are technically
more difficult to treat in detail. They share the essential features of
the $SL(2, R)_k/U(1)$ coset at critical level by the decoupling of gravity,
and possibly other fields that depend on the choice of $H$, as their Virasoro
central charge also becomes infinite when $k = g^{\vee}$. It is rather unfortunate,
however, that the quantum theory of non-abelian parafermions is less developed
than the abelian case, which prevents us to have explicit expressions in terms
of free fields and compute their operator product expansion at critical level.
Thus, we only have circumstantial evidence for their behavior in the ultra-quantum
limit, and as a result we have no explicit construction of the higher spin
$W$-algebras that arise on the world-sheet in the general case. Actually, it
might be possible to obtain an alternative definition of the parafermion
currents, which is only valid at critical values of $k$, without having to
go through their exact formulation for arbitrary values of $k$. In any case,
we think that the resulting $W$-algebras will contain $W_{\infty}^{(3)}$ among
their generators, and possibly many others that depend on the details of the particular
coset model. We also hope that the formal BRST analysis of the infinite
world-sheet symmetries at critical level, which was performed in detail
for the $SL(2,R)_2/U(1)$ model, thus rendering
$W_{\infty}^{(3)}$ free of anomalies, can be
made more systematic in future formulations of the problem, and it can be generalized
to groups of higher rank. The existing results indicate that there is a more general
framework at work, which underlies the systematic study of tensionless limit, but it
still remains unknown.

There are also other general problems that remain largely unexplored in the present work
and deserve special attention in the future. Here, we only
present some rough ideas that emerge from the necessity to find a more
fundamental definition of the tensionless limit and connect it with some
more traditional and better understood structures. The details will be studied
separately and appear elsewhere.

First, it will be interesting to apply the representation theory of
$W_{\infty}$-type algebras to the non-conformal symmetries that arise on the
world-sheet of the tensionless quantum models. We already know the general
theory of quasi-finite representations that consist of highest weight
representations with only a finite number of non-zero weights for all
higher spin operators, $W_0^s|h> = h_s |h>$, and $W_n^s |h> = 0$ for all
$n>0$, \cite{bak2, radul2, matsuo}.
It is quite interesting that all quasi-finite representations can be obtained
by free field realizations, \cite{bak2, matsuo},
as for the higher spin truncation of $W_{\infty}$ that arises
in gauged WZW models at critical level. As a result, the character formulae
are expected to be rather simple.
Such representations could also be used to assign higher spin charges to all quantum
states that become degenerate in the tensionless limit, and they
deserve further study in order to sharpen our current understanding of the
whole subject. In this framework, we might also be able to explore the topological
character of the quantum theory defined by the coset $SL(2, R)_k/U(1)$ at
critical level, and its generalizations thereof.

Second, there are some intriguing mathematical constructions based on Langlands
duality for current algebras that relate small with large $\alpha^{\prime}$
expansions on the corresponding dual faces, since $(k-g^{\vee})^{-1}$ is
replaced by $\tilde{k} - \tilde{g}^{\vee}$ by duality, \cite{frenkel1, frenkel2}.
This duality, which is essentially a current
algebra generalization of the usual electric-magnetic duality for the zero mode
(global) algebras, \cite{nuyts},
should be properly understood in WZW models in order to
reformulate the tensionless limit in more accessible terms. Thus, electric and
magnetic charges for loop groups, which are naturally defined for a certain class
of periodic instanton configurations in four dimensions, \cite{murray},
might turn out to play a very important role in future constructions towards a dual
formulation of the problem.
It should be emphasized, however, that we are
only interested in the
case of non-compact groups, as the compact models exhibit no tensionless limit.
In any
case, we expect that these methods can be applied directly to gauged WZW models
and lead to a new non-perturbative formulation, where the tensionless limit
can be studied more systematically in all generality.
Also, in this context, it will be interesting to understand the relation between
the world-sheet symmetries of the dual models, when appropriately defined
by Langlands duality, and generalize earlier mathematical work on the subject,
\cite{frenkel1, frenkel2}. The results should
be able to explain why the black-hole coset $SL(2, R)_k/U(1)$ exhibits
a symmetry of $W_{\infty}$ type in both limits, $k \rightarrow 2$ and $k \rightarrow \infty$,
modulo the Virasoro algebra.

Third, the possible connections with non-commutative geometry
should be put on a firm basis following the general ideas outlined
in reference \cite{jurg}. In particular, WZW models at large
values of the level $k$ are naturally related to the usual
commutative geometry in target space, whereas $\alpha^{\prime}
\sim 1/k$ corrections may be viewed as inducing quantum
deformations that lead to non-commutative structures. For
non-compact groups $G_k$ there is a critical value of the level,
$k = g^{\vee}$, which resembles the infinite non-commutativity
limit, and therefore it becomes tractable in many respects. In
this case the notion of space-time becomes very fuzzy, as points
become totally delocalized, and a new formalism is required to
make sense of the underlying structures from a more fundamental
viewpoint. Experience with other non-commutative field theories
might prove useful in this respect, and the structure of the
$SL(2, R)_k/U(1)$ model at critical level might resemble the
quantum Hall effect in the infinite non-commutativity limit, when
the strength of the external magnetic field becomes infinite and
the Hamiltonian vanishes; recall in this case that the Landau
levels become degenerate with zero energy, as the energy levels of
the system are proportional to the fundamental frequency $\omega
\sim 1/B$, which tends to zero. In any case, we expect that
non-commutative geometry could be used further in order to provide
an intrinsic definition of the tensionless limit of quantum string
theory. Thus, the physics of very strong gravitational fields and
the meaning of space-time singularities should be revisited in
this context.

In conclusion, the tensionless limit of two-dimensional conformal field theories,
and their relevance to the ultimate formulation of string theory beyond the
effective field theory description, are interesting problems that deserve
better attention. The present work indicates that many surprising things
emerge along this path, and new ideas are certainly required in order to put forward
some of the results in a systematic way. Also, there might be other tensionless models
which are defined by different methods, without ever arising as limiting cases
of two-dimensional conformal field theories. The universality of this limit and
its fundamental definition in a background independent way remain open problems.

\vskip1cm

\centerline{\bf Acknowledgments}
\no
This work was supported in part by the European Research and Training Networks
``Superstring Theory" (HPRN-CT-2000-00122) and ``Quantum Structure of Space-time"
(HPRN-CT-2000-00131), as well as the Greek State Foundation Award
``Quantum Fields and Strings" (IKYDA-2001-22) and NATO Collaborative
Linkage Grant ``Algebraic and Geometric Aspects of Conformal Field Theories
and Superstrings" (PST.CLG.978785). One of us (C.S.) is also thankful to the research
committee of the University of Patras for
a graduate student fellowship ``C. Caratheodory"
under contract number 2453; he also wishes to acknowledge support of the
Japanese Ministry of Education, Culture, Sports, Science and Technology
(Monbukagakusho) during his stay in Osaka, where this work was completed.
Finally, we thank C. Bachas, C. Kounnas, U. Lindstrom,
G. Savvidy and K. Sfetsos for useful discussions.

\newpage

\centerline{\bf Note added}

It was pointed out by the referee (and other colleagues) that one should rather expect 
an alternative interpretation of the tensionless limit as a gauge theory of higher 
spins with huge gauge symmetry and that any consistent theory of massless 
higher spin fields should also involve gravity. This picture emerges at the free level in 
Minkowski space (see, for instance, \cite{augo1}, and references therein), and it 
has been further extended to (A)dS backgrounds using the BRST formalism, 
as in reference \cite{augo2}. Thus, if such picture persists in all tensionless models
it will raise a puzzle that certainly calls for attention in connection with our results.

We do not have a definite answer to offer at this moment but only a few general 
comments that indicate the differences with other works and the means of 
investigation. First, it should be noted that the construction of $W_{\infty}^{(3)}$ 
proves the existence of higher spin operator algebras in two dimensions without 
Virasoro generators. This result might be specific to two-dimensional world-sheet 
symmetries, but this is precisely were most of our work is confined. Of course, 
we also have a contracted Virasoro algebra, as in flat space, which can be used at
will, but it is completely decoupled from the remaining symmetries of our model.
Second, we do not have a target space interpretation of the $W_{\infty}^{(3)}$ 
symmetry in order to explore directly the connection with the gauge theory of 
higher spin fields and their space-time field equations, if appropriate, as in 
references \cite{augo1, augo2}. 

However, using the BRST formalism for the algebra $W_{\infty}^{(3)}$ it is 
possible to construct Lagrangians of the form 
${\cal L} \sim <\Phi | {\cal Q} | \Phi >$ and address the problem in all 
generality, but the results can be rather involved. Yet, it may prove instructive 
to compare the results with the similar construction based on BRST formalism for 
the contracted Virasoro algebra, as it is usually done in flat space for the 
higher spin triplets. Such comparison may very well separate the topological from 
the gauge theory aspects of higher spin fields in the tensionless limit of gauged 
WZW models and provide us with a definite answer. On the technical side, the 
role of the many null states that arise in the representation theory of non-compact 
current algebras at critical level also needs to be understood in this context in 
order to build tensors of general type. They rest on special properties of the 
$SL(2, R)$ current algebra at $k=2$, which are not shared by the oscillators 
at generic level. A few steps were already taken by other authors, \cite{ulf3}, in 
an attempt to reproduce the analogue of Fronsdal's conditions at critical level,
but the results are still inconclusive: a huge gauge symmetry makes its appearance,
as in flat space, but the transversality condition cannot be obtained from the 
contracted Virasoro constraints.

The above issues constitute open problems for future work and indicate that 
tensionless WZW models have additional features which are not shared by other 
models. They should be compared with other instances that involve topological 
modes of higher spin fields, as in two-dimensional string theory.

\end{document}